\documentclass[11pt]{article}
\pdfoutput=1
\usepackage{epsfig}
\usepackage{latexsym}
\usepackage{graphicx}
\usepackage{color}
\usepackage{amsmath,amssymb}
\usepackage{cite}
\usepackage{cancel}

\usepackage{afterpage}
\usepackage{framed}
\usepackage{multirow}
\usepackage{seqsplit}
\usepackage{enumitem} 

\usepackage{amsmath, amsthm, amsfonts, amssymb, latexsym}

\usepackage{upgreek}
\usepackage[caption=false]{subfig}

\usepackage{url}
\usepackage{ifpdf}

\def\be{\begin{equation}}
\def\ee{\end{equation}}
\def\ba{\begin{eqnarray}}
\def\ea{\end{eqnarray}}

\def\ltap{\;\centeron{\raise.35ex\hbox{$<$}}{\lower.65ex\hbox{$\sim$}}\;}
\def\gtap{\;\centeron{\raise.35ex\hbox{$>$}}{\lower.65ex\hbox{$\sim$}}\;}

\newcommand{\bea}{\begin{eqnarray}}
\newcommand{\eea}{\end{eqnarray}}

\makeatletter
\def\section{\@startsection {section}{1}{\z@}{-3.5ex plus -1ex minus -.2ex}{2.3ex plus .2ex}{\large\bf}}
\def\subsection{\@startsection{subsection}{2}{\z@}{-3.25ex plus -1ex
minus -.2ex}{1.5ex plus .2ex}{\normalsize\bf}}
\makeatother
\newcommand{\captionfonts}{\small}
\makeatletter  % Allow the use of @ in command names
\long\def\@makecaption#1#2{%
  \vskip\abovecaptionskip
  \sbox\@tempboxa{{\captionfonts #1: #2}}%
  \ifdim \wd\@tempboxa >\hsize
    {\captionfonts #1: #2\par}
  \else
    \hbox to\hsize{\hfil\box\@tempboxa\hfil}%
  \fi
  \vskip\belowcaptionskip}
\makeatother   % Cancel the effect of \makeatletter
\catcode`@=11
\def\marginnote#1{}
\newcount\hour
\newcount\minute
\newtoks\amorpm
\hour=\time\divide\hour
by60
\minute=\time{\multiply\hour by60 \global\advance\minute
by-\hour}
\edef\standardtime{{\ifnum\hour<12 \global\amorpm={am}
\else\global\amorpm={pm}\advance\hour by-12 \fi
 \ifnum\hour=0
\hour=12 \fi
 \number\hour:\ifnum\minute<10
0\fi\number\minute\the\amorpm}}
\edef\militarytime{\number\hour:\ifnum\minute<10
0\fi\number\minute}
\def\draftlabel#1{{\@bsphack\if@filesw
{\let\thepage\relax
 \xdef\@gtempa{\write\@auxout{\string
\newlabel{#1}{{\@currentlabel}{\thepage}}}}}\@gtempa
 \if@nobreak
\ifvmode\nobreak\fi\fi\fi\@esphack}
\gdef\@eqnlabel{#1}}
\def\@eqnlabel{}
\def\@vacuum{}
\def\draftmarginnote#1{\marginpar{\raggedright\scriptsize\tt#1}}
\def\draft{\oddsidemargin
0.0truein
 \def\@oddfoot{\sl preliminary draft \hfil
\rm\thepage\hfil\sl\today\quad\militarytime}
 \let\@evenfoot\@oddfoot
\overfullrule 3pt
 \let\label=\draftlabel
\let\marginnote=\draftmarginnote
\def\@eqnnum{(\theequation)\rlap{\kern\marginparsep\tt\@eqnlabel}
\global\let\@eqnlabel\@vacuum}
}
\catcode`@=12
\newcommand{\beq}{\begin{eqnarray}}
\newcommand{\eeq}{\end{eqnarray}}

\newcommand{\gsim}{\raisebox{-0.13cm}{~\shortstack{$>$ \\[-0.07cm]
     $\sim$}}~}

\begin{document}

\thispagestyle{empty}

\begin{center}

\begin{center}

{\LARGE \bf \mbox{Singlet $\!$Extensions of the Standard $\!$Model at
  $\!$LHC Run~$\!$2:}\\[0.2cm]
Benchmarks and Comparison with the NMSSM} \\ [0.2cm]
\end{center}
\vspace{1.4cm}

\renewcommand{\thefootnote}{\fnsymbol{footnote}}
{\bf Raul Costa$^{\,1,\,2\,}$}\footnote{E-mail: \texttt{rauljcosta@ua.pt}},
{\bf Margarete M\"uhlleitner$^{\,3}\,$}\footnote{E-mail: \texttt{margarete.muehlleitner@kit.edu}},
{\bf Marco O. P. Sampaio$^{\,2,\,4\,}$}\footnote{E-mail:
  \texttt{msampaio@ua.pt}} \\ and
{\bf Rui Santos$^{\,1,\,5\,}$}\footnote{E-mail: \texttt{rasantos@fc.ul.pt}}
\\

\vspace{1.cm}

${}^1\!\!$
{\em Centro de F\'{\i}sica Te\'{o}rica e Computacional,
    Faculdade de Ci\^{e}ncias,
    Universidade de Lisboa,} \\
{\em Campo Grande, Edif\'{\i}cio C8 1749-016 Lisboa, Portugal}
\\
${}^2\!\!$
{\em Departamento de F\'\i sica da Universidade de Aveiro,} \\
{\em Campus de Santiago, 3810-183 Aveiro, Portugal}
\\
${}^3\!\!$
{\em Institute for Theoretical Physics, Karlsruhe Institute of
    Technology,} \\
{\em 76128 Karlsruhe, Germany}
\\
${}^4\!\!$
{\em CIDMA - Center for Research $\&$ Development in Mathematics and Applications,} \\
{\em Campus de Santiago, 3810-183 Aveiro, Portugal}
\\
${}^5\!\!$
{\em {ISEL - 
 Instituto Superior de Engenharia de Lisboa,\\
 Instituto Polit\'ecnico de Lisboa 
 1959-007 Lisboa, Portugal}
}\\

\end{center}

\begin{abstract}
The Complex singlet extension of the Standard Model (CxSM) is the
simplest extension that provides scenarios for Higgs pair production
with different masses. The model has two interesting phases: the
dark matter phase, with a Standard Model-like Higgs boson, a new scalar and a
dark matter candidate; and the broken phase, with all three neutral
scalars mixing. In the latter phase Higgs decays into a pair of two
different Higgs bosons are possible.

In this study we analyse Higgs-to-Higgs decays in the framework of
singlet extensions of the Standard Model (SM), with focus on the
CxSM. After demonstrating that scenarios with large rates for such chain decays
are possible we perform a comparison between the NMSSM and the
CxSM. We find that, based on Higgs-to-Higgs decays, the only
possibility to distinguish the two models at the LHC run~2 is through
final states with two different scalars. This conclusion builds a
strong case for searches for final states with two different scalars
at the LHC run~2.

Finally, we propose a set of benchmark points for the real and complex
singlet extensions to be tested at the LHC run~2. They
have been chosen such that the discovery prospects of the involved
scalars are maximised and they fulfil the dark matter
constraints. Furthermore, for some of the points the theory is stable up
to high energy scales. For the computation of the decay widths and
branching ratios we developed the Fortran code {\tt sHDECAY}, which is
based on the implementation of the real and complex singlet extensions
of the SM in {\tt HDECAY}.

\end{abstract}    
\newpage
  
%%%%%%%%%%%%%%%%%%%%%%%%%%%%%%%%%%%%%%%%%%%%%%%%%%%%%%%
\section{\label{sec:intro} Introduction}

The ATLAS~\cite{Aad:2012tfa} and CMS~\cite{Chatrchyan:2012ufa} collaborations at 
the Large Hadron Collider (LHC) have established the existence of a new scalar 
particle identified with the Higgs boson. The investigations are now focused on 
pinning down the pattern of electroweak symmetry breaking (EWSB), which
already started 
with the measurements of the Higgs couplings 
and will
continue with the search for physical processes that could give us
some direct insight on the structure of the potential. These are the
processes involving triple scalar vertices 
\cite{Djouadi:1999rca,Muhlleitner:2000jj,Baglio:2012np} from which the
resonant ones are the most promising to lead to a detection during the
next LHC runs.  
So far no such resonant decay has been observed. Both the ATLAS and
the CMS collaboration have searched for double Higgs final states
produced by a narrow resonance 
decaying into $b \bar b b \bar b$~\cite{Khachatryan:2015yea,ATLAS:2014rxa}
and $\gamma \gamma b \bar b$~\cite{Aad:2014yja,CMS:2014ipa}. These and
other final states in double Higgs production, will be one of the
priorities in Higgs physics during the LHC run 2. Furthermore, the
Higgs couplings extracted so far 
from the 7 and 8 TeV data show no large deviations from the Standard
Model (SM) expectations and point towards a very SM-like Higgs sector.

In the SM, the double Higgs production cross section is rather small and
even at the high luminosity stage of the LHC it will be extremely
challenging to measure the triple Higgs coupling. However, in many
beyond the SM (BSM) scenarios, resonant decays are possible due to the
existence of other scalar particles in the theory. This increases the
prospects for observing a final state with two scalars. The simplest
extension where a di-Higgs final state would be detectable
at the LHC is the singlet extension of the SM, where a hypercharge
zero singlet is added to the scalar field content of the model. When
the singlet is real one either obtains a new scalar mixing with the
Higgs boson or the minimal model for dark
matter~\cite{Silveira:1985rk, McDonald:1993ex, Burgess:2000yq, Bento:2000ah,Davoudiasl:2004be,Kusenko:2006rh, vanderBij:2006ne, He:2008qm,Gonderinger:2009jp, Mambrini:2011ik, He:2011gc, Gonderinger:2012rd,Cline:2013gha, Gabrielli:2013hma,Robens:2015gla,Profumo:2014opa}. The model can also accommodate
electroweak baryogenesis by allowing a strong first-order phase
transition during the era of EWSB~\cite{Menon:2004wv, Huber:2006wf,  
  Profumo:2007wc,Barger:2011vm, Espinosa:2011ax}, if the singlet is
complex. In this case the spectrum features three Higgs bosons,
either all visible or one
being associated with dark matter. In the present work we will focus mainly
on the complex version of the singlet extension although we will also discuss
the case of a real singlet. The collider phenomenology of the singlet
model, which can lead to some distinctive signatures at the LHC, has been previously
discussed in~\cite{Datta:1997fx, Schabinger:2005ei,
  BahatTreidel:2006kx, Barger:2006sk, Barger:2007im, Barger:2008jx,
  O'Connell:2006wi,Gupta:2011gd, Ahriche:2013vqa, Chen:2014ask}.    
 
Both the real and the complex singlet extensions, in the minimal
versions that we will discuss, can have at least two phases. Namely, a
$\mathbb{Z}_2$-symmetric phase with a dark matter candidate and a
broken phase where the singlet component(s) mix with the neutral
scalar field of the SM doublet. Therefore these models can be used as
simple benchmarks for resonant double Higgs production $pp \to H \to hh$, where $H$ generically denotes a new heavy scalar and $h$ the
SM-like Higgs boson. Furthermore, the broken phase of the complex model
also allows for a decay of a heavy scalar $h_3$ into two other scalars
$h_1$, $h_2$ of different mass, $h_3 \to h_2 + h_1$. This particular 
decay is common in other extensions of the SM, such as the Complex
Two-Higgs Doublet Model (C2HDM) and the Next-to-Minimal Supersymmetric
extension of the SM
(NMSSM)~\cite{Fayet:1974pd,Barbieri:1982eh,Dine:1981rt,Nilles:1982dy,Frere:1983ag,Derendinger:1983bz,Ellis:1988er,Drees:1988fc,Ellwanger:1993xa,Ellwanger:1995ru,Ellwanger:1996gw,Elliott:1994ht,King:1995vk,Franke:1995tc,Maniatis:2009re,Ellwanger:2009dp}.  
In the NMSSM a complex superfield $\hat{S}$ is added to the Minimal
Supersymmetric Model field content allowing for a dynamical solution of the $\mu$
problem when the singlet field acquires a non-vanishing vacuum
expectation value (VEV).  
NMSSM Higgs sectors that feature light Higgs states in the spectrum
can lead to sizeable decay widths for Higgs decays into a pair of
lighter Higgs bosons. Higgs-to-Higgs decays in the NMSSM have been
studied in 
\cite{Nhung:2013lpa,Munir:2013dya,King:2014xwa,Wu:2015nba,Buttazzo:2015bka,Muhlleitner:2015dua}
and several recent studies have investigated the production of Higgs
pairs in the NMSSM 
\cite{Nhung:2013lpa,Ellwanger:2013ova,Han:2013sga,King:2014xwa,Cao:2014kya}. These
processes involve the trilinear Higgs self-coupling, which is known in
the NMSSM including the full one-loop corrections
\cite{Nhung:2013lpa}  and the two-loop ${\cal O}(\alpha_t \alpha_s)$ corrections
\cite{Muhlleitner:2015dua} both in the real and in the complex NMSSM.
A recent study within the C2HDM on all possible CP-violating scalar decays and the
chances of detecting direct CP violation at the LHC can be found
in~\cite{Fontes:2015xva}.  It is interesting to point out that decays
of the type $h_i \to h_j Z$ are forbidden 
in the singlet extension because the model is CP-conserving \textit{and} has
no CP-odd states. In the real NMSSM (and also in the CP-conserving
2HDM) the decay is possible provided CP($h_i$) $= -$ CP($h_j$) while
in CP-violating models, like the C2HDM and the complex NMSSM, $h_i \to
h_j Z$ is always possible if kinematically allowed. Note that combinations 
of scalar decays into pairs of scalars together with $h_i \to h_j Z$ allow us not only
to probe the CP nature of the model but can also be used as a tool to
identify the CP quantum numbers of the scalars of the
theory~\cite{Fontes:2015xva}. Finally there is an interesting scenario
where all three scalars $h_i$ are detected in $h_i \to ZZ$. If the potential
is CP-conserving then all three scalars are CP-even. This
is exactly what happens in the CxSM and in the NMSSM
but not in other simple extension of the the SM like the 2HDM\footnote{Note
however that if CP-violation is introduced via the Yukawa Lagrangian,
the pseudoscalar can decay to $ZZ$ although in principle with a very small rate 
as shown in~\cite{Arhrib:2006rx} for a particular realisation of the 2HDM.}.

In order to calculate the branching ratios of the singlet models we
have modified {\tt HDECAY}~\cite{Djouadi:1997yw,Butterworth:2010ym} 
to include the new vertices. This new tool, {\tt sHDECAY}, can be used 
to calculate all branching ratios and widths in both the real and the complex
singlet extensions of the SM that we discuss. In each extension, both the symmetric phase and the
broken phase were considered leading to a total of four models. The detailed 
description of {\tt sHDECAY} can be found in
appendix~\ref{sec:app_SHDECAY}\footnote{The program {\tt sHDECAY} can
  be downloaded from http://www.itp.kit.edu/$\thicksim$maggie/sHDECAY/.}.     

In this study we define benchmark points for the LHC run~2 both for
the complex singlet model (CxSM) and the real singlet model (RxSM). We
are guided by phenomenological and theoretical goals. The first is to
maximize at least one of the Higgs-to-Higgs decays. 
The second is related to the stability of the theory at higher orders.
As discussed in~\cite{Costa:2014qga}, the radiative stability of the
model, at two loops, up to a high-energy scale, combined with the
constraint that the 125 GeV Higgs boson found at the LHC is in the
spectrum, forces the new scalar to be heavy. When we include all
experimental constraints and measurements from collider
data, dark matter direct detection experiments and from the Planck
satellite and in addition force stability at least up to the Grand Unification Theory (GUT) scale, we find that the lower bound on the new scalar is about 170
GeV~\cite{Costa:2014qga}. Finally, whenever a dark matter candidate
exists we force it to fully saturate the dark matter relic density
measured by Planck as well as to make it consistent with the latest
direct detection bounds.   

The renormalisation of the real singlet model has recently been addressed 
in~\cite{Kanemura:2015fra, Bojarski:2015kra}. In~\cite{Kanemura:2015fra}
the focus was on the modifications of the couplings of the SM-like Higgs
to fermions and gauge bosons while in~\cite{Bojarski:2015kra}
the goal was to determine the electroweak
corrections to the heavy Higgs decay, $H$, into the two light
ones, $h$. In the former
it was found that the electroweak corrections are at most of the order of 1 \%
and this maximal value is attained in the limit where the theory
decouples and becomes  indistinguishable from the SM.
In the latter it was found that the corrections to the triple scalar vertex ($Hhh$)
are small, typically of a few percent, once all theoretical and
experimental constraints 
are taken into account. Furthermore it was concluded that electroweak corrections
are stable for changes of the renormalisation scheme. There are so far no
calculations of electroweak corrections in the complex singlet extension of the SM.

We end the introduction with a word of caution regarding interference
effects in the real singlet model. The problem of interference in
BSM scenarios has been raised in~\cite{Maina:2015ela} for
the particular case of a real singlet extension of the SM. As shown
in~\cite{Maina:2015ela}, the SM-like Higgs contribution to the process
$gg \to h^*, H^{(*)} \to ZZ \to 4l$ is non-negligible outside the non-SM
scalar ($H$) peak region and interference effects can be quite
large. These effects were shown to range from
${\cal O}$(10\%) to ${\cal O}$(1) for the integrated cross sections at the LHC at a
center-of-mass energy of 8 TeV~\cite{Kauer:2015hia}. Moreover, these
effects also modify the line shape of the heavier scalar. However, as shown 
in~\cite{Kauer:2015hia} judicious kinematical cuts can be used in the
analysis to reduce the interference  effects to ${\cal O}$(10\%). 
Recently, interference effects were studied in the process
$gg \to h^*, H^{(*)} \to hh$ at next-to-leading order (NLO) in 
QCD~\cite{Dawson:2015haa}. It was found that the interference effects
distort the double Higgs invariant mass distributions. Depending on
the heavier Higgs mass value, they can either decrease by 30\% or increase
by 20\%. Further, it was shown that the NLO QCD corrections are
large and can significantly distort kinematic distributions near
the resonance peak. This means that any experimental analysis to be performed in
the future should take these effects into account. 

The paper is organized as follows. In Sect.~\ref{sec:models} we
define the singlet models, CxSM in Sect.~\ref{sec:CxSM} and RxSM in
Sect.~\ref{sec:RxSM}, and we describe, in Sect.~\ref{sec:Cons}, the
constraints that are 
applied. In the experimental constraints particular emphasis is put on
the LHC~run~1 data. The NMSSM is briefly introduced in Sect.~\ref{sec:nmssmintro}
together with the constraints that are used in the scan. 
In Sect.~\ref{sec:res} we present our numerical analysis and results. First, the
possibility of SM Higgs boson production from chain decays in the CxSM
and RxSM after run~1 is discussed in \ref{sec:chainprod}. Afterwards
Higgs-to-Higgs decays at run~2 are investigated in \ref{Sec:LHC_Run2}
and the related features of the RxSM are compared to those of the CxSM
in Sect.~\ref{sec:comprxsmcxsm} before moving on to the comparison of
the CxSM with the NMSSM, which is performed in
Sect.~\ref{sec:cxsmnmssmcomp}. Finally in
Sect.~\ref{Sec:LHC_Run2_benchs} benchmark points are  
presented for the CxSM and the RxSM for the LHC run~2.
Our conclusions are given in Section~\ref{sec:concl}.
In appendix~\ref{sec:app_SHDECAY} the implementation of the singlet
models in {\tt sHDECAY} is described and sample input and output files are
given. In appendix~\ref{sec:app_FeynmanRules} we list 
the expressions for the scalar vertices in the singlet extensions used
in the scalar decays into pairs of scalars.

%%%%%%%%%%%%%%%%%%%%%%%%%%%%%%%%%%%%%%%%%%%%%%%%%%%%%%%%%
\section{Models and Applied Constraints} \label{sec:models}

In this section we define the reference complex and real 
singlet models. These will be analysed to define benchmarks for the
various allowed kinematical situations at the LHC run~2, with
a center-of-mass (c.m.)~energy  of 13~TeV. We review how EWSB proceeds
consistently with the observed Higgs boson and define the couplings 
of the scalars of each theory with the SM particles as well as within the scalar
sector. We then describe the various constraints that we apply to these singlet
models. Subsequently the NMSSM is briefly introduced, and the ranges
of the parameters used in the NMSSM scan are described, together with
the applied constraints. 
%%%%%%%%%%%%%%%%%%%%%%%%%%%%%%%%%%%%%%%%%%%%%%%%%%%%%%%%%%%
\subsection{The CxSM}
\label{sec:CxSM} 

The main model discussed is this work is a simple extension of the SM 
where a complex singlet field 
\beq
\mathbb{S}=S+i A \;,
\eeq 
with hypercharge zero, is added do the SM field content. All
interactions are determined by the scalar potential, which can be seen
as a model with a  
$U(1)$ global symmetry that is broken softly. The most
general renormalisable scalar potential, with soft breaking terms with mass dimension up to two, is given by
\begin{multline}
V_{\rm CxSM}=\dfrac{m^2}{2}H^\dagger H+\dfrac{\lambda}{4}(H^\dagger H)^2+\dfrac{\delta_2}{2}H^\dagger H |\mathbb{S}|^2+\dfrac{b_2}{2}|\mathbb{S}|^2+
\dfrac{d_2}{4}|\mathbb{S}|^4+\left(\dfrac{b_1}{4}\mathbb{S}^2+a_1\mathbb{S}+c.c.\right)
\, ,  \label{eq:V_CxSM} 
\end{multline}
where the soft breaking terms are shown in parenthesis and the doublet
and complex singlet are, respectively, 
\begin{equation}\label{eq:vacua_CxSM}
H=\dfrac{1}{\sqrt{2}}\left(\begin{array}{c} G^+ \\
    v+h+iG^0\end{array}\right) \quad \mbox{and} \quad
\mathbb{S}=\dfrac{1}{\sqrt{2}}\left[v_S+s+i(v_A+ a)\right] \;.
\end{equation}
Here $v\approx 246\;\mathrm{GeV}$ is the SM Higgs VEV, and $v_S$ and $v_A$
are, respectively, the real and imaginary parts of the complex singlet field VEV.  

In~\cite{Coimbra:2013qq}, the various phases of the model were
discussed. In order to classify them it is convenient to treat the
real and the imaginary components of the complex singlet as
independent, which is equivalent to building a model with two real
singlet fields. We focus on a version of the model that is obtained
by requiring a $\mathbb{Z}_2$ symmetry for the imaginary component $A$
(this is equivalent to imposing a symmetry under
$\mathbb{S}\rightarrow \mathbb{S}^*$).  
As a consequence of this symmetry, the soft breaking couplings must be
both real, i.e. $a_1\in\mathbb{R}$ and $b_1\in\mathbb{R}$. Observe
that $m, \lambda, \delta_2, b_2$ and $d_2$ must be real parameters for
the potential to be real. 

By analysing the minimum conditions one finds two possible phases that
are consistent with EWSB triggering the Higgs mechanism. They are:  
\begin{itemize}
\item $v_A=0$ and $v_S\neq 0$, in which case mixing between the doublet
field $h$ and the real component $s$ of the singlet field occurs, while
the  imaginary component $A\equiv a$ becomes a dark matter candidate. 
We call this the symmetric or \textit{dark matter phase}.

\item $v_S\neq 0~{\rm and}~ v_A\neq 0$, 
which we call the \textit{broken phase}, with no dark matter
candidate and mixing among all scalars. 
\end{itemize}
The model phases are summarized in table~\ref{tab:phases}.
\begin{table}[h!]
\begin{center}
\begin{tabular}{||  c || c | c ||}
\hline			
  Phase & Scalar content & VEVs at global minimum   \\
\hline 
\hline		
Symmetric (dark) &   2 mixed + 1  dark &  $\left<S\right>\neq 0$ and
$\left<A\right>=0$ \\ 
\hline 
Broken ($\cancel{\mathbb{Z}}_2$ ) & 3 mixed & 
$\left<S\right>\neq 0$ and
$\left<A\right>\neq 0$\\
\hline  
\end{tabular}
\end{center}
\caption{Phase classification for the version of the CxSM 
with the $\mathbb{Z}_2$ symmetry on the imaginary component of the
singlet, $A \to -A$.} 
\label{tab:phases}
\end{table}

As discussed in~\cite{Coimbra:2013qq}, simpler models can be obtained 
with the same field content by imposing extra symmetries
on the potential. The exact $U(1)$-symmetric potential has $a_1=b_1=0$, leading
to either one or to two dark 
matter candidates depending on the pattern of symmetry breaking. One
can also impose     
a separate $\mathbb{Z}_2$ symmetry for $S$ and $A$ that forces $a_1=0$ 
and $b_1\in \mathbb{R}$ and, again, gives rise to the possibility of
having one or two dark matter candidates. 
However, from the phenomenological point of view, the model presented
here covers all possible scenarios in terms of the accessible physical
processes to be probed at the LHC in a model with three scalars. In
fact, even simpler models, like the real singlet extension of the SM
discussed below, have some similarities with this version of the CxSM
in what concerns the planned searches for the next LHC runs. However,
typically, the allowed parameter space of  different models can also
be quite different, implying that the number of expected events may
allow the exclusion of one model but not of another qualitatively
similar one. 

%%%%%%%%%%%%%%%%%%%%%%%%%%%%%%%%%%%%%%%%%%%%%%%%%%%%%%%%%%
\subsubsection*{Physical states and couplings}
To obtain the couplings of the scalars to the SM particles, we define the mass
eigenstates as  $h_i$ ($i=1,2,3$). They are obtained from the gauge  
eigenstates $h,s$ and $a$ through the mixing matrix~$R$
\be
\left(
\begin{array}{c}
h_1\\
h_2\\
h_3
\end{array}
\right)
= R
\left(
\begin{array}{c}
h\\
s\\
a
\end{array}
\right) \, ,
\label{h_as_eta}
\ee
with
\be
R\, {\cal M}^2\, R^T = \textrm{diag} \left(m_1^2, m_2^2, m_3^2 \right),
\ee
and $m_1 \leq m_2 \leq m_3$ denoting the masses of the neutral Higgs particles.
The mixing matrix $R$ is parametrized as
\be
R =
\left(
\begin{array}{ccc}
c_1 c_2 & s_1 c_2 & s_2\\
-(c_1 s_2 s_3 + s_1 c_3) & c_1 c_3 - s_1 s_2 s_3  & c_2 s_3\\
- c_1 s_2 c_3 + s_1 s_3 & -(c_1 s_3 + s_1 s_2 c_3) & c_2 c_3
\end{array}
\right)
\label{matrixR}
\ee
with $s_i \equiv \sin{\alpha_i}$ and
$c_i \equiv \cos{\alpha_i}$ ($i = 1, 2, 3$) and
\be
- \pi/2 < \alpha_i \leq \pi/2 \;.
\label{range_alpha}
\ee
In the dark matter phase, $\alpha_2 =\alpha_3 = 0$, the $a$ field coincides with $A$ and it is the dark
matter candidate, which does not mix with $h$ nor with $s$.  

The couplings of $h_i$ to the SM particles are all modified by the
same matrix element $R_{i1}$. This means that, for any SM coupling
$\lambda_{h_{SM}}^{(p)}$ where $p$ runs over all SM fermions and  
gauge bosons, the corresponding coupling in the singlet model for the
scalar $h_i$ is given by  
\begin{equation}\label{eq:couplings_SM}
\lambda_{i}^{(p)}=R_{i1}\lambda_{h_{SM}}^{(p)}\; .
\end{equation}
The coupling modification factor is hence independent of the specific
SM particle to which the coupling corresponds.

In the dark matter phase the same arguments apply with $R_{i1} =
(R_{11} , R_{21}, 0)$. The state $i=3$ then corresponds to the dark
matter candidate, which does not couple to any of the SM fermions and
gauge bosons. 

The couplings that remain to be defined are the ones of the scalar sector.
They are read directly from the scalar potential,
Eq.~\eqref{eq:V_CxSM}, after replacing the expansion about the vacua,
Eq.~\eqref{eq:vacua_CxSM}. For the purpose of our analysis we only
need the scalar triple couplings. We define them according to the
following normalisation ($i,j,k=1,2,3$):
\begin{equation}\label{eq:cubic_norm} 
V_{h_{\rm cubic}}=\frac{1}{3!}g_{ijk}h_{i}h_{j}h_{k} \; .
\end{equation}
These couplings determine, up to the phase space factor, the leading
order expressions for the Higgs-to-Higgs decay widths. In
appendix~\ref{sec:selfCxSM} we provide the
expressions for these couplings as well as for the quartic couplings, for completeness. 

%%%%%%%%%%%%%%%%%%%%%%%%%%%%%%%%%%%%%%%%%%%%%%%%%%%%%%%%%%
\subsubsection*{Parameters of the model}
In Sect.~\ref{sec:res} we will present scans over the parameter space
of this model and we will identify benchmark points that represent
various qualitatively different regions in this space. Regardless of
the phase of the model under consideration, the model always has seven
parameters. The particular choice of independent parameters
is just a matter of convenience.  
For the broken phase we choose the set
$\{\alpha_1,\alpha_2,\alpha_3,v,v_S,m_1,m_3\}$ and for the dark phase
the set $\{\alpha_1,v,v_S,a_1,m_1,m_2, m_3\equiv m_A\}$. 
In our scans, all other dependent parameters are determined internally
by the {\tt ScannerS} program~\cite{Coimbra:2013qq,ScannerS} according to the minimum conditions for the vacuum corresponding to the given phase. This program provides tools to automatise the parameter space scans of generic scalar extensions of the SM. It also contains generic modules for testing local vacuum stability, to detect symmetries, and library interfaces to the codes we have used to implement the constraints (described later in Sect.~\ref{sec:Cons}).  Apart from
$v$, which is obtained internally from the Fermi constant $G_F$, {\tt
  sHDECAY} uses the same sets of input parameters and all other dependent
parameters are calculated internally from these. 
In Table~\ref{scan_table_CxSM} we present the values and ranges of
parameters that were allowed in the scan, before applying
constraints. Note that in the broken phase the mass that is not an
input is obtained from the following relation 
\begin{equation}\label{Eq:mass_not_input}
m_2^2=-\dfrac{m_1^2m_3^2R_{21}R_{22}}{m_3^2R_{11}R_{12} + m_1^2R_{31}R_{32}} \; .
\end{equation}
\begin{table}
\begin{center}
\begin{tabular}{|r||c c|c c|}
\hline
\multicolumn{1}{|c||}{\multirow{2}{*}{Input parameter}} & \multicolumn{2}{c|}{Broken  phase}                    \\  
\multicolumn{1}{|c||}{}                                & Min                   & Max                                     \\ \hline \hline
$m_{h_{125}}$ (GeV)                                           & 125.1                 & 125.1                             \\ \hline
$m_{h_{\rm other}}$ (GeV)                         & 30                    & 1000                                     \\ \hline
$v$ (GeV)                                           & 246.22                   & 246.22                                      \\ \hline
$v_S$ (GeV)                                           & 1                     & 1000                                      \\  \hline
$\alpha_1$  & $-\pi/2$                      & $\pi/2$                                                                                   \\ \hline
$\alpha_2$ & $-\pi/2$                      & $\pi/2$                                                                                   \\ \hline
$\alpha_3$  & $-\pi/2$                      & $\pi/2$                                                                                  \\ \hline
\end{tabular} \; \; \; \; \; \begin{tabular}{|r||c c|c c|}
\hline
\multicolumn{1}{|c||}{\multirow{2}{*}{Input parameter}} & \multicolumn{2}{c|}{Dark  phase}                    \\  
\multicolumn{1}{|c||}{}                                & Min                   & Max                                     \\ \hline \hline
$m_{h_{125}}$ (GeV)                                           & 125.1                 & 125.1                             \\ \hline
$m_{h_{\rm other}}$ (GeV)                         & 30                    & 1000                                     \\ \hline
$m_{A}$ (GeV)                         & 30                    & 1000                                     \\ \hline
$v$ (GeV)                                           & 246.22                   & 246.22                                      \\ \hline
$v_S$ (GeV)                                           & 1                     & 1000                                      \\  \hline
$\alpha_1$  & $-\pi/2$                      & $\pi/2$                                                                                   \\ \hline
$a_1 ({\rm GeV}^3)$  & $-10^8$                      & $0$                                                                                   \\ \hline
\end{tabular}
\end{center}
\caption{{\em Ranges of input parameters
    used for each phase of the 
    CxSM:}  In both phases, we denote one of the visible states by
  $m_{h_{125}}$, which is the SM-like Higgs state, and its value is
  fixed to the experimental value. In the broken phase (left), $m_{h_{\rm
      other}}$ denotes one of the other visible scalars, and the
  remaining one is obtained from the input using Eq.~\eqref{Eq:mass_not_input}.
  In the dark phase (right), all three masses are input parameters,
  and $m_A$ refers to the dark matter scalar. 
\label{scan_table_CxSM}
}
\end{table}

%%%%%%%%%%%%%%%%%%%%%%%%%%%%%%%%%%%%%%%%%%%%%%%%%%%%%%%%%%
\subsection{The RxSM}
\label{sec:RxSM}
In this study our main focus is the CxSM model in which all
possibilities exist  for the decays of a scalar into two different or
identical scalars, in a theory with three CP-even scalars. The
special case when the two scalars in the final state of the decay are
identical is, however, also allowed in simpler models with just two
scalar degrees of freedom. The simplest such model is the real singlet
model (RxSM). In fact, the broken phase of the CxSM, the dark phase of
the CxSM and the broken RxSM all share this possibility. In order to
compare these models in our analysis we now briefly summarise the
RxSM.  

This model is obtained by adding a real singlet $S$ with a
$\mathbb{Z}_2$  symmetry ($S\rightarrow -S$) to the SM. Then the most
general renormalisable potential reads  
\begin{equation}
V_{\rm RxSM}=\dfrac{m^2}{2}H^\dagger H+\dfrac{\lambda}{4}(H^\dagger
H)^2+\dfrac{\lambda_{HS}}{2}H^\dagger H S^2 +\dfrac{m^2_S}{2} S^2 + 
\dfrac{\lambda_S}{4!}S^4 \, ,  \label{eq:V_RxSM}
\end{equation}
 where $m,\lambda,\lambda_{HS},m_S$ and $\lambda_S$ are real. Electroweak
 symmetry breaking, with a vacuum consistent with 
the Higgs mechanism occurs when the following vacuum
expectation values are chosen: 
\begin{equation}\label{eq:vacua:RxSM}
H=\dfrac{1}{\sqrt{2}}\left(\begin{array}{c} G^+ \\
    v+h+iG^0\end{array}\right) \quad \mbox{and}  \quad S =v_S + s \;.
\end{equation}
Again $v \approx 246\;\mathrm{GeV}$ is the SM Higgs VEV, and $v_S$ is
the singlet VEV. In this model, we are primarily interested in the
broken phase, $v_S\neq 0$, where we use the notation $m_1$ and $m_2$ for
the scalar states, ordered in mass, that mix two-by-two. The mixing
matrix is precisely the same as the sub-block responsible for the
mixing in the dark phase of the complex singlet when $\alpha_1\equiv
\alpha$ and $\alpha_2=\alpha_3=0$.  

In the symmetric phase, only the observed Higgs boson with mass
$m_1\equiv m_{h_{125}}$ is visible. For the dark matter candidate we
have the mass $m_2\equiv m_D$. In the broken phase, on the contrary, both
scalars are visible with one of them corresponding to the SM-like
Higgs boson. In the following we will focus on the broken phase since
it allows for the Higgs-to-Higgs decays that we want to use in the
comparison with the CxSM.

Regarding couplings everything is similar to the CxSM, i.e. the
couplings of the various scalars are controlled by the same rule,
Eq.~\eqref{eq:couplings_SM}, with the reduced two-by-two mixing
matrix. The couplings among the scalars are again directly read from
the potential of the model, Eq.~\eqref{eq:V_RxSM}, after
expanding about the vacua,
Eq.~\eqref{eq:vacua:RxSM}, and they are given in 
Appendix~\ref{sec:selfRxSM}.

\begin{table}[tb!]
\begin{center}
\begin{tabular}{|r||c c|}
\hline
\multicolumn{1}{|c||}{\multirow{2}{*}{Scan parameter}} & \multicolumn{2}{c|}{Broken  phase}                    \\  
\multicolumn{1}{|c||}{}                                & Min                   & Max                   \\ \hline \hline
$m_{h_{125}}$ (GeV)                                           & 125.1                & 125.1                         \\ \hline
$m_{h_{\rm other}}$ (GeV)                         & 30                    & 1000                                     \\ \hline
$v$ (GeV)                                           & 246.22                   & 246.22                   \\ \hline
$v_S$ (GeV)                                           & 1                     & 1000                              \\ \hline

$\alpha$  & $-\pi/2$                      & $\pi/2$                                                                                \\ \hline
\end{tabular}
\end{center}
\caption{{\em Ranges of scan parameters used for
    the broken phase of the RxSM:} The mass of one of the two visible scalars has
    been fixed to the  measured Higgs mass value $m_{h_{125}}$,
    whereas the other mass $m_{h_{\rm other}}$ is scanned over.} 
\label{scan_table_RxSM}
\end{table}
Finally, for the scans over the parameter space of this model we note
that the model has five independent parameters, which are chosen as 
$\{\alpha_1\equiv \alpha,m_1,m_2,v,v_S\}$. In the implementation
of the decays for the broken phase in {\tt sHDECAY} we choose the
same set, with the SM VEV, $v$, determined from the
Fermi constant $G_F$. All other dependent 
parameters are determined internally by {\tt sHDECAY}, and by {\tt ScannerS}
according to the minimum conditions for the
vacuum corresponding to the given phase. In
Table~\ref{scan_table_RxSM} we present the values and ranges of
parameters that were allowed in the scan before applying constraints.  

For completeness we note that the input parameters for the implementation of the dark matter phase in {\tt sHDECAY}, which we do not analyse here, are chosen as
$\left\{m_1\equiv m_{h_{125}},m_2\equiv m_D,\lambda_S,m^2_S,v\right\}$
 ($v$ is again
determined from $G_F$). 

%%%%%%%%%%%%%%%%%%%%%%%%%%%%%%%%%%%%%%%%%%%%%%%%%%%%%%%%%%
\subsection{Constraints Applied to the Singlet Models}\label{sec:Cons}

To restrict the parameter space of the singlet models we apply various
theoretical and phenomenological constraints. These have been
described in~\cite{Coimbra:2013qq} and recently updated
in~\cite{Costa:2014qga}. We will only discuss them briefly expanding
mostly on the constraints from collider physics, which is the focus of
our discussion on benchmarks for the LHC run~2. 

\subsubsection*{Theoretical constraints}
We apply both to the CxSM and the RxSM the constraints that: i) the
potential must be bounded from below; ii) the vacuum we have chosen
must be a global minimum and iii) that perturbative unitarity
holds. The first two are implemented in the {\tt ScannerS}
code~\cite{ScannerS,Coimbra:2013qq} using the relevant inequalities
for the CxSM and the RxSM. Perturbative unitarity is tested using a
general internal numerical procedure in \textsc{ScannerS}, which can
perform this test for any model, see~\cite{ScannerS,Coimbra:2013qq}. 

\subsubsection*{Dark matter constraints}
In the dark phase of the CxSM, we compute the relic density $\Omega_A
h^2$ with {\tt micrOMEGAS}~\cite{Belanger:2014hqa} and use it to
exclude parameters space points for which $\Omega_A h^2$ is larger
than $\Omega_ch^2+3\sigma$, where $\Omega_{c}h^2=0.1199\pm 0.0027$ is the combination of the measurements from the WMAP and Planck
satellites~\cite{Ade:2013zuv,Hinshaw:2012aka} and $\sigma$ denotes the 
standard deviation. As for bounds on the direct detection of
dark matter, we compute the spin-independent scattering cross section
of weakly interacting massive particles (WIMPS) on nucleons also with
{\tt micrOMEGAS} with the procedure described in~\cite{Coimbra:2013qq} and 
the points are rejected if the cross section is larger than the upper bound
set by the \textsc{LUX2013} collaboration~\cite{Akerib:2013tjd}. 

\subsubsection*{Electroweak precision observables}
We also apply a $95\%$ exclusion limit from the electroweak precision
observables $S,T,U$~\cite{Maksymyk:1993zm,Peskin:1991sw}. We follow
precisely the procedure reviewed in~\cite{Costa:2014qga} and
  implemented in the {\tt ScannerS} code and therefore do not repeat
the discussion here.  

\subsubsection*{Collider constraints}

The strongest phenomenological constraints on models with an extended
Higgs sector typically arise from collider data, most importantly from
the LHC data. On one hand, one must ensure that one of the scalar
states matches the observed signals for a Higgs boson with a mass of
$\simeq 125$~GeV. The masses of the other two scalars can either be
both larger, both smaller and one larger and one smaller than that of the
discovered Higgs. 
 
On the other hand, all other new scalars must be 
consistent with the exclusion bounds from searches at various
colliders, namely the Tevatron, LEP and LHC searches. The strongest
constraints on the parameter space come from the measurements of the
rates of the 125 GeV Higgs performed by ATLAS and CMS during the LHC
run~1. The experimental bounds are imposed with the help of 
{\tt ScannerS} making use of its external interfaces with other
codes. We have applied $95\%$ C.L.~exclusion limits using  
{\tt HiggsBounds}~\cite{Bechtle:2013wla}. As for consistency with 
the Higgs signal measurements we test the global signal strength of
the $125$~GeV Higgs boson with the latest combination of the ATLAS
and the CMS LHC run 1 datasets, i.e.~$\mu_{125}=1.09\pm
0.11$~\cite{ATLASplusCMS_mus}. In the
remainder of the analysis, for the singlet models, we fix the SM-like
Higgs mass to $125.1$~GeV, which is the central value\footnote{The
  reported value with the experimental errors is  $m_{h_{SM}}=125.09\pm
  0.21 {\rm(stat)} \pm 0.11 {\rm(syst)}$~GeV.} of the ATLAS/CMS
combination reported in~\cite{Aad:2015zhl}.  Note, also, that in our scans we
required the non-SM-like Higgs masses to deviate by at least 3.5~GeV
from 125.1~GeV, in order to avoid degenerate Higgs signals.
 
{\tt HiggsBounds} computes internally various experimental quantities
such as the signal rates 
\begin{equation}
\mu_{h_i} =\dfrac{\sigma_{\rm New}(h_i){\rm BR}_{\rm
    New}\left(h_i\rightarrow X_{\rm SM}\right)}{\sigma_{\rm
    SM}(h_{i}){\rm BR}_{\rm SM} \left(h_{i}\rightarrow X_{\rm
      SM}\right)} \, \,  \; .
\label{mu}
\end{equation}
Here $\sigma_{\rm New}(h_i)$ denotes the production cross section
  of the Higgs boson, $h_i$, in the new model under consideration and
  $\sigma_{\rm SM}(h_i)$ the SM production cross section of a Higgs boson with
  the same mass. Similarly, ${\rm BR}_{\rm New}$ is the
  branching ratio for $h_i$ in the new model to decay into a final state with SM
  particles $X_{\rm SM}$, and ${\rm BR}_{\rm SM}$
  the corresponding SM quantity for a Higgs boson with the same mass.
For the computation of the rates {\tt HiggsBounds} requires as input for all scalars:
the ratios of the cross 
sections for the various production modes, the branching ratios and the total
decay widths. In singlet models, at leading order (and also at higher order in QCD), the
cross section ratios are all simply given by the suppression factor squared,
$R_{i1}^2$, see Eq.~\eqref{eq:couplings_SM}. 
Regarding the branching ratios and total decay  widths we have used
the new implementation of the CxSM and the RxSM 
in~{\tt sHDECAY} as described in appendix~\ref{sec:app_SHDECAY}.

In the analyses of Sects.~\ref{Sec:LHC_Run2}
and~\ref{Sec:LHC_Run2_benchs} the $h_i$ production cross sections through
gluon fusion, $gg \to h_i$, are needed at 13~TeV c.m.~energy. We have
computed these both with the programs {\tt
  SusHi}~\cite{Harlander:2012pb} and {\tt HIGLU}~\cite{Spira:1995mt}
at next-to-next-to-leading order (NNLO) QCD and with higher order
electroweak corrections consistently turned off. The results of the
two computations agreed within the numerical integration errors and within parton density
function uncertainties. In the analysis we have used the results
from~{\tt HIGLU}.

%%%%%%%%%%%%%%%%%%%%%%%%%%%%%%%%%%%%%%%%%%%%%%%%%%%%%%%%%%%
\subsection{The NMSSM \label{sec:nmssmintro}}
Supersymmetric models require the introduction of at least two Higgs
doublets. In the NMSSM this minimal Higgs sector is extended by an
additional complex superfield $\hat{S}$. After EWSB the NMSSM Higgs
sector then features seven Higgs bosons. These are, in the
CP-conserving case, three neutral CP-even, two neutral CP-odd and two
charged Higgs bosons. The NMSSM Higgs potential is derived from the
superpotential, the soft SUSY breaking terms and the $D$-term
contributions. The scale-invariant NMSSM superpotential in terms of
the hatted superfields reads 
\beq
{\cal W} = \lambda \widehat{S} \widehat{H}_u \widehat{H}_d +
\frac{\kappa}{3} \, \widehat{S}^3 + h_t
\widehat{Q}_3\widehat{H}_u\widehat{t}_R^c - h_b \widehat{Q}_3
\widehat{H}_d\widehat{b}_R^c  - h_\tau \widehat{L}_3 \widehat{H}_d
\widehat{\tau}_R^c \; ,
\label{eq:superpot}
\eeq
where we have included only the third generation fermions. While the
first term replaces the $\mu$-term $\mu \hat{H}_u \hat{H}_d$ of the
MSSM superpotential, the second term, cubic in the singlet superfield,
breaks the Peccei-Quinn symmetry so that no massless axion can
appear. The last three terms represent the Yukawa interactions. The
soft SUSY breaking Lagrangian gets contributions from
the scalar mass parameters for the Higgs and the sfermion fields. In
terms of the fields corresponding to the complex scalar components of
the superfields the Lagrangian reads
\beq
\label{eq:Lmass}
 -{\cal L}_{\mathrm{mass}} &=&
 m_{H_u}^2 | H_u |^2 + m_{H_d}^2 | H_d|^2 + m_{S}^2| S |^2 \nonumber \\
  &+& m_{{\tilde Q}_3}^2|{\tilde Q}_3^2| + m_{\tilde t_R}^2 |{\tilde t}_R^2|
 +  m_{\tilde b_R}^2|{\tilde b}_R^2| +m_{{\tilde L}_3}^2|{\tilde L}_3^2| +
 m_{\tilde  \tau_R}^2|{\tilde \tau}_R^2|\; .
\eeq
The contributions from the trilinear soft SUSY breaking interactions
between the sfermions and the Higgs fields are comprised in 
\beq
\label{eq:Trimass}
-{\cal L}_{\mathrm{tril}}=  \lambda A_\lambda H_u H_d S + \frac{1}{3}
\kappa  A_\kappa S^3 + h_t A_t \tilde Q_3 H_u \tilde t_R^c - h_b A_b
\tilde Q_3 H_d \tilde b_R^c - h_\tau A_\tau \tilde L_3 H_d \tilde \tau_R^c
+ \mathrm{h.c.}
\eeq
The soft SUSY breaking Lagrangian that contains the gaugino mass
parameters is given by
\beq
-{\cal L}_\mathrm{gauginos}= \frac{1}{2} \bigg[ M_1 \tilde{B}
\tilde{B}+M_2 \sum_{a=1}^3 \tilde{W}^a \tilde{W}_a +
M_3 \sum_{a=1}^8 \tilde{G}^a \tilde{G}_a  \ + \ {\rm h.c.}
\bigg].
\eeq
We will allow for non-universal soft terms at the GUT scale. After
EWSB the Higgs doublet and singlet fields acquire non-vanishing
VEVs. By exploiting the three minimisation 
conditions of the scalar potential, the soft SUSY breaking masses squared for
$H_u$, $H_d$ and $S$ in ${\cal L}_{\mathrm{mass}}$ are traded for
their tadpole parameters. The Higgs mass matrices for the three
scalar, two pseudoscalar and the charged Higgs bosons are obtained
from the tree-level scalar potential after expanding the Higgs fields
about their VEVs $v_u$, $v_d$ and $v_s$, which we choose to be real and positive,
\beq
H_d = \left( \begin{array}{c} (v_d + h_d + i a_d)/\sqrt{2} \\
   h_d^- \end{array} \right) \;, \quad
H_u = \left( \begin{array}{c} h_u^+ \\ (v_u + h_u + i a_u)/\sqrt{2}
 \end{array} \right) \;, \quad
S= \frac{v_s+h_s+ia_s}{\sqrt{2}} \;.
\eeq
The $3 \times 3$ mass matrix squared, $M_S^2$, for the
CP-even Higgs fields is diagonalised through a rotation matrix ${\cal R}^S$ yielding the CP-even mass eigenstates $H_i$ ($i=1,2,3$),
\beq
(H_1, H_2, H_3)^T = {\cal R}^S (h_d,h_u,h_s)^T \;.
\label{eq:scalarrot}
\eeq
The $H_i$ are ordered by ascending mass, $M_{H_1} \le M_{H_2} \le
M_{H_3}$. In order to obtain the CP-odd mass eigenstates, $A_1$ and
$A_2$, first a rotation ${\cal R}^G$ is performed to separate the
massless Goldstone boson $G$, which is followed by a rotation ${\cal R}^P$ to
obtain the mass eigenstates
\beq
(A_1,A_2,G)^T = {\cal R}^P {\cal R}^G (a_d,a_u,a_s)^T \;,
\label{eq:pseudorot}
\eeq
which are ordered such that $M_{A_1} \le M_{A_2}$. 
 
The tree-level NMSSM Higgs sector can be parametrised by the six parameters
\beq
\lambda\ , \ \kappa\ , \ A_{\lambda} \ , \ A_{\kappa}, \
\tan \beta =v_u/ v_d \quad \mathrm{and}
\quad \mu_\mathrm{eff} = \lambda v_s/\sqrt{2}\; .
\eeq
We choose the sign conventions
for $\lambda$ and $\tan\beta$ such that they are positive, whereas
$\kappa$, $A_\lambda$, $A_\kappa$ and $\mu_{\mathrm{eff}}$ can have
both signs. In order to make realistic predictions for the Higgs
masses we have taken into account the higher order corrections to the
masses. Through these corrections also the soft SUSY breaking mass terms for
the scalars and the gauginos as well as the trilinear soft SUSY
breaking couplings enter the Higgs sector. 

In the NMSSM several Higgs-to-Higgs decays are possible if
kinematically allowed. In the CP-conserving case the heavier
pseudoscalar $A_2$ can decay into the lighter one $A_1$ and one of the
two lighter Higgs bosons $H_1$ or $H_2$\footnote{The heavy
  pseudoscalar $A_2$ and the heavy scalar $H_3$ are in most scenarios
  degenerate in mass.}. The heavier CP-even Higgs bosons can decay
into various combinations of a pair of lighter CP-even Higgs bosons or into a
pair of light pseudoscalars. Hence, in principle we can have, 
\beq
\begin{array}{lllllllll}
A_2 &\to& A_1 + H_1 \;,& \qquad A_2 &\to& A_1 + H_2 \\
H_{2,3} &\to& H_1 + H_1 \;,& \qquad H_3 &\to& H_1 + H_2 \;,& \qquad H_3
&\to& H_2 H_2 \\
H_{1,2,3} &\to& A_1 A_1 \;.
\end{array}
\eeq
Not all of these decays lead to measurable rates,
however. Furthermore, if the decay widths of the SM-like Higgs boson into
lighter Higgs pairs are too large, the compatibility with the measured
rates is lost. In order to investigate what rates in the various SM
final states can be expected from Higgs-to-Higgs decays an extensive
scan in the NMSSM parameter spaces needs to be performed. We have generated a new sample of NMSSM scenarios, following a procedure similar to an earlier publication~\cite{King:2014xwa}, to obtain the decay rates into Higgs pairs. The
scan ranges and the applied criteria shall be briefly discussed here and we refer the reader to~\cite{King:2014xwa} for further details. 

For the scan, $\tan\beta$, the effective $\mu$ parameter and the
NMSSM specific parameters $\lambda,\kappa,A_\lambda$ and $A_\kappa$
have been varied in the ranges
\beq 
1 \le \tan\beta \le 30 \; , \qquad 0 \le \lambda \le 0.7 \; , \qquad
-0.7 \le \kappa \le 0.7 \;,
\eeq
\beq
-2 \mbox{ TeV} \le A_\lambda \le 2 \mbox{ TeV} \; , \;
-2 \mbox{ TeV} \le A_\kappa \le 2 \mbox{ TeV} \; , \;
-1 \mbox{ TeV} \le \mu_{\text{eff}} \le 1 \mbox{ TeV} \;.
\label{eq:cond2}
\eeq
Perturbativity constraints have been taken into account by applying
the rough constraint
\beq
\sqrt{\lambda^2+\kappa^2} \le 0.7 \;.
\eeq
We have varied the trilinear soft SUSY breaking couplings of the up-
and down-type quarks and the charged leptons, $A_U, A_D$ and $A_L$
with $U\equiv u,c,t, D\equiv d,s,b$ and $L\equiv e,\mu,\tau$,
independently in the range
\beq
-2 \mbox{ TeV} \le A_U, A_D, A_L \le 2 \mbox{ TeV} \,.
\label{eq:cond3}
\eeq
For the soft SUSY breaking right- and left-handed masses of
the third generation we choose
\beq
600 \mbox{ GeV} \le M_{\tilde{t}_R} = M_{\tilde{Q}_3} \le 3 \mbox{
 TeV} \; , \;
600 \mbox{ GeV} \le M_{\tilde{\tau}_R} = M_{\tilde{L}_3} \le 3 \mbox{
 TeV} \; , \;
M_{\tilde{b}_R} = 3 \mbox{ TeV} \;.
\label{eq:cond4}
\eeq
For the first two generations we take
\beq
M_{\tilde{u}_R,\tilde{c}_R} =
M_{\tilde{d}_R,\tilde{s}_R}=M_{\tilde{Q}_{1,2}}=M_{\tilde{e}_R,\tilde{\mu}_R}=
M_{\tilde{L}_{1,2}} = 3 \mbox{ TeV} \;.
\label{eq:cond5}
\eeq
The gaugino soft SUSY breaking masses are chosen to be positive and
varied in the ranges 
\beq
100 \mbox{ GeV} \le M_1 \le 1 \mbox{ TeV} \;, \;
200 \mbox{ GeV} \le M_2 \le 1 \mbox{ TeV} \;, \;
1.3 \mbox{ TeV} \le M_3 \le 3 \mbox{ TeV} \;.
\label{eq:cond6}
\eeq
The chosen parameter ranges entail particle masses that are
compatible with the exclusion limits  on SUSY particle masses 
\cite{Aad:2012tx,Aad:2012yr,Aad:2013ija,Aad:2014qaa,Aad:2014wea,Aad:2014bva,Aad:2014kra,Aad:2014nra,Chatrchyan:2013xna,Chatrchyan:2013fea,Chatrchyan:2013mya,Khachatryan:2014doa,Khachatryan:2015pwa,Khachatryan:2015vra}
and with the lower bound on the charged Higgs mass
\cite{TheATLAScollaboration:2013wia,CMS:mxa}.  
With the help of the program package {\tt NMSSMTools}
\cite{Ellwanger:2004xm,Ellwanger:2005dv,Ellwanger:2006rn} we checked
for the constraints from low-energy observables and computed the input necessary for \texttt{HiggsBounds} to check for consistency with the LEP, Tevatron and the LHC run~1 latest exclusion limits from searches for new Higgs bosons. Details are given on the webpage of the
program\footnote{http://www.th.u-psud.fr/NMHDECAY/nmssmtools.html}. Via
the interface with {\tt micrOMEGAS} \cite{Belanger:2001fz,Belanger:2004yn,Belanger:2008sj,Belanger:2010gh} also the compatibility
with the upper bound of the relic abundance of the lightest neutralino
as the NMSSM dark matter candidate with the latest PLANCK results was
verified~\cite{Ade:2013zuv}. 
Among the points that are compatible with the {\tt NMSSMTools} constraints, only
those are kept that feature a SM-like Higgs boson with mass between
124 and 126 GeV. This can be either $H_1$ or $H_2$ and will be called
$h$ in the following. In addition, according to our criteria, a Higgs boson is
SM-like if its signal rates into the $WW$, $ZZ$, $\gamma\gamma$,
$b\bar{b}$ and $\tau\tau$ final states are within 2 times the
1$\sigma$ interval around the respective best fit value. We use the
latest combined signal rates and errors reported by ATLAS and CMS
in~\cite{ATLASplusCMS_mus}, in a 6-parameter fit. In
this fit, the five rates mentioned above are for fermion mediated production, so 
in their computation we approximate the inclusive
production cross section by the dominant production mechanisms, gluon
fusion and $b\bar{b}$ annihilation. The sixth parameter of the fit is the ratio between the rate for vector mediated production and the fermion mediated production, which we also check to be within $2\sigma$. 
The cross section for gluon fusion is obtained by
multiplying the SM gluon fusion cross section with the ratio between
the NMSSM Higgs decay width into gluons and the corresponding SM decay
width at the same mass value. The two latter rates are obtained from {\tt
  NMSSMTools} at NLO QCD. The SM cross section was calculated at NNLO
QCD with~{\tt HIGLU} \cite{Spira:1995mt}. For the SM-like Higgs boson
this procedure approximates the NMSSM Higgs cross section, computed
at NNLO QCD with {\tt HIGLU}, to better than 1\%. As for the cross section for $b\bar{b}$ annihilation
%, which is only important in the large $\tan\beta$ region where the $b\bar{b}$ coupling to scalars can be enhanced, 
we multiply the SM $b\bar{b}$ annihilation cross section with the effective squared $b\bar{b}$ coupling obtained from {\tt NMSSMTools}. For the cross section values we use the data from~\cite{Ferreira:2014dya}, which was produced with the code~{\tt SusHi}~\cite{Harlander:2012pb}. Note that, similarly
to the CxSM analysis, we have excluded degenerate cases where other
  non-SM-like Higgs bosons would contribute to the SM-like Higgs
  signal, by requiring that the masses of the non-SM-like Higgs bosons
  deviate by at least 3.5 GeV from 125.1 GeV.
Finally, note that in~\cite{King:2014xwa} the branching ratios obtained
with {\tt NMSSMTools} were cross-checked against the ones calculated with {\tt 
  NMSSMCALC} \cite{Baglio:2013iia,King:2015oxa}. There are differences
due to the treatment of the radiative corrections to the Higgs boson
masses\footnote{In {\tt 
    NMSSMTools} the full one-loop and the two-loop ${\cal O}(\alpha_s
  (\alpha_b+\alpha_t))$ corrections at vanishing external
  momentum \cite{Degrassi:2009yq} are included. {\tt NMSSMCALC} provides
  both for the real and the complex NMSSM the full one-loop
  corrections including the momentum dependence in a mixed
  $\overline{\mbox{DR}}-$on-shell renormalisation scheme
  \cite{Ender:2011qh,Graf:2012hh,Muhlleitner:2014vsa} and 
  the two-loop ${\cal O}(\alpha_s \alpha_t)$ corrections at vanishing
  external momentum. For the latter, in the (s)top sector the user can
  choose between 
  on-shell and $\overline{\mbox{DR}}$ renormalisation conditions. For
  a comparison of the codes, see also \cite{Staub:2015aea}.}, as well
as due to the more sophisticated and up-to-date inclusion of the
dominant higher order corrections to the decay widths and the
consideration of off-shell effects in {\tt NMSSMCALC}. The overall
picture, however, remains unchanged. 
%%%%%%%%%%%%%%%%%%%%%%%%%%%%%%%%%%%%%%%%%%%%%%%%%%%%%%%%%%
\section{Numerical Analysis}
\label{sec:res}
In this section, we analyse the consequences of the Higgs-to-Higgs decays on the allowed parameter space after applying the constraints described in Sects.~\ref{sec:Cons} and~\ref{sec:nmssmintro}. First we analyse the importance of chain decays for the production of the SM-like Higgs boson (Sect.~\ref{sec:chainprod}), then we discuss the prospects of using such Higgs-to-Higgs decays to distinguish between the various models (Sect.~\ref{Sec:LHC_Run2}) and, finally, we provide a set of benchmark points for the singlet models in Sect.~\ref{Sec:LHC_Run2_benchs}. 

%%%%%%%%%%%%%%%%%%%%%%%%%%%%%%%%%%%%%%%%%%%%%%%%%%%%%%%%%%
\subsection{Higgs Boson Production from Chain Decays in 
the CxSM and RxSM}
\label{sec:chainprod}
In models with extended Higgs sectors, where additional scalar states
exist that can interact with the SM Higgs boson, the signal rates for a
given SM final state $ X_{\rm SM}$, Eq.~\eqref{mu}, do not
account for the total process leading to $X_{\rm SM}$ through Higgs
boson decays. This is the case for the singlet models, where the same
final state may be reached through an intermediate step with a heavy
scalar, $h_i$, decaying into two lighter scalars $h_j,h_k$ (which may
be different or not), that finally decay into a final state $X_{\rm SM}$. 

In the following we will investigate the question: Can such
resonant decays of a heavy Higgs boson into a final state containing a
reconstructed SM-like Higgs boson (which is the one being observed at
the LHC) compete with the direct production of the observed
Higgs state~\cite{Arhrib:2013oia}? We will call the former production
mechanism {\it chain} production. For a given
channel $X_{\rm SM}$ in which 
the SM-like Higgs is observed, the total rate is then defined as
\begin{equation} 
\mu_{h_{125}}^T\equiv \mu_{h_{125}}+\mu_{h_{125}}^{\rm C} \;,
\end{equation}
where 
\begin{equation}\label{eq:chain_contributions}
\mu_{h_{125}}^{\rm C}\equiv \sum_i \dfrac{\sigma_{\rm
    New}(h_{i})}{\sigma_{\rm SM}(h_{125})}N_{h_i,h_{125}} \dfrac{{\rm
    BR}_{\rm New}\left(h_{125}\rightarrow X_{\rm SM}\right)}{{\rm
    BR}_{\rm SM} \left(h_{125}\rightarrow X_{\rm SM}\right)}  
 \, \,  \; .
\end{equation}
Here $N_{h_i,h_{125}}$ is the expected number of $h_{125}$ Higgs
bosons produced in the decay of $h_i$, which, for a model
with up to 3 scalars, is given by
\begin{eqnarray}
N_{h_i,h_{125}}&\equiv& \sum_{n=1}^4n\times P_{h_i,n\times h_{125}}\; ,
\end{eqnarray}
where $P_{h_i,n\times h_{125}}$ are the probabilities of producing $n$
of the observed Higgs boson $h_{125}$ from the decay of $h_i$, 
\begin{eqnarray}
P_{h_i,1\times h_{125}}&=&{\rm BR}_{\rm New} \left(h_{i}\rightarrow
  h_{j}+h_{125}\right)\times(1-{\rm BR}_{\rm New}
\left(h_{j}\rightarrow h_{125}+h_{125}\right)) \nonumber\\ 
P_{h_i,2\times h_{125}}&=&{\rm BR}_{\rm New} \left(h_{i}\rightarrow
  h_{125}+h_{125}\right)+2{\rm BR}_{\rm New} \left(h_{i}\rightarrow
  h_{j}+h_{j}\right)\times\nonumber\\ 
&&\times(1-{\rm BR}_{\rm New} \left(h_{j}\rightarrow
  h_{125}+h_{125}\right))\times {\rm BR}_{\rm New}
\left(h_{j}\rightarrow h_{125}+h_{125}\right) \\ 
P_{h_i,3\times h_{125}}&=&{\rm BR}_{\rm New} \left(h_{i}\rightarrow
  h_{j}+h_{125}\right)\times{\rm BR}_{\rm New} \left(h_{j}\rightarrow
  h_{125}+h_{125}\right) \nonumber\\ 
P_{h_i,4\times h_{125}}&=&{\rm BR}_{\rm New} \left(h_{i}\rightarrow
  h_{j}+h_{j}\right)\times {\rm BR}_{\rm New} \left(h_{j}\rightarrow
  h_{125}+h_{125}\right)^2 \nonumber \;,
\end{eqnarray} 
with $i\ne j$ and $m_{h_j} < m_{h_i}$. Replacing in
Eq.~(\ref{eq:chain_contributions}) the branching ratios by the partial
and total widths and exploiting the fact that the production cross
section and the direct decay widths of $h_i$ into SM final states are both
modified by $R_{i1}^2$, compared to the SM, we arrive at
\begin{equation}\label{RED_EQ}
\mu_{h_{125}}^{\rm C}\simeq \sum_i R_{i1}^2\frac{\sigma_{\rm SM}(h_i)}{\sigma_{\rm SM}(h_{125})}N_{h_i,h_{125}}
\sum_{X_{\rm SM}}{\rm BR}_{\rm New}(h_{125}\rightarrow X_{\rm SM})
\;. 
\end{equation} It is important to note, from Eq.~\eqref{RED_EQ}, that $\mu_{h_{125}}^{\rm C}$ is in fact independent of the SM final state $X_{SM}$. This simplifies the analysis because we can study this global quantity instead of each final state individually. 

In the remainder of this section we compare the relative importance of
direct production to chain production in the singlet models
 through the ratio
$\mu_{h_{125}}^{\rm C} /\mu_{h_{125}}^T$.  This measures the
fraction of Higgs events from chain production with respect
  to the total number of Higgs events. The allowed parameter space for the CxSM model after LHC run~1 was
recently analysed in detail in~\cite{Costa:2014qga}. In this study it
was found that, in the  
broken phase, all kinematical possibilities are still allowed,
i.e.~the observed Higgs boson can be any of the three scalars
and the corresponding couplings may deviate considerably from the SM. 
Similarly, in the dark phase of the CxSM or the broken phase of the
RxSM, all kinematically different situations are still allowed for the
two visible scalars.

Here we point out that even for the interpretation of the LHC run~1
data, it is important to take into account the contributions from
chain decays in the measurements of the SM Higgs signal
rates. In some cases, these contribution can
be up to $15\%$ of the total 
signal rate. In singlet models, where the direct decay signal rate is
simply suppressed relative to the SM limit, this can lead to viable
points that would naively be excluded if this contribution is not
included. On the other hand it is interesting to investigate which points
are still allowed by the LHC run 1 data and, simultaneously, have the
largest possible chain decay contributions. In this way, a new
heavy scalar may be found indirectly at run 2 in decays into a SM Higgs pair
(in any of the singlet models we discuss) or  
into a SM Higgs with a new light scalar (in the broken phase of the CxSM).

\begin{figure}[hb!]
\centering
\mbox{\includegraphics[width=0.51\linewidth]{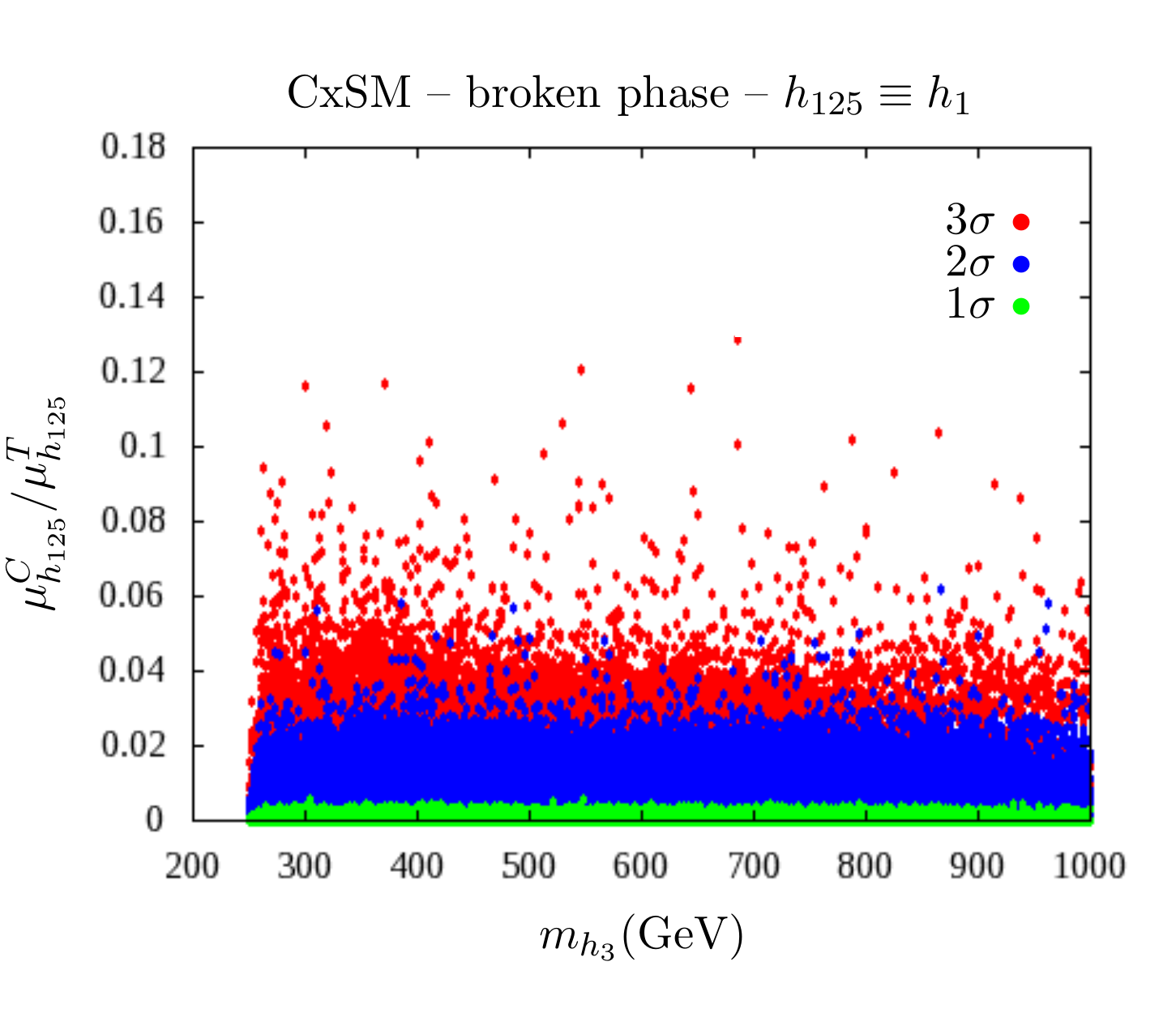}\includegraphics[width=0.51\linewidth]{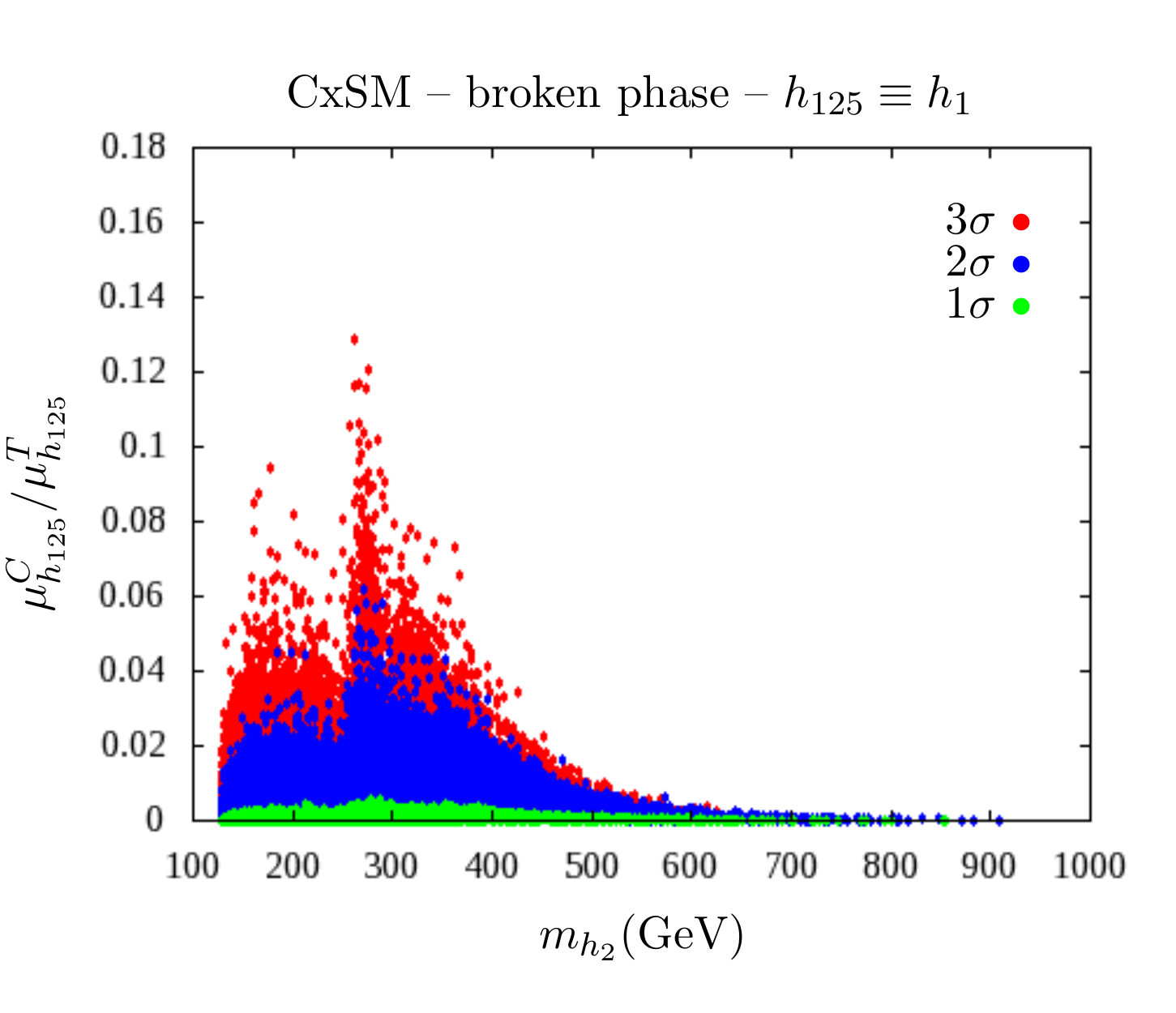}} \\
\mbox{\includegraphics[width=0.51\linewidth]{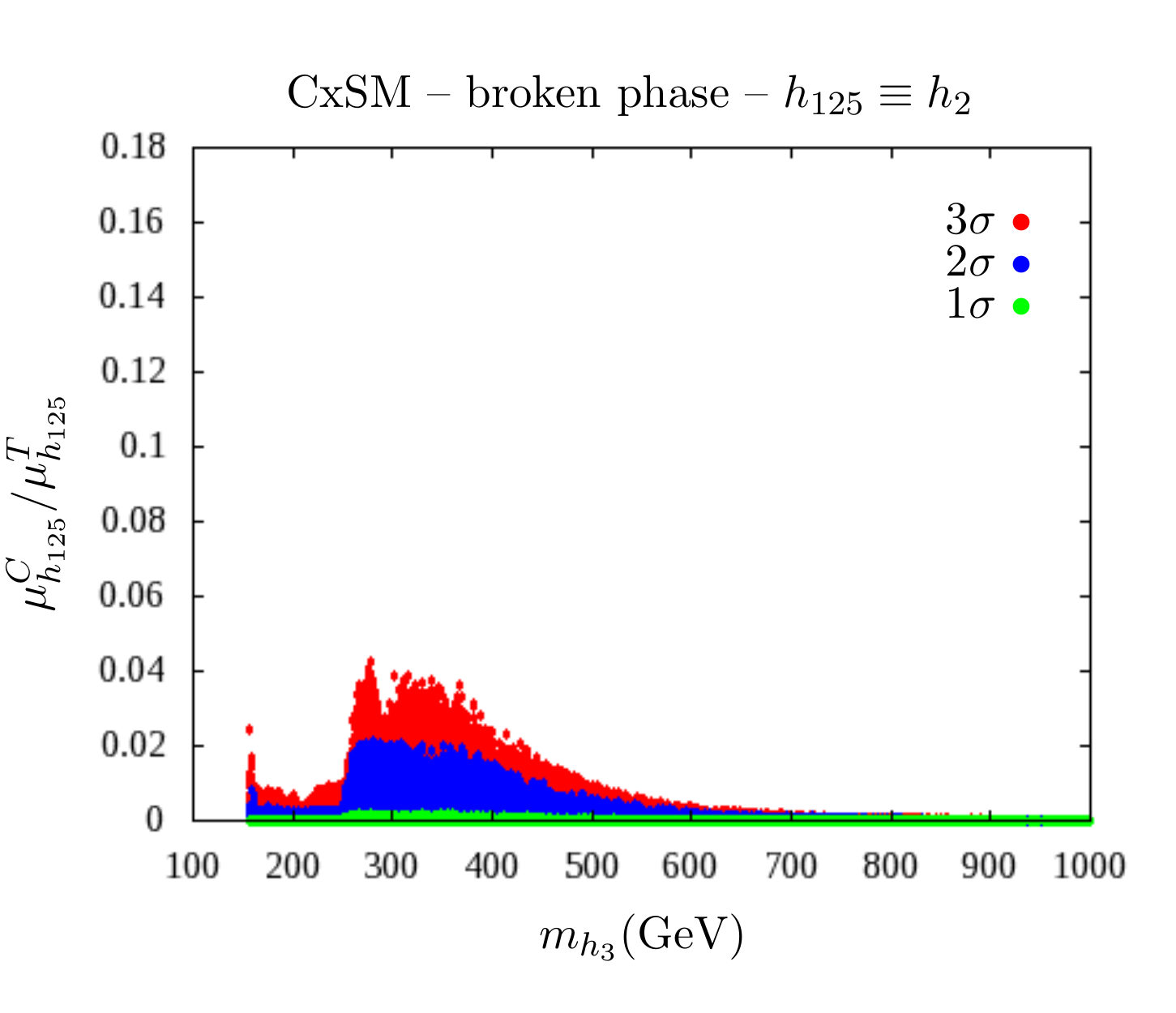}\includegraphics[width=0.51\linewidth]{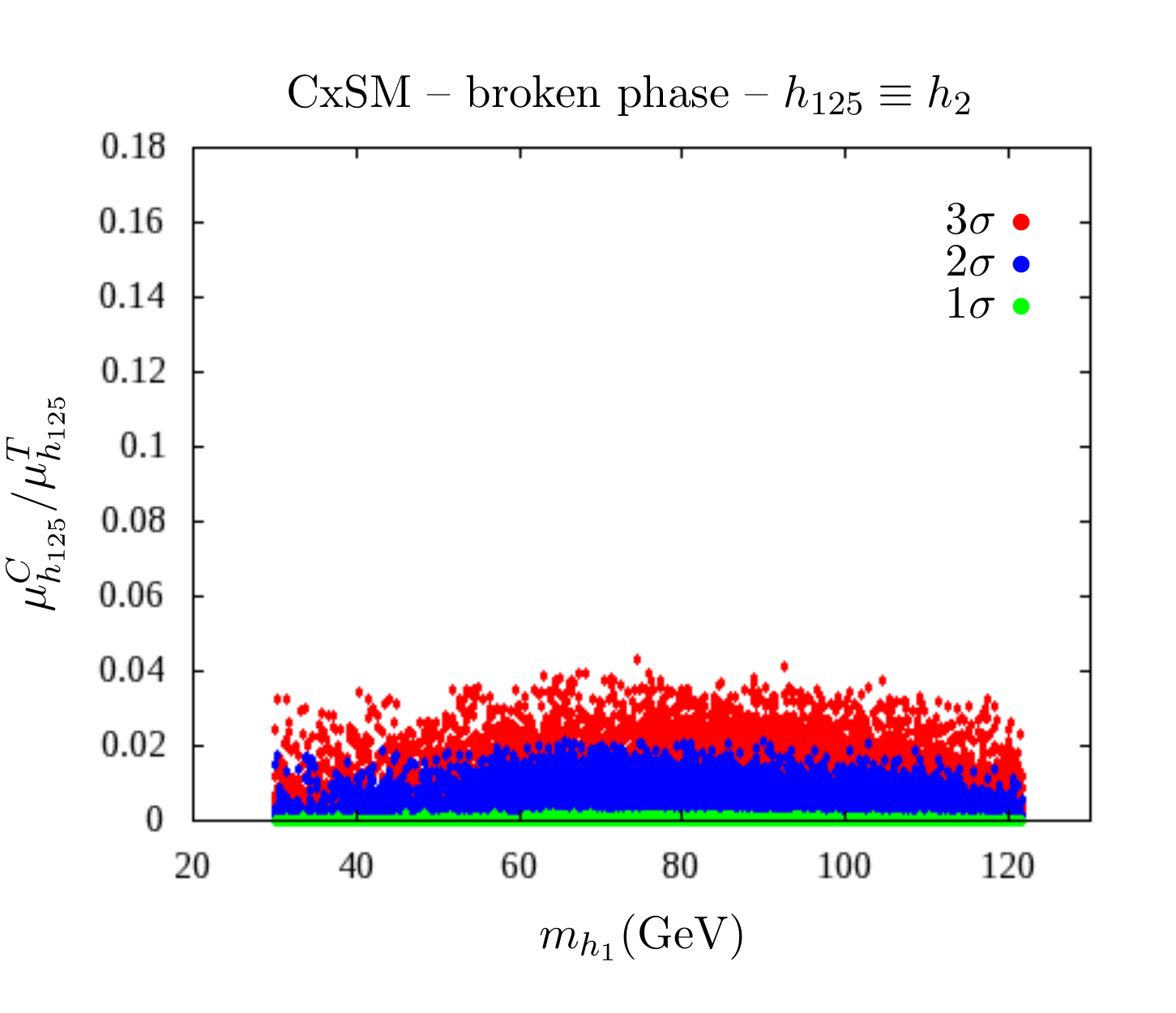}}  
\caption{The fraction of chain decays to the total signal rate as a
  function of $m_{h_3}$ (left) and of the mass of the non-SM-like
  Higgs boson (right) for the case where $h_{125} \equiv h_1$ (upper)
  and $h_{125} \equiv h_2$ (lower). The coloured points are
  overlaid on top of each other in the order $3\sigma$, $2\sigma$,
  $1\sigma$. Each layer corresponds to points that fall inside the
  region where the total signal rate $\mu^{T}_{h_{125}}$ is within
  $3\sigma$ (red), $2\sigma$ (blue) and $1\sigma$ (green) from the
  best fit point of the global signal rate using all LHC run 1 data
  from the ATLAS and CMS
  combination~\cite{ATLASplusCMS_mus}. 
}
\label{fig:chain_vs_pvalue_CxSM_broken}
\end{figure}
In Fig.~\ref{fig:chain_vs_pvalue_CxSM_broken} we present a sample of
points generated for the broken phase of the CxSM in the scenario
where the observed Higgs boson is the lightest scalar (top panels) and the
next-to-lightest scalar (bottom panels). 
We present projections against the masses of the two new scalars that
can be involved in the chain decay contributions.  We
overlay three layers of points for which the total signal rate $\mu_{h_{125}}^T$ is,
respectively, within $3\sigma$ (red), $2\sigma$ (blue)  and $1\sigma$
(green) of the LHC run~1 best fit point for the global signal
strength. The vertical axis shows the fraction of the signal rate
that is due to the chain decay contribution. In
the upper plots, where $h_1 \equiv h_{125}$, chain decays become
possible above $m_{h_3}=250$~GeV, {\it
  cf.}~Fig.~\ref{fig:chain_vs_pvalue_CxSM_broken} (upper left).
In the upper right plot points exist when $m_{h_2} \gtrsim
129$~GeV due to the minimum required distance of 3.5~GeV
from $m_{h_{125}}$.
In the lower two plots, $h_2\equiv h_{125}$, so that with the lowest
$h_1$ mass value of 30~GeV in our scan, {\it
  cf.}~Fig.~\ref{fig:chain_vs_pvalue_CxSM_broken} (lower right), chain
decays come into the game for $m_{h_3} \gtrsim 155$~GeV. It can be
inferred from the top panels, that when the Higgs boson is the
lightest scalar, the chain decay contribution can be up to $\sim
6\%$ ($\sim 12\%$) at $2\sigma$ ($3\sigma$). We have also
produced plots where we apply the $2\sigma$ ($3\sigma$) cut
on the rate for direct $h_{125}$ production and decay, i.e. on
$\mu_{h_{125}}$ instead of $\mu_{h_{125}}^T$. In that case we observe
a reduction of the maximum allowed $\mu_{h_{125}}^C/\mu_{h_{125}}^T$
and the numbers will change to $\sim 4\%$ ($\sim 9\%$). This means
that imposing a constraint on the parameter space without including
the chain decay contribution would be too strong and points with
larger chain decays at the LHC run~2 would in fact still be
allowed. The kink observed in 
  Fig.~\ref{fig:chain_vs_pvalue_CxSM_broken} (upper right) at
  $m_{h_2}=250$~GeV is due to the opening of the decay channel $h_2 \to h_{125} + h_{125}$ which now also adds to the chain decay contributions
  from $h_3 \to h_{125} + h_{125}$ and $h_3 \to h_{125} + h_2$.
The observed tail for large masses stems from the constraints from
electroweak precision observables (EWPOs). In this
region both $m_{h_2}$ and $m_{h_3}$ are large so that EWPOs force the 
factors $R_{i1}^2$ for the SM couplings of the two heavy
Higgs bosons to be small, thus suppressing 
the contribution from chain decays. In the upper left plot on the
other hand, the points for large $m_{h_3}$ also include the cases
where $m_{h_2}$ can be small, which can then have a larger modification 
factor $R_{21}^2$ without being in conflict with EW precision
data.  Overall, the shape of the three different $\sigma$ regions in the
various panels is the result of an interplay between the kinematics
and the applied constraints. To a certain extent the structure can be
directly related to the exclusion curves from the collider
searches imposed by {\tt HiggsBounds}. This is e.g.~the case for
the peaks at $m_{h_2}\sim 170$~GeV and $m_{h_2}\sim
280$~GeV, as we have checked explicitly by generating a sample of points without
imposing collider constraints.\footnote{A similar behaviour was found in
  Fig.~8 (top and bottom right) of Ref.~\cite{Costa:2014qga}.}

\begin{figure}[tb!]
\centering
\mbox{\includegraphics[width=0.51\linewidth]{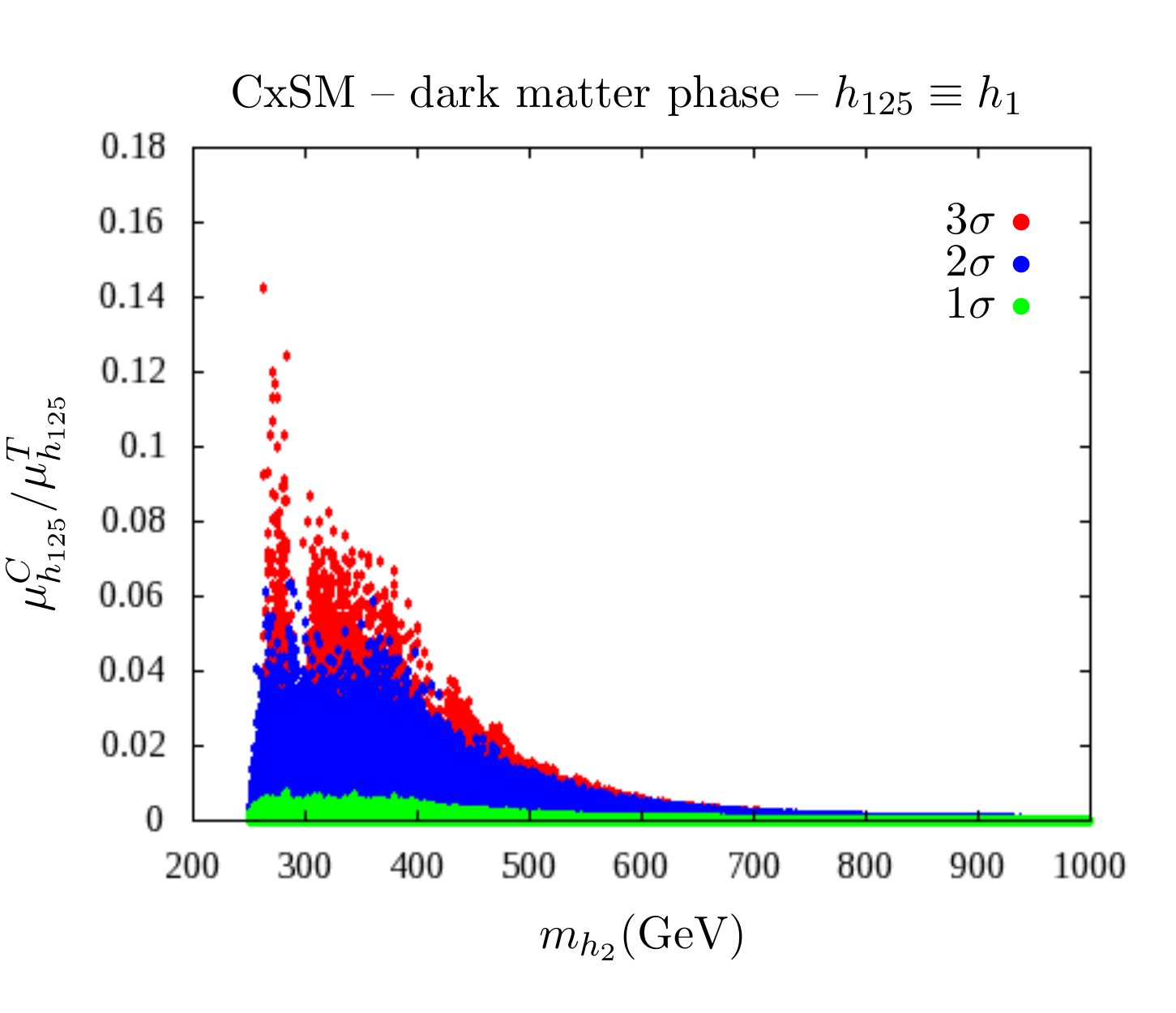}\includegraphics[width=0.51\linewidth]{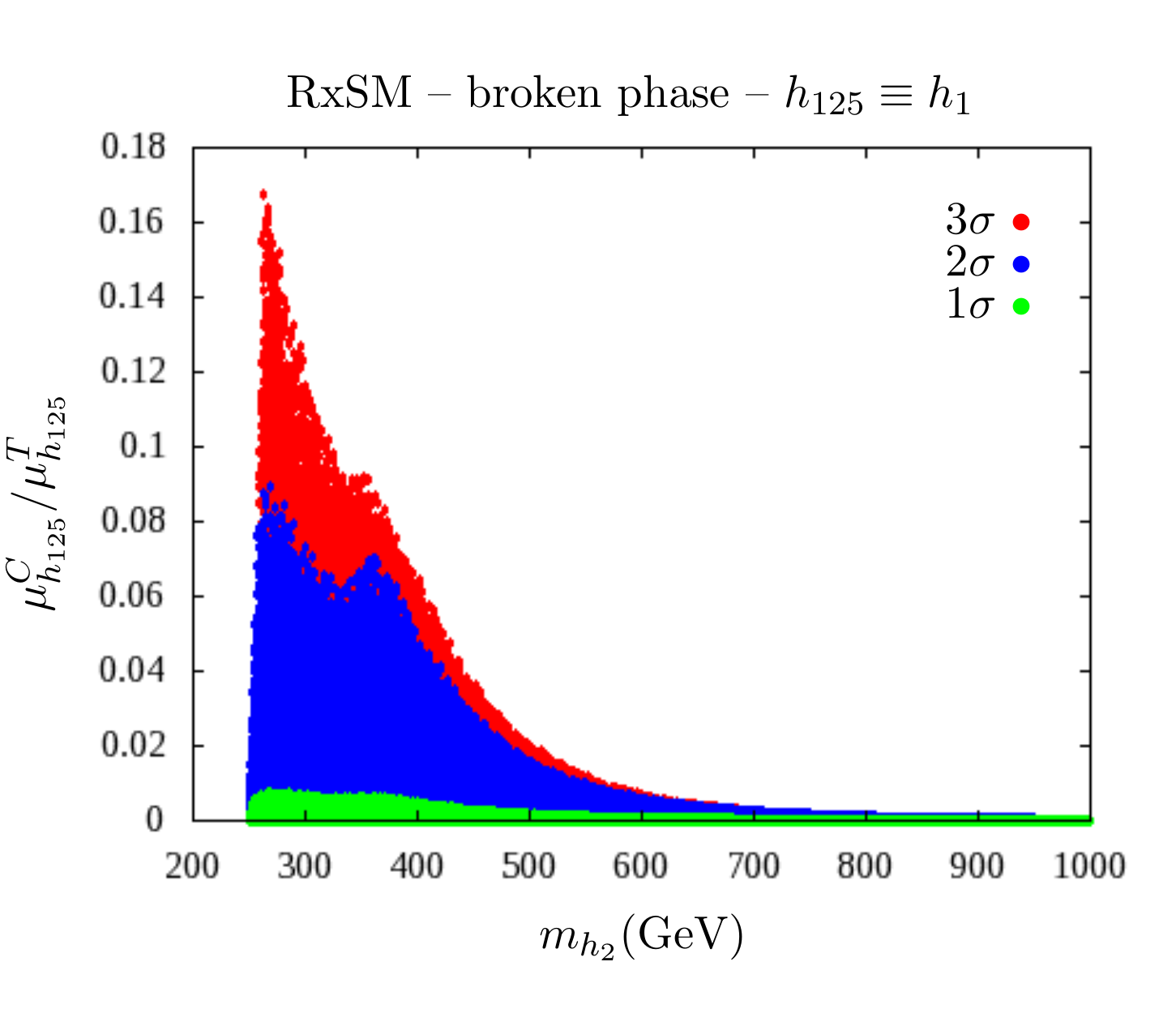}} \\
\caption{The fraction of chain decays to the total signal rate in the
  dark phase of the 
    CxSM (left) and in the broken phase of the RxSM
    (right): The colour code is the same as in
  Fig.~\ref{fig:chain_vs_pvalue_CxSM_broken}. } 
\label{fig:chain_vs_pvalue_CxSM_dark_vs_RxSM_broken}
\end{figure}
In the bottom plots of Fig.~\ref{fig:chain_vs_pvalue_CxSM_broken},
where $h_{125}$ is the next-to-lightest Higgs boson, the relative
fraction of the chain contribution on the total rate is much smaller
with at most $\sim 2\%$ ($\sim 4\%$) for the $2\sigma$ ($3\sigma$)
region. The reason is that here only the decays $h_3 \to h_1
+h_2, h_2 + h_2$ contribute to the chain decays, while in the scenario
with $h_{125} \equiv h_1$ also the $h_2 \to h_1+ h_1$ decay is
possible. The strong increase at $m_{h_3}= 250$~GeV in the lower left
panel is due to the opening of the decay $h_3 \to h_2 +h_2$, and the 
decrease in the number of points for large values of $m_{h_3}$ is due
to EWPOs. The remainder of the shapes of the three $\sigma$ region
again is explained by the interplay between kinematics and applied
constraints, namely the exclusion curves from the Higgs data and their
particular shape.

In Fig.~\ref{fig:chain_vs_pvalue_CxSM_dark_vs_RxSM_broken} we display
the fraction of chain decays for the scenarios
where the Higgs is the lightest visible scalar in the dark phase of
the CxSM (left) and in the broken phase of the RxSM (right). Both models show
a similar behaviour. Above the kinematic threshold for the chain decay,
at $m_{h_2}=250$~GeV, the fraction of chain decays increases to about 14\% in
the CxSM (left plot) and 17\% in the RxSM (right plot) for $m_{h_2}
\approx 270$~GeV and a total signal rate within $3\sigma$ of the
measured value of the global signal rate $\mu$. For the total rate within
$2\sigma$ these numbers decrease to $\sim 7\%$ (left plot), respectively,
$\sim 9\%$ (right plot). As before, the
suppression of the fraction of chain decays at large masses $m_{h_2}$
can be attributed to EW precision constraints, and the second peak at
larger $h_2$ masses together with the overall shape is explained by the
combination of kinematics and constraints, in particular the Higgs
exclusion curves.  

Note, that in the CxSM the density of
points is lower simply because the parameter space is higher
dimensional and the independent parameters that were used to sample it
do not necessarily generate a uniform distribution of points in terms
of the fraction of chain decays. The samples have been generated in
both cases with a few million points. 

%%%%%%%%%%%%%%%%%%%%%%%%%%%%%%%%%%%%%%%%%%%%%%%%%%%%%%%%%%%
\subsection{Higgs-to-Higgs Decays at the LHC Run2}
\label{Sec:LHC_Run2}
Many extensions of the SM allow for the decay of a Higgs boson into two lighter
Higgs states of different masses. Such a decay is not necessarily a sign of 
CP violation. In fact, in the broken phase of the CxSM,
although all three scalars mix, they all retain the quantum numbers of
the scalar in the doublet, i.e.~they are all even under a CP
transformation. On the contrary, in the C2HDM it could be a signal of
CP violation. Furthermore, even in CP-conserving models, the CP number
of the new heavy scalars may not be accessible at an early stage if
they are to be found at the LHC run 2. Thus, models that have a
different theoretical structure and Higgs spectrum may in fact look
very similar at an early stage of discovery. 

In section \ref{sec:cxsmnmssmcomp} we will perform a comparison between the
broken phase of the CxSM and the NMSSM, which contains
similar possibilities in
terms of scalar decays into scalars, if the CP numbers are not
measured. We focus on resonant decays, which allow to probe the scalar
couplings of the theory, and, in many scenarios, may provide an
alternative discovery channel for the new scalars.  First, however, we
compare the Higgs-to-Higgs decay rates at the LHC run 2 for the real
singlet extension in its broken phase with the CxSM. The results of this 
  comparison will then allow us to draw meaningful conclusions in the subsequent
  comparison between the NMSSM and CxSM-broken.

%%%%%%%%%%%%%%%%%%%%%%%%%%%%%%%%%%%%%%%%%%%%%%%%%%%%%%%%%
\subsubsection{Comparison between the RxSM-broken and the Two Phases
  of the CxSM \label{sec:comprxsmcxsm}}
\begin{figure}[ht]
\centering
\mbox{\includegraphics[width=0.51\linewidth]{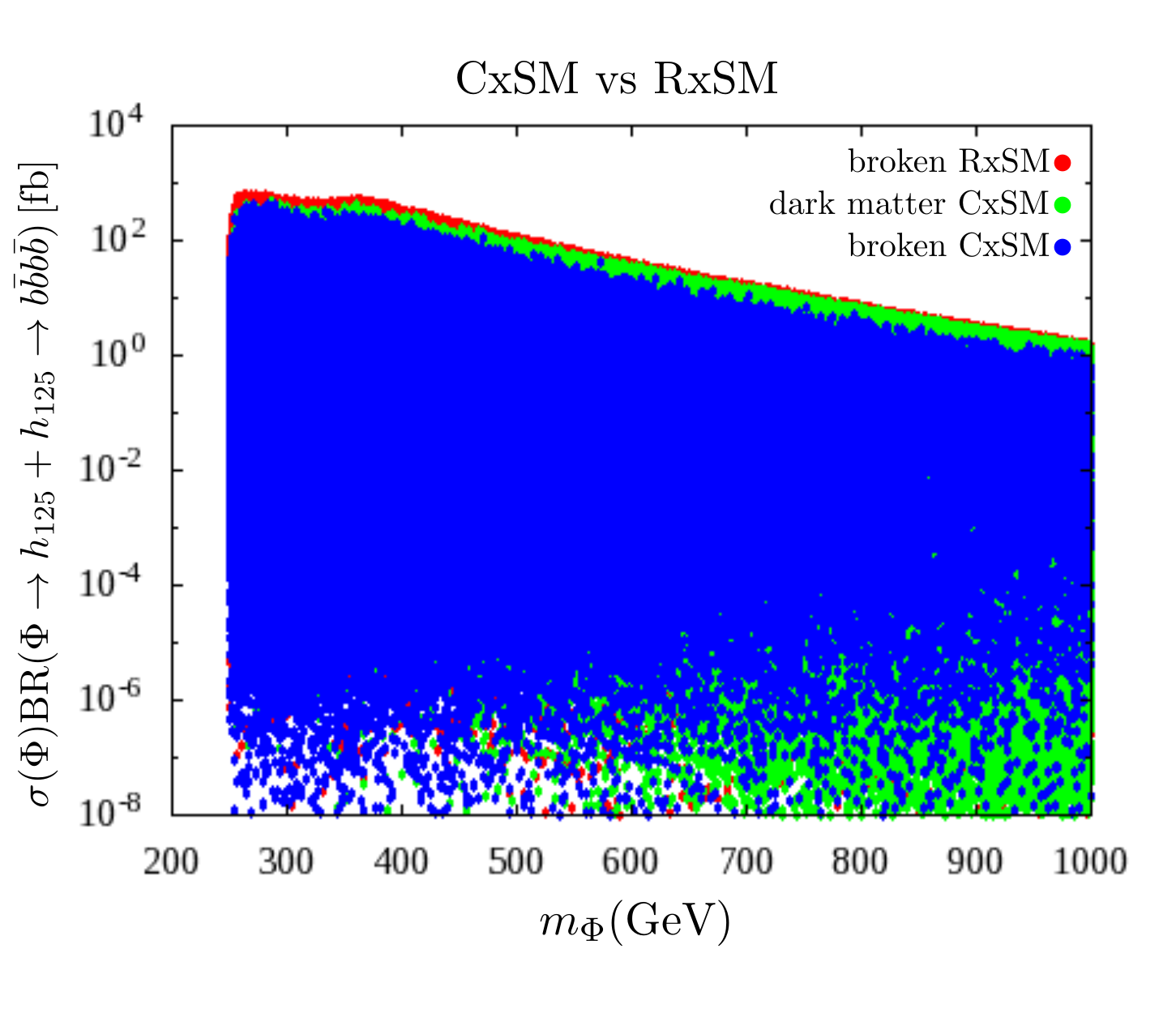}\includegraphics[width=0.51\linewidth]{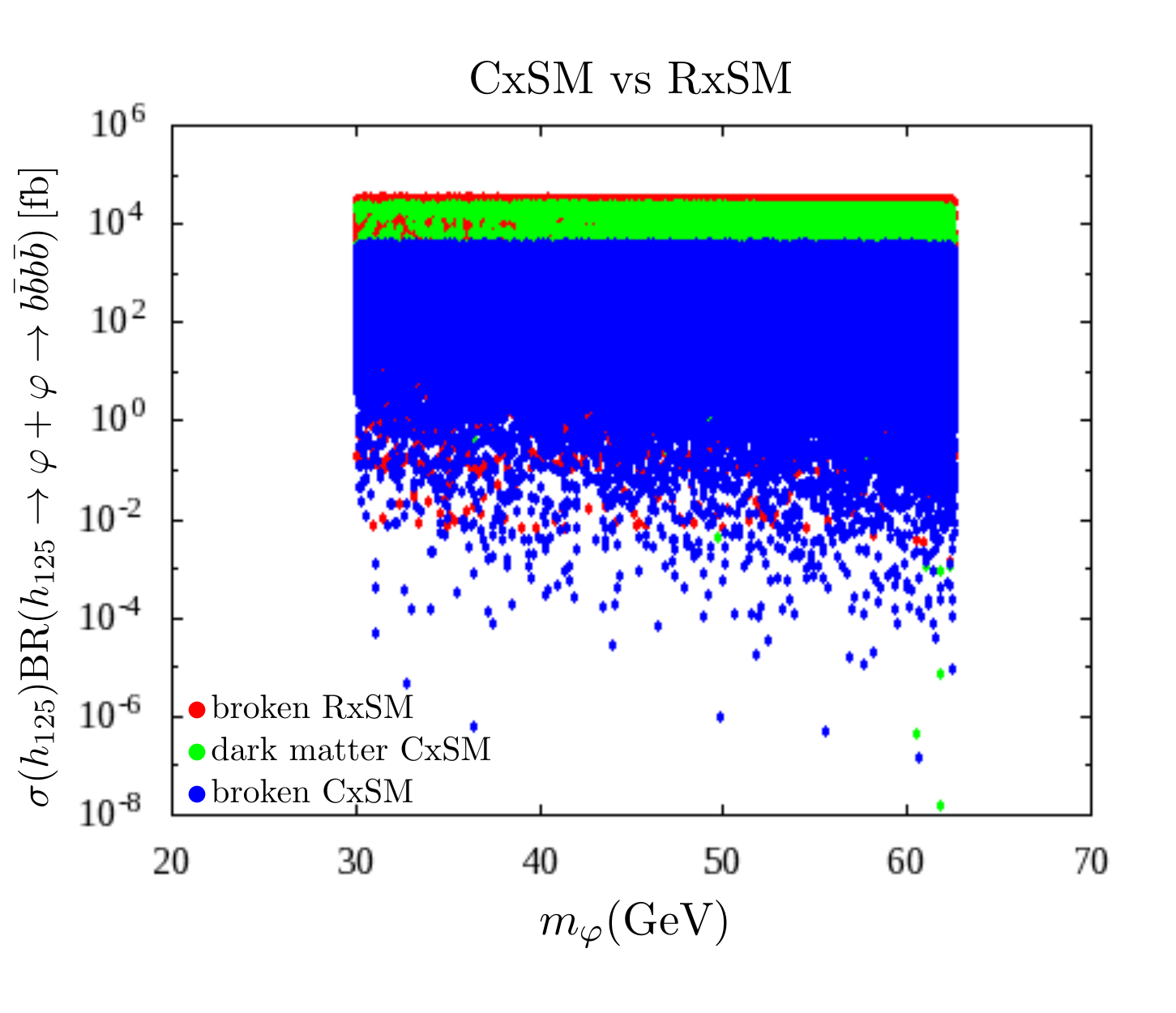}}
\caption[hum]{The $4b$ final state rates for a heavier Higgs
    $\Phi$ decaying into two SM-like bosons $h_{125}$ (left) and for
    the case where $h_{125}$ decays into a pair of lighter bosons
    $\varphi$. The production process is gluon fusion at a c.m.~energy
    of $\sqrt{s}=13$~TeV. Blue points: CxSM-broken; 
    green: CxSM-dark; and red: RxSM-broken.
   }
\label{fig:RxSMvsCxSMdarkAndBroken}
\end{figure}

We discuss the broken phase of the RxSM because it is the only one
allowing for Higgs-to-Higgs decays in this model. Depending on the
mass of the non-SM-like additional Higgs boson, we have two
possible scenarios for the comparison of RxSM-broken and CxSM in its
symmetric (dark) and its broken phase:
\begin{enumerate}
\item[1)] {\em {Scalar decaying into two SM-like Higgs bosons:}} This
  case is realized if the non-SM-like Higgs state $\Phi$ is heavy
  enough to decay into a pair of SM-like Higgs states $h_{125}$. The
  real singlet case has to be compared to all possible decays of the
  complex case, which are the same in the dark matter phase of
  the CxSM. In the 
  broken phase, however, we have the possibilities $\Phi \equiv
  h_{3,2} \to h_1 + h_1$ for $h_1 \equiv h_{125}$ and $\Phi \equiv h_3
  \to h_2 + h_2$ for $h_2 \equiv h_{125}$. With the $b$-quark final
  state representing the dominant decay channel, we compare for
  simplicity only $4b$ final states. 
\item[2)] {\em {SM-like Higgs boson decaying into two identical Higgs
      bosons:}} Here the non-SM-like Higgs is lighter than the SM-like
  Higgs boson, and we denote it by $\varphi$. If it is light enough
  the decay $h_{125} \to \varphi + \varphi$ is possible. This is the
  only case that can be realized in RxSM-broken and CxSM-dark. In
  contrast, the decay possibilities of CxSM-broken are given by 
  $h_3 \to \varphi + \varphi$ with $\varphi = h_1$ or $h_2$ 
  for $h_3 \equiv h_{125}$, and $h_2 \to \varphi + \varphi$ with
  $\varphi \equiv h_1$ for $h_2 \equiv h_{125}$. For simplicity again $4b$ final
  states are investigated. 
\end{enumerate}
Figure~\ref{fig:RxSMvsCxSMdarkAndBroken} (left) shows the decay rates
for case 1) in the broken phase (blue points) and the dark matter phase (green)
of the CxSM, and in the RxSM-broken (red). It demonstrates, that the maximum rates in the
RxSM-broken exceed those of the CxSM, although the differences are not
large. 

The results for case 2) are displayed in
Fig.~\ref{fig:RxSMvsCxSMdarkAndBroken} (right). Again the maximum
rates in RxSM-broken are larger than those in the CxSM. In the dark
matter phase the maximum rates are not much smaller, while the
largest rates achievable in CxSM-broken lie one order of magnitude
below those of the RxSM. We have verified that the
larger rates allowed for the models with two-by-two mixing (CxSM-dark
and RxSM) result from their different vacuum structure. The larger
rates can be traced back to the branching ratio for the Higgs-to-Higgs
decay, ${\rm BR}(h_{125}\to \varphi+\varphi)$, which differs, from
model to model, in its allowed parameter space for the new scalar
couplings of the theory.

%%%%%%%%%%%%%%%%%%%%%%%%%%%%%%%%%%%%%%%%%%%%%%%%%%%%%%%%%%%
\subsubsection{Comparison of CxSM-broken with the NMSSM} \label{sec:cxsmnmssmcomp}
We now turn to the comparison of the Higgs-to-Higgs decay rates that
can be achieved in the broken phase of the complex singlet extension
with those of the NMSSM.  We will focus, in our discussion, on the comparison of final state signatures that are common to both models. We will not consider additional decay channels, that are
possible for the NMSSM Higgs bosons, such as decays into lighter SUSY
particles, namely neutralinos or charginos, or into a gauge and Higgs
boson pair. These decays would add to the 
distinction of the CxSM-broken from the NMSSM, if they can be
identified as such. Depending on the NMSSM parameter points, various
such decay possibilities may be possible and will require a dedicated
analysis that is beyond the scope of this paper. Our approach
here is a different one. 
We assume we are in a situation where we have found so far only a
subset of the Higgs bosons common to both models, where we have not
observed any non-SM final state signatures yet and where we do not have
information yet on the CP properties of the decaying Higgs
boson. Additionally, we assume that we do not observe any final state
signatures that are unique in either of the models.\footnote{In short,
we assume the difficult situation in which no obvious non-SM signal nor unique
signal for either of the models has been observed.} 
We then ask the question: If one focusses on Higgs-to-Higgs decays only in final states that are common to both models, will it be
possible to tell the CxSM-broken from the NMSSM based on the total
rates? 
The discovery of additional
non-CxSM Higgs bosons and the observation of non-CxSM final state
signatures would add new information but not limit our 
findings with respect to the distinction of the models based on the
signatures that we investigate here. Our analysis,
  provided the answer is positive, can therefore be seen as a trigger
  for further studies in the future taking into account other decay
  processes.

To organise the discussion we distinguish four cases, which are in
principle different from the point of view of the 
experimental searches. The cases that are possible both in the
CxSM-broken and in the NMSSM are:
\begin{enumerate}
\item[1)] {\em Scalar decaying into two SM-like Higgs bosons:} 
Both in the CxSM-broken and in the NMSSM this Higgs-to-Higgs decay is
possible for scenarios where the SM-like Higgs boson is the lightest
($h_1\equiv h_{125}$) or next-to-lightest Higgs boson ($h_2\equiv
h_{125}$). The corresponding decays are then $h_{3,2}\to h_{1} +
h_{1}$ in the former case and only $h_3\rightarrow h_2 + h_2$ in the
latter case.\footnote{For simplicity, we use here and in the
  following, where appropriate, the notation with small
$h$ both for the CxSM-broken and the NMSSM. Otherwise, whenever extra
channels are available in the NMSSM, we specify them with upper case
notation.}
  Since the SM-like Higgs boson dominantly decays into
  $b$-quarks, we concentrate on the $4b$ final state. The approximate
  rate for $\tau$ lepton 
  final states can be obtained by multiplying each SM-like Higgs boson
  decay into $b$-quarks by $1/10$. 

\item[2)] {\em {Higgs decaying into one SM-like Higgs
        boson and a new Higgs state}:} 
The resonant decay channels that give rise to these final
states in both models are $h_3\to h_1+h_2$ with
$h_{125}\equiv h_1$ or $h_2$. In addition, in the NMSSM, the channel
$A_2\to A_1 + h_{125}$ provides a similar signature if the CP numbers are
not measured. The SM-like Higgs boson
  dominantly decays into a 
  pair of $b$-quarks. For the new scalar produced in association, in
  the low mass region the most important decay channel is the one into
  $b$-quarks followed by the decay into $\tau$
  leptons.\footnote{Decays into muons and lighter quarks can
    dominate for very light Higgs bosons with masses below the
    $b$-quark threshold. As the scans for the singlet models do not
    include such light Higgs bosons this possibility does not apply for the scenarios
    presented here.}
Decays into photons might become interesting due to their clean
signature. 
In the high mass regions, if kinematically allowed, the 
  decays into pairs of massive vector bosons $V\equiv W,Z$ and of top
  quarks become relevant. 
    In the NMSSM the importance of
  the various decays depends on the 
  value of $\tan\beta$ and the amount of the singlet component in the
  Higgs mass eigenstate,  
  whereas in the singlet models it only depends on the
  singlet admixture to the doublet state. 
In order to simplify the discussion, we resort to the $4b$ final state.
We explicitly checked other possible final states to be sure that they
do not change the conclusion of our analysis in the following. 

\item[3)] {\em SM-like Higgs boson decaying into two 
    light Higgs states:} If the SM-like Higgs is not the lightest
  Higgs boson, then it can itself decay into a lighter Higgs boson pair. 
The new decay channels into Higgs bosons add to the 
total width of the SM-like state so that its branching ratios
and hence production rates are changed. Care has to be taken not to
violate the bounds from the LHC Higgs data. In the NMSSM, the masses
are computed from the input parameters and are subject to
supersymmetric relations, so that such scenarios are not easily
realised. Both in the CxSM-broken and the NMSSM the decays
$h_{125} \to h_1 + h_1$ are possible. In the CxSM we additionally have
for $h_{125} \equiv h_3$ the decays $h_{125} \to h_1+h_1, h_1+h_2$ and
$h_2 + h_2$  while the NMSSM features in
addition $h_{125} \to A_1 + A_1$ decays with $h_{125} \equiv h_1$ or
$h_2$. Given the lower masses of the Higgs boson pair, final states
involving bottom quarks, $\tau$ leptons and photons are to be investigated here. 

\item[4)] {\em New Higgs boson decaying into a pair of
        non-SM-like identical Higgs bosons:} 
The decays that play a
role here for the two models are $h_3 \to h_1 + h_1 \; (h_3 \to h_2 + h_2)$
in case $h_{125} \equiv h_2 \; (h_1)$. In the NMSSM also $h_3 \to A_1
+ A_1$ is possible as well as $h_2 \to A_1 + A_1 \; (h_1 \to A_1 + A_1)$ for
$h_{125} \equiv h_1 \; (h_2)$. In search channels
  where a non-SM-like Higgs boson 
  is produced in the decay of a heavier non-SM-like Higgs boson we
  have a large variety of final state signatures. Leaving apart the
  decay channels not present in the CxSM-broken, the final state Higgs
  bosons can decay into massive gauge boson or top quark pairs, if
  they are heavy enough. Otherwise final states with $b$-quarks,
  $\tau$ leptons and photons become interesting.
As we found that the various final states do not change our following
conclusions, for simplicity we again focus on the 
$4b$ channel in the comparison with the NMSSM.  
\end{enumerate}  

Note that, in the CxSM, we do not have the possibility of a heavier Higgs decaying into a pair of non-SM-like Higgs bosons that are not identical, so
that this case does not appear in the above list.

\begin{figure}[b!]
\centering
\mbox{\includegraphics[width=0.51\linewidth]{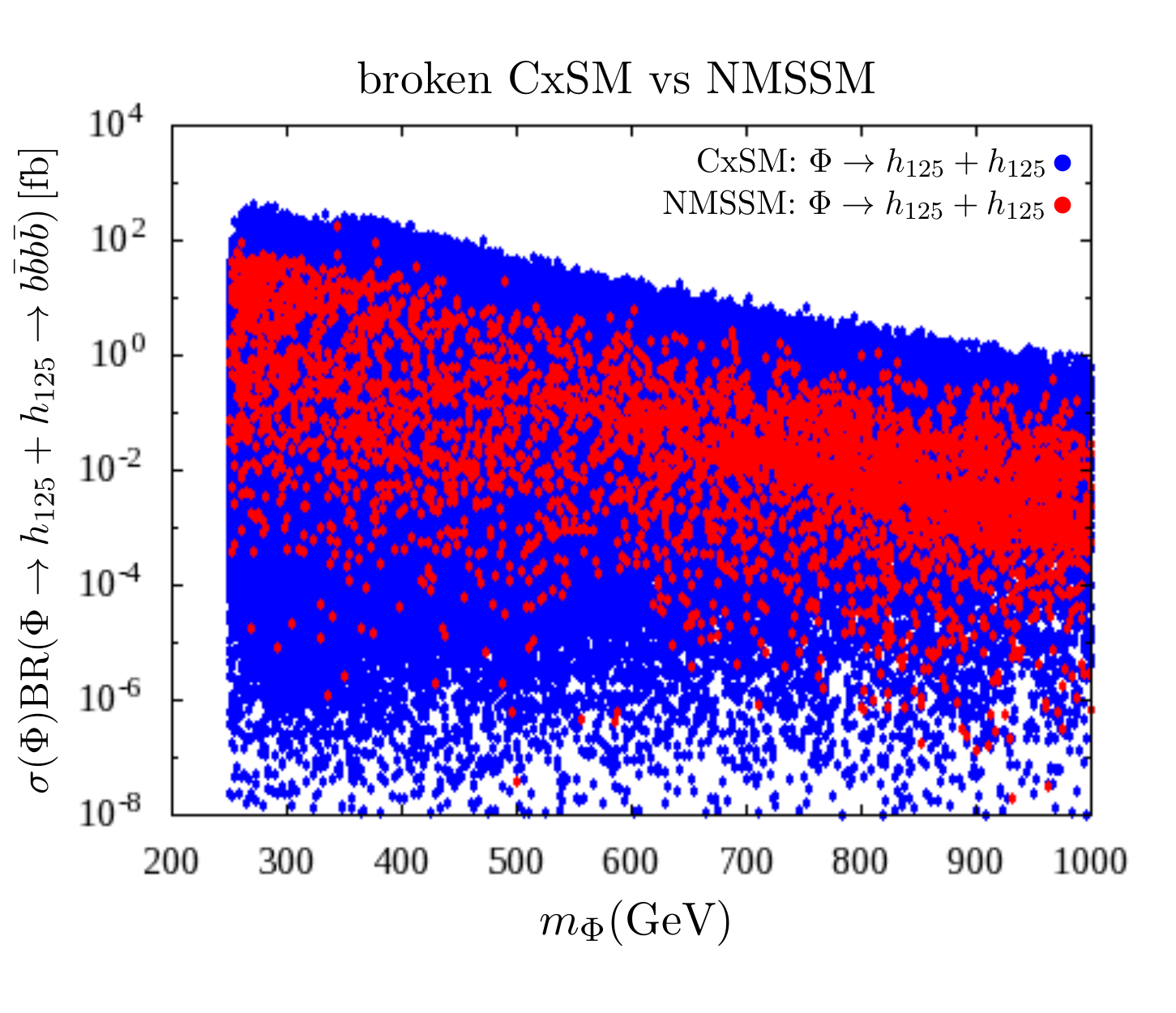}} 
\caption[hum]{The 4$b$ final state rates for the production
  of a heavy Higgs boson $\Phi=h_{3,2} (h_3)$ decaying into two
  SM-like Higgs states $h_{125} \equiv h_1 (h_2)$, that subsequently
  decay into $b$-quarks, in the 
  CxSM-broken (blue points) and the NMSSM (red points).}
\label{fig:NMSSM_CxSM_comparison_case1}
\end{figure}
Figure~\ref{fig:NMSSM_CxSM_comparison_case1} shows the
rates corresponding for case 1). A
heavier Higgs boson $\Phi$ is produced and decays into a pair
of SM-like Higgs bosons $h_{125}$ that subsequently decay into a
$b$-quark pair each. The red (blue) points represent all decays that are
possible in the NMSSM (CxSM-broken), i.e.~$\Phi = h_{3,2}$ for $h_{125} \equiv h_1$
and $\Phi = h_3$ for $h_{125} \equiv h_2$. As can be inferred from the
plot, such decay chains do not allow for a distinction of the two
models. The maximum possible rates in the NMSSM
can be as high as in the CxSM. The differences in the lowest possible
rates for the two layers are due to the difference in the density of
the samples. Such small rates, however, are not accessible
experimentally. Note finally that here and in the following plots the
inclusion of all possible Higgs-to-Higgs decay channels in each of the
models is indeed essential. Otherwise a fraction of points might be 
missed and a possible distinction of the models might be falsely
mimicked.

\begin{figure}[hb!]
\centering
\mbox{\includegraphics[width=0.51\linewidth]{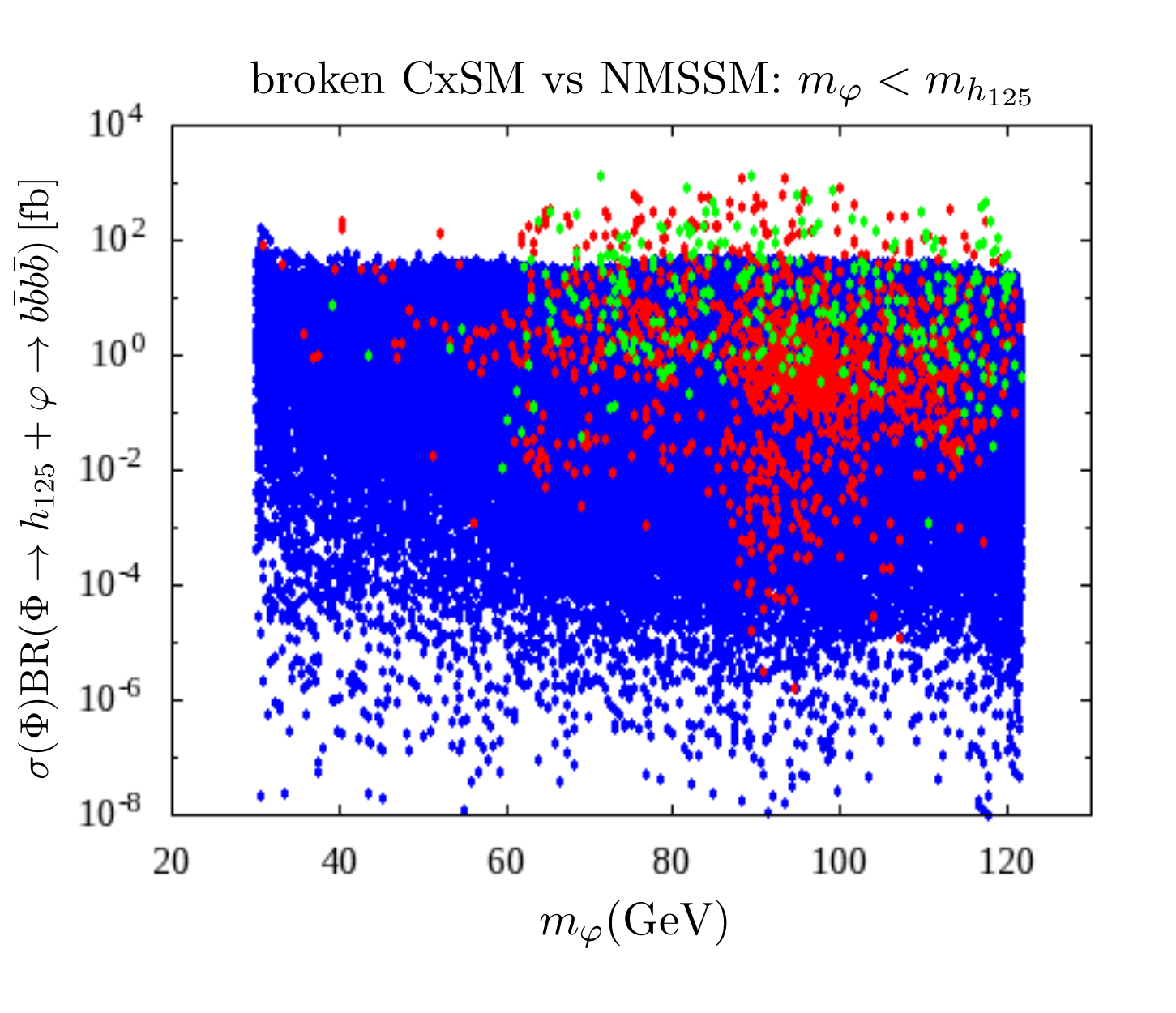}\includegraphics[width=0.51\linewidth]{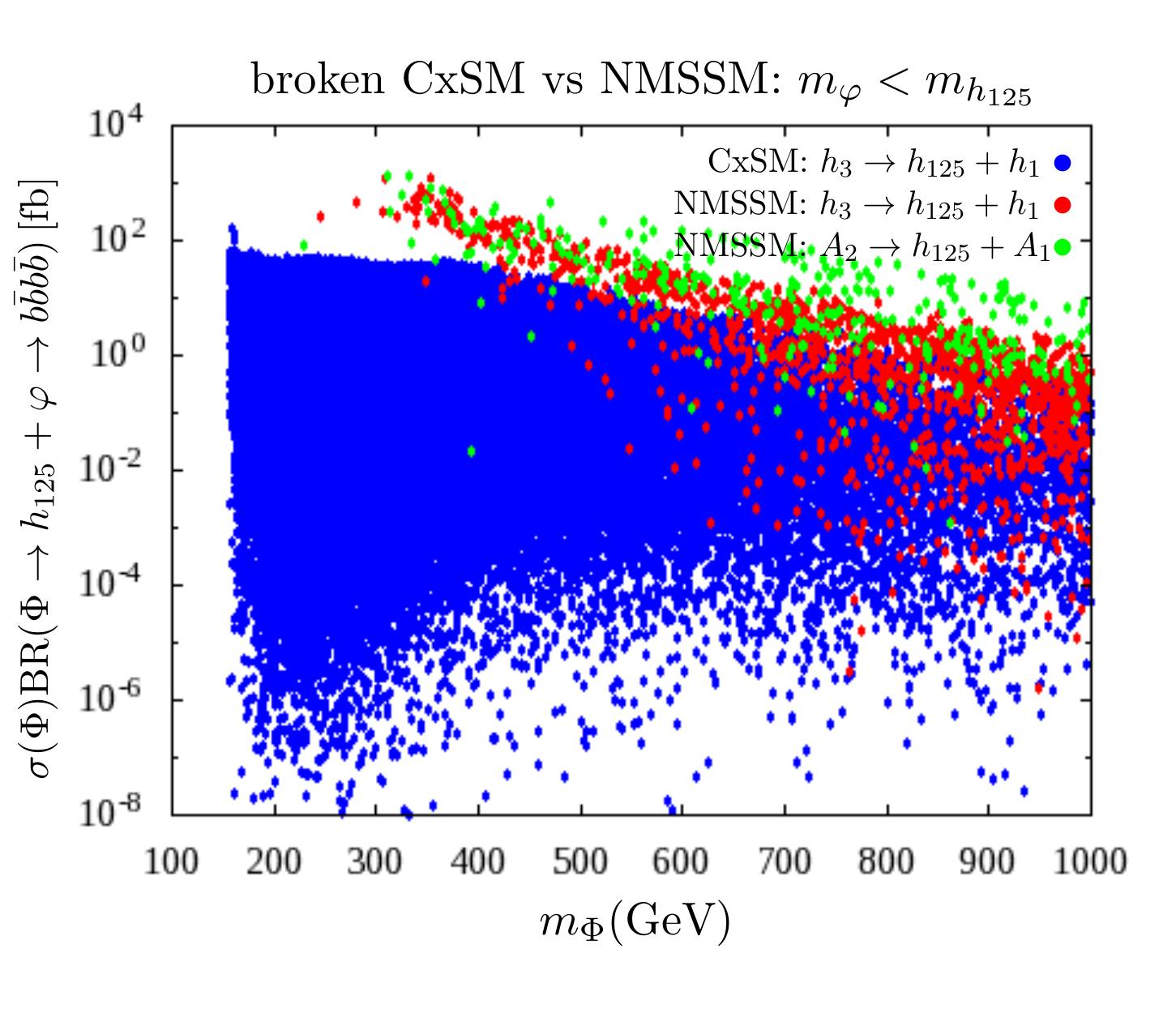}} 
\caption[hum]{The 4$b$ final state rates for the production
  of a heavy Higgs boson $\Phi$ decaying into a
  SM-like Higgs state $h_{125}$ and a non-SM-like
  light Higgs boson $\varphi$ with $m_\varphi < m_{h_{125}}$, that subsequently
  decay into $b$-quarks. Left (right) plot: as a function of $m_\varphi$
  ($m_\Phi$). Blue (CxSM-broken) and red 
  (NMSSM) points: $\Phi \equiv h_3$, $h_{125} \equiv h_2$ and
  $\varphi \equiv h_1$; green points (NMSSM): $\Phi \equiv A_2$,
  $h_{125} \equiv h_{1,2}$ and $\varphi \equiv A_1$.}
\label{fig:NMSSM_CxSM_comparison_case2}
\end{figure}
The rates for case 2) are displayed in
Fig.~\ref{fig:NMSSM_CxSM_comparison_case2}. Here a heavier Higgs
$\Phi$ decays into $h_{125}$ and a non-SM-like lighter Higgs boson
$\varphi$, where the plot strictly refers to scenarios with
$m_\varphi < m_{h_{125}}$. Thus the blue points (CxSM) and red points (NMSSM) represent the cases 
$\Phi= h_3$ and $\varphi = h_1$ with $h_2 \equiv
h_{125}$. Additionally, in the NMSSM, the green points have to be included
to cover the case $\Phi = A_2$, $\varphi=A_1$ with $h_{125}
\equiv h_1$ or $h_2$. The possible rates in the two models are shown
as a function of $m_\varphi$ (left plot) and as a function of $m_\Phi$
(right plot). The overall lower density of points in the NMSSM is due
to its higher dimensional parameter space which, combined with its
more involved structure, limits the computational speed in the generation of the samples. 
The performed scan starts at 30~GeV for the lightest Higgs boson mass,
so that the Higgs-to-Higgs decays set in at $m_\Phi = 155$~GeV (blue
points, right plot). There are no points above
$m_\varphi = 121.5$~GeV due to 
the imposed minimum mass difference of 3.5~GeV from the SM-like Higgs
boson mass to avoid degenerate Higgs
signals. The left plot shows that the masses of the $A_1$, in the NMSSM, are mostly larger than about 60~GeV. While a more extensive scan of the NMSSM
could yield additional possible scenarios, light Higgs masses
allow for $h_{125}$ decays into Higgs pairs, that move the $h_{125}$
signal rates out of the allowed experimental range. This explains why
there are less NMSSM points for very small masses $m_\varphi$.
The figure clearly
demonstrates that the maximum achieved rates of the NMSSM in this
decay chain can be 
enhanced by up to two orders of magnitude compared to the CxSM over
the whole mass ranges of $m_\varphi$ and $m_\Phi$ where they are
possible. The observation 
of a much larger rate than expected in the CxSM-broken in the decay of a
heavy Higgs boson into a SM-like Higgs and a lighter Higgs state would
therefore be a hint to a different model, in this case the NMSSM. 

The enhancement of these points can be traced back
to larger values for the production cross sections of the heavy Higgs
boson, for the Higgs-to-Higgs decay branching ratio and for the
branching ratios of the lighter Higgs bosons into $b$-quark pairs in
the NMSSM compared to the CxSM. In the NMSSM the larger production
cross sections are on the one hand due to pseudoscalar Higgs production. In gluon fusion these yield larger cross sections than for
scalars, provided that the top Yukawa couplings are similar, in scenarios where the top loops dominate.  Additionally, the
investigation of the top Yukawa coupling, 
which is the most important one for gluon fusion for
  not too large values of $\tan\beta$, shows that in the NMSSM for the
enhanced points it is close to the SM coupling or even somewhat larger. In comparison with these NMSSM top-Yukawa couplings, in the CxSM the top-Yukawa couplings are suppressed for all parameter points. This is because in the CxSM all Higgs couplings to SM
  particles can at most reach SM values and are typically smaller, for
  the new Higgs bosons, due to the sum rule $\sum_iR_{i1}^2=1$. This rule
  assigns most of the coupling to the observed Higgs
  boson and therefore its coupling factor can not deviate too much from 1. We also have NMSSM scenarios where the bottom Yukawa coupling is larger by one to two orders of magnitude in
the NMSSM compared to the CxSM, This enhancement is due to
$\tan\beta$, for which we allow values between 1 and 30. In this case, where the top Yukawa coupling is suppressed, the Higgs bosons are dominantly produced in $b\bar{b}$ annihilation. 
The behaviour of the branching
ratios can be best understood by first recalling that in the CxSM the
branching ratios for each Higgs boson are equal to those of a SM Higgs
boson with the same mass, in case no Higgs-to-Higgs decays are present. This is because all Higgs
couplings to SM particles are modified by the same factor. If Higgs-to-Higgs decays are allowed the branching ratios even drop below the corresponding SM value. In the NMSSM, however, the branching ratio
${\rm BR}(\varphi\to b \bar{b})$ is close to 1. Despite $\varphi$ being the
singlet-like $A_1$ or $H_1$ it dominantly decays into $b\bar{b}$ as
the SUSY particles of our scan are too heavy to allow for $\varphi$ to decay into them. As for the branching ratio ${\rm BR}(h_{125}\to b\bar{b})$, in the CxSM it can reach at most the SM value of around\footnote{Note that we have
    consistently neglected EW corrections.}
0.59,
while in NMSSM this branching ratio can be of up to about 0.7 due to enhanced couplings to the $b$-quarks.
Note, in
particular, that the uncertainty in the experimental value of the SM
Higgs boson rates into $b\bar{b}$ in the NMSSM still allows for
significant deviations of the SM-like Higgs coupling to bottom quarks
from the SM value. Finally, the branching ratio ${\rm BR}(\Phi \to h_{125}
\varphi)$ for the enhanced points is larger in the NMSSM compared to
the singlet case. This can be either due to the involved trilinear Higgs
self-coupling or due to a larger phase space. The self-coupling
$\lambda_{\Phi h_{125} \varphi}$ depends on a combination of the  
NMSSM specific parameters $\lambda,
  \kappa, A_\kappa, A_\lambda, v_s$, the Higgs mixing matrix elements
  and $\tan\beta$. Through the Higgs mixing matrix
    elements it also depends on soft SUSY breaking masses and
    trilinear couplings, as we include higher order corrections in 
the NMSSM Higgs masses and mixing matrix elements. Therefore a 
distinct NMSSM parameter region or 
  specific combination of parameters that is responsible for the
  enhanced self-couplings compared to the CxSM cannot be
  identified. The same holds for a possible different mass
  configuration of the involved Higgs bosons, with the
    higher-order corrected masses depending on NMSSM specific and
  SUSY breaking parameters. 

\begin{figure}[b!]
\centering
\mbox{\includegraphics[width=0.51\linewidth]{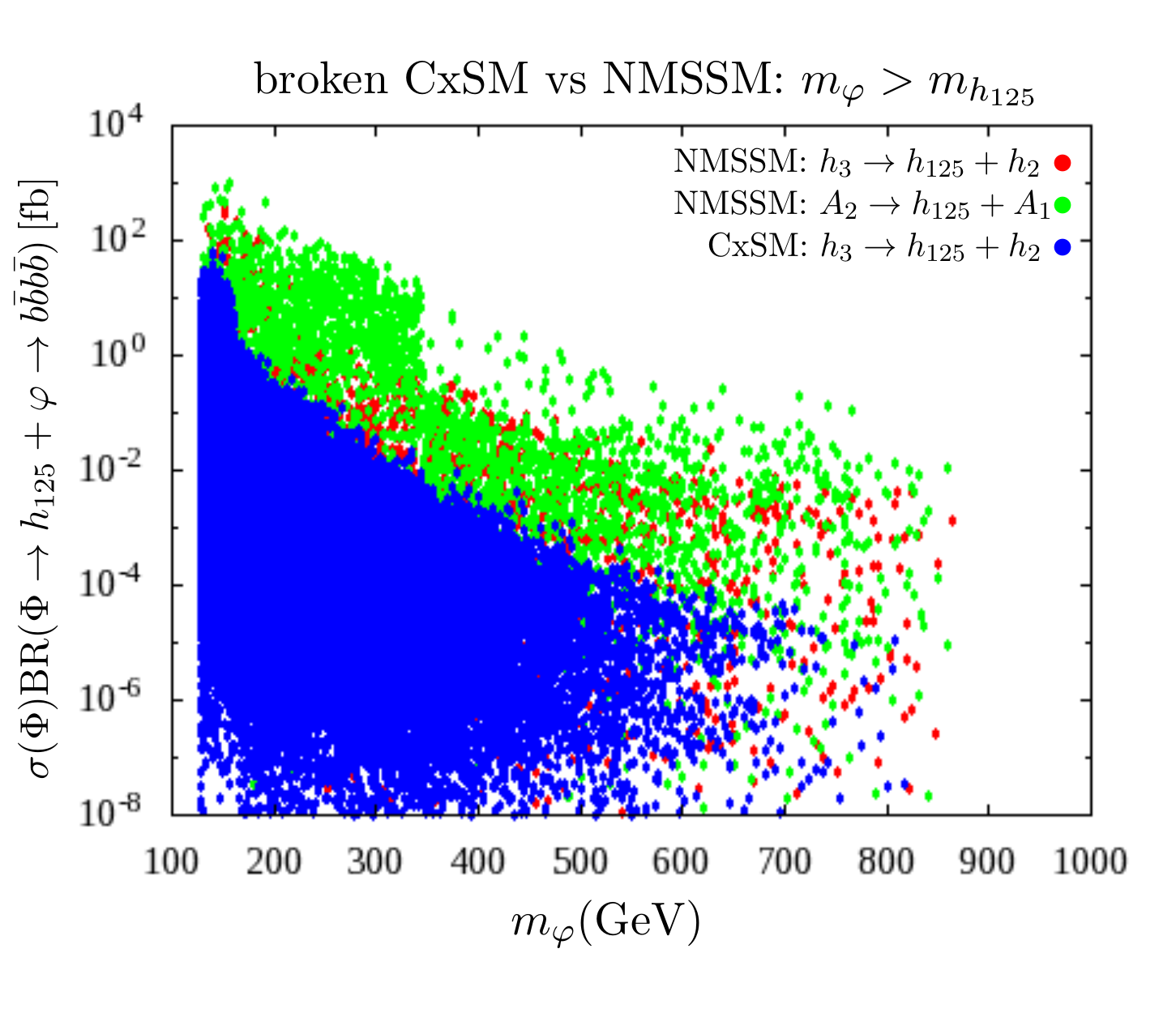}\includegraphics[width=0.51\linewidth]{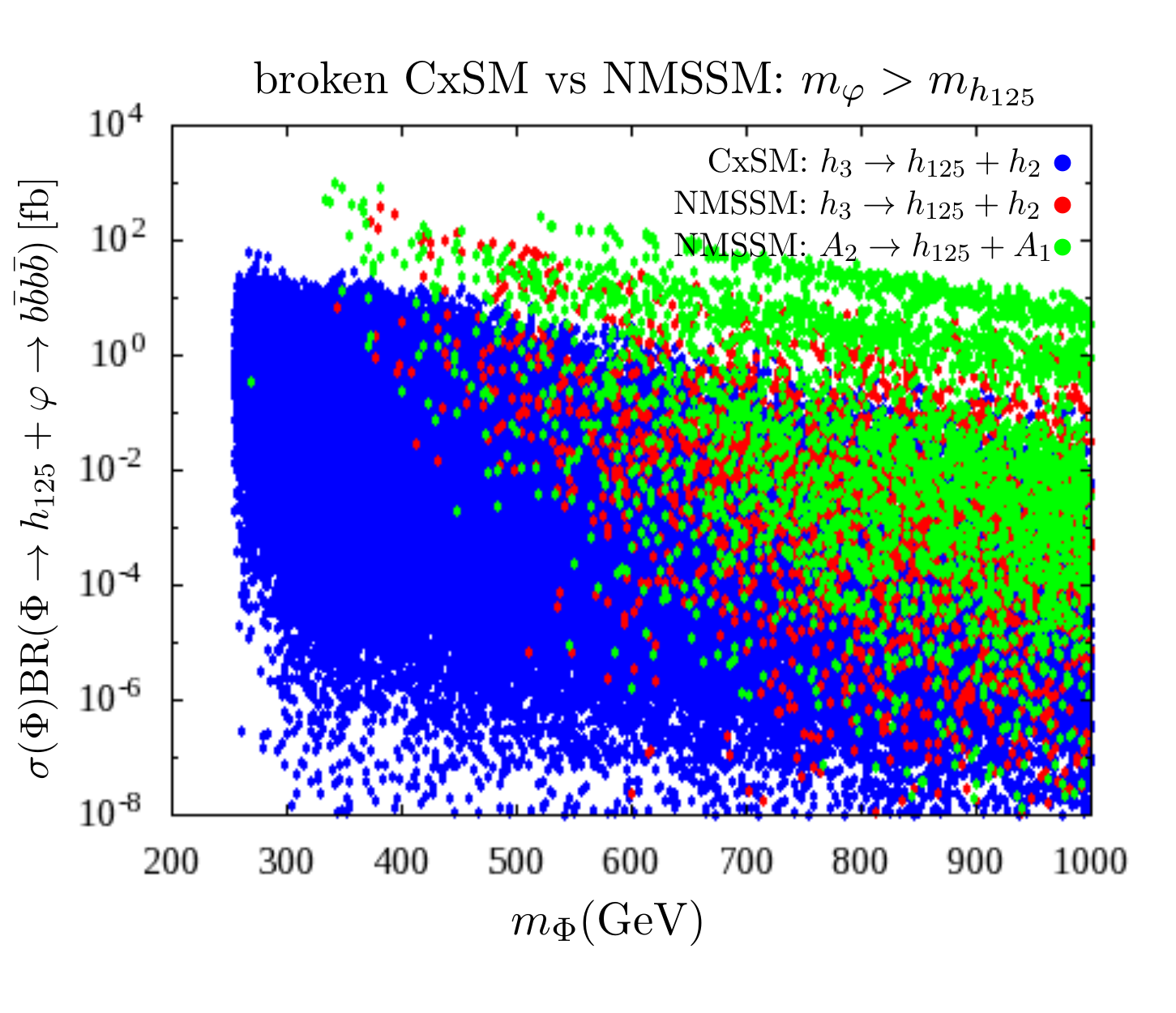}}  
\caption[hum]{The 4$b$ final state rates for the production
  of a heavy Higgs boson $\Phi$ decaying into a
  SM-like Higgs state $h_{125}$ and a non-SM-like
  light Higgs boson $\varphi$ with $m_\varphi > m_{h_{125}}$, that subsequently
  decay into $b$-quarks. Left (right) plot: as a function of $m_\varphi$
  ($m_\Phi$). Blue (CxSM-broken) and red 
  (NMSSM) points: $\Phi \equiv h_3$, $h_{125} \equiv h_1$ and
  $\varphi \equiv h_2$; green points (NMSSM): $\Phi \equiv A_2$,
  $h_{125} \equiv h_{1,2}$ and $\varphi \equiv A_1$.
}
\label{fig:NMSSM_CxSM_comparison_case2b}
\end{figure}
Figure~\ref{fig:NMSSM_CxSM_comparison_case2b} also refers to case 2),
but this time the mass of the non-SM-like Higgs boson $\varphi$ is
larger than $m_{h_{125}}$. With this mass configuration the blue and red points
of the CxSM and NMSSM, respectively, represent the cases 
$\Phi= h_3$ and $\varphi = h_2$ with $h_1 \equiv h_{125}$. In the NMSSM we also have the possibilities $\Phi = A_2$,
$\varphi=A_1$ and $h_{125}\equiv h_1$ or $h_2$, which are covered 
by the green points. In the left plot the points set in at 
$m_\varphi=128.5$~GeV due to the imposed minimum mass distance from
$m_{h_{125}}$. 
The upper limits on $m_\varphi$ in both models are
due to the bounds on the input parameters chosen for the
scans. In the right plot the 
points start at the kinematic lower limit
of 253.5~GeV.
Both plots clearly demonstrate, that the NMSSM
Higgs-to-Higgs decay rates exceed those of the CxSM-broken in the
whole mass range of $m_\varphi$, respectively $m_\Phi$, by up to two
orders of magnitude and allow for a distinction of the models if the
largest possible signal rates in the NMSSM are discovered. The enhancement can be understood by looking at the production
cross section of the heavy Higgs boson and at the various involved
branching ratios. The enhanced production compared to the CxSM case is
again most important in pseudoscalar NMSSM Higgs production, but also the
$H_3$ production can be somewhat enhanced, because in the singlet case the
heavy Higgs couplings to tops and bottoms are 
suppressed compared to the SM, while in the NMSSM we can have points where the top Yukawa coupling can be close to the SM value or even a bit larger and we can also have other points where the bottom Yukawa coupling can be much larger than the corresponding SM coupling while the top Yukawa coupling is suppressed. We have verified that there are scenarios where the top Yukawa coupling provides the dominant contribution to the cross section enhancement relative to the CxSM and other scenarios where it is the bottom Yukawa coupling.  
In the region where
the NMSSM points yield larger rates, the branching ratio ${\rm BR}(\Phi
\to h_{125} \varphi)$ can be larger, but there are also cases, where
the singlet case and the NMSSM lead to similar branching ratios. However,
it turns out that the value of ${\rm BR}(\varphi \to b\bar{b})$ in the NMSSM clearly exceeds that of the
singlet case. For masses below the top pair threshold, it can reach
values close to 1 in the NMSSM. In the singlet case it reaches at most the
value of a SM Higgs boson with the same mass, or lower, when the decay into
$h_{125} h_{125}$ is kinematically possible. In the NMSSM
below the top pair threshold the only important decay is the one into
$b$-quark pairs, as the SUSY particles from our scan are too heavy and
decays into massive gauge bosons are forbidden for $\varphi = A_2$
because it is CP-odd, and for $\varphi = H_3$ because of the coupling
sum rules for the scalar Higgs couplings to gauge bosons. Above the
top pair threshold $\varphi$ can then also decay into $t\bar{t}$. For
large values of $\tan\beta$, however, the branching ratio into $b\bar{b}$
can reach values close to 1. In the cases where
also decays into lighter SUSY particles, Higgs pairs or
Higgs and gauge boson pairs are kinematically possible, the branching
ratio into $b\bar{b}$ drops below 1, but is still larger than the corresponding
branching ratio in the CxSM.
As for the larger $h_{125}$ branching ratio into
$b\bar{b}$ we have discussed this already in the context of
Fig.~\ref{fig:NMSSM_CxSM_comparison_case2}. 

 The clear
difference in the maximal values of the rates between CxSM-broken and
NMSSM is less pronounced in the $2b2W$ final state, in particular for
small masses $m_\varphi$ and $m_\Phi$. This is obvious, as here it is
the decay $A_2 \to h_{125} + A_1$ that leads to larger rates (green
points) than the CxSM counterpart. As the pseudoscalar $A_1$, however,
does not decay into massive gauge bosons, these rates go down in the
$2b2W$ final state. 

The investigation of case 3) (not plotted here), where a SM-like Higgs
decays into two light Higgs states with subsequent decays into SM
particles, shows that the maximum possible NMSSM rates are 
slightly larger than the maximum rates achieved in the
CxSM-broken. This decay chain also takes place in RxSM-broken with
maximum rates somewhat exceeding those of the NMSSM. The maximum
possible rates are clearly limited by the observed rates for
$h_{125}$. An enhanced rate of the light $\varphi$ into SM particles
leaves some room to further increase the rate without violating the
experimental measurements of the $h_{125}$ rates. In the NMSSM, however,
the light scalar 
states $h_1$ and $A_1$ are mostly singlet-like with suppressed
couplings to SM particles. The maximum possible rates in the NMSSM
therefore cannot be expected to exceed those of RxSM-broken by far, so
that the distinction of these two models (and hence also of the NMSSM
and CxSM-broken) based on SM-like Higgs decays into lighter Higgs
states is not possible.  

In case 4) (also not plotted here) with a heavy Higgs boson decaying into a pair of
non-SM-like Higgs bosons $\varphi$ with $m_\varphi > m_{h_{125}}$ the
maximum possible NMSSM rates exceed those of the CxSM in the $4b$ final
state for certain kinematic configurations where $m_\varphi \gsim
170$~GeV. With maximum values of about 1~fb,  they are, however, too
small to be exploited. On the other hand the rates of the $4W/4Z$ final state 
yields much larger rates in both models. However, the maximum rates 
in the CxSM and NMSSM are for these final states comparable making impossible
to distinguish them. In the opposite case, where $m_\varphi < m_{h_{125}}$, in the
dominant $4b$ final state the maximum rates are again comparable and
no distinction of the models is possible.

To finalise the discussion we remark that our conclusions do not
depend on the approximation of Higgs pair production from the
decay of a resonantly produced heavy Higgs boson. In order to verify
this we computed for all possible di-Higgs final states the complete
Higgs pair production processes from gluon fusion which take into
account both the resonant and the continuum diagrams. For the
computation of these processes we implemented our 
models ({\it i.e.}~the NMSSM and the complex singlet extension CxSM)
into the existing code {\tt HPAIR}\footnote{See M.~Spira's website, http://tiger.web.psi.ch/proglist.html.}. This Fortran code, initially designed for
the SM and the MSSM, calculates Higgs pair production at NLO QCD
accuracy. We found that the difference between the resonant approximation and
the full calculation, including the continuum, is typically of the
order of $\sim 10\%$ in the region of the plots with large rates
where we can distinguish the CxSM from the NMSSM. However, it should be noted
that in a full experimental analysis, the resonant production gives an extra handle
in reducing the background relative to the continuum. This extra
handle will surely make the signal significance in the resonant case 
much larger than the continuum one. In fact, although we are presenting here total cross sections,
the signal events are mainly concentrated in the bins around the peak of the resonant scalar.
This kinematical cut around the peak, even if broad, is only possible in the resonant case.   
Thus our conclusions are unaffected by the use of the approximation.

%%%%%%%%%%%%%%%%%%%%%%%%%%%%%%%%%%%%%%%%%%%%%%%%%%%%%%%%%%
\subsection{Benchmark Points for the LHC Run 2}
\label{Sec:LHC_Run2_benchs}

In this section we provide a set of benchmark points with focus on the
CxSM. At the end, we also discuss the RxSM which is simpler with
respect to the possibilities for Higgs-to-Higgs decays.

As discussed in section~\ref{sec:models}, in these singlet models, at
leading order in the electroweak corrections, the production cross
sections and decay widths into SM particles are given by multiplying
the corresponding SM result with the mixing matrix element squared,
$R_{i1}^2$, cf. Eq.~\eqref{eq:couplings_SM}. One can show that the
signal strength factor is approximately given by
\begin{equation}
\mu_{i} \simeq R_{ih}^2\sum_{X_{\rm SM}}{\rm BR}_{\rm New}(h_i\rightarrow X_{\rm SM}) \;.
\end{equation}  
There is hence only one common signal strength factor regardless of
the decay channel into SM particles. This provides a useful measure for
the deviation of the model point from the SM, and it is given by the
squared overlap of the scalar state with the SM Higgs fluctuation
(i.e. the mixing matrix element) multiplied by the sum of the
branching ratios into SM particles. If there are no Higgs-to-Higgs decays
the latter factor is one.

In the next subsections we present several benchmark points. These
were chosen as to cover various physical situations.  From a
phenomenological perspective, we are interested in maximising the
visibility of the new scalars in the LHC run~2 and in covering, simultaneously, 
many kinematically different possibilities. In particular we are
interested in scenarios where $\mu^C_{h_{125}}/\mu^T_{h_{125}}$ is as
large as possible, so that we can observe the new scalars produced in
chains containing the SM-like Higgs boson at the LHC run~2, while
preserving consistency with the LHC run~1 measurements. Thus, in many
of the points presented below 
we have tried to maximise the cross section for Higgs-to-Higgs
decays. At the same time we required the rates of the SM-like Higgs to
be within $2\sigma$ or at most $3\sigma$ from the global signal
strength provided by the combination of the ATLAS and CMS data from
the LHC run~1. We require at least $3\sigma$ consistency but in many
cases we find points that satisfy the required properties within
$2\sigma$. 

From the theoretical viewpoint, whenever possible, we choose points
for which the model remains stable up to a large cutoff scale
$\mu$. Here $\mu$ is the cutoff scale at which the couplings of the
theory reach a Landau pole or the scalar potential develops a
runaway direction. This scale was obtained through a renormalisation
group analysis as detailed in~\cite{Costa:2014qga} and it will be
indicated by the quantity $\log_{10}(\mu/{\rm GeV})$. In the dark
matter phase of the CxSM we also require that the dark matter relic
density predicted by the model, $\Omega_A h^2$, is consistent with the
combination of the measurements of the cosmic microwave 
background (CMB) from the WMAP and Planck
satellites~\cite{Ade:2013zuv,Hinshaw:2012aka}. We choose points where
$\Omega_A h^2$ is within 3$\sigma$ of the central value of the
combination, which is $\Omega_{c}h^2=0.1199\pm 0.0027$.   

%%%%%%%%%%%%%%%%%%%%%%%%%%%%%%%%%%%%%%%%%%%%%%%%%%%%%%%%%
\subsubsection{CxSM Broken Phase}
In Tables~\ref{tabTh:CxSM_Broken} and~\ref{tabPh:CxSM_Broken} we show
a sample of various kinematically allowed set-ups for the three mixing
scalars of the broken phase of the CxSM. The first, 
Table~\ref{tabTh:CxSM_Broken}, contains the parameters that define
the chosen benchmark points and the production rates of the lightest
and next-to-lightest Higgs bosons $h_1$ and $h_2$ in the various final
states. The corresponding values for $h_3$ are listed in 
Table~\ref{tabPh:CxSM_Broken}. For $h_2$ and $h_3$ the tables contain, in
particular, the Higgs-to-Higgs decay rates. We also give
the signal rates $\mu_{h_i}$ ($i=1,2,3$) as defined in
Eq.~(\ref{mu}). We have chosen two points where the SM-like Higgs is
the lightest Higgs boson (CxSM.B1 and CxSM.B2); two 
points where it is the next-to-lightest Higgs boson (CxSM.B3 and 
CxSM.B4); and one point where it is the heaviest (CxSM.B5). An  
interesting feature is that there are points for which the model
remains stable up to a large cutoff scale $\mu$, as shown by 
$\log_{10}(\mu/{\rm GeV})$ in the last row of
Table~\ref{tabPh:CxSM_Broken}. In particular the benchmark points
CxSM.B3 and CxSM.B4 are theoretically very interesting, since the new
heavy scalar here stabilises the theory up to a scale close to or
above the GUT scale of $10^{16}$ GeV.  

Most points were chosen such that the cross sections for the indirect
decay channels of the new scalars can compete with the direct
decays. In particular, in most cases, we have tried to maximise
$h_3\rightarrow h_2 + h_1$ where all three scalars could be observed at
once.  We have furthermore chosen points with large cross sections for the
new scalars, so that they can also be detected directly in their
decays. Since each point represents a different kinematic situation we
discuss the main features of each of them separately: 
\begin{table}[t!]
\begin{center}
\footnotesize
\begin{tabular}{|c|c|c|c|c|c|}
\hline
 	 & CxSM.B1 	 & CxSM.B2 	 & CxSM.B3 	 & CxSM.B4 	 & CxSM.B5\\ \hline \hline
$\star$ $m_{h1}$ (GeV) 	 & $125.1$ 	 & $125.1$ 	 & $57.83$ 	 & $86.79$ 	 & $33.17$\\
$\phantom{\star}$ $m_{h_2}$ (GeV) 	 & $260.6$ 	 & $228$ 	 & $125.1$ 	 & $125.1$ 	 & $64.99$\\
$\star$ $m_{h_3}$ (GeV) 	 & $449.6$ 	 & $311.3$ 	 & $299$ 	 & $291.8$ 	 & $125.1$\\
$\star$ $\alpha_1$ 	 & $-0.04375$ 	 & $0.05125$ 	 & $-1.102$ 	 & $-1.075$ 	 & $1.211$\\
$\star$ $\alpha_2$  	 & $0.4151$ 	 & $-0.4969$ 	 & $1.136$ 	 & $0.8628$ 	 & $-1.319$\\
$\star$ $\alpha_3$ 	 & $-0.6983$ 	 & $-0.5059$ 	 & $-0.02393$ 	 & $-0.0184$ 	 & $1.118$\\
$\star$ $v_S$ (GeV) 	 & $185.3$ 	 & $52.3$ 	 & $376.9$ 	 & $241.9$ 	 & $483.2$\\
$\phantom{\star}$ $v_A$ (GeV) 	 & $371.3$ 	 & $201.6$ 	 & $236.3$ 	 & $286.1$ 	 & $857.8$\\
$\lambda$ 	 & $1.148$ 	 & $1.018$ 	 & $0.869$ 	 & $0.764$ 	 & $0.5086$\\
$\delta_2$ 	 & $-0.9988$ 	 & $1.158$ 	 & $-0.4875$ 	 & $-0.4971$ 	 & $0.01418$\\
$d_2$ 	 & $1.819$ 	 & $3.46$ 	 & $0.6656$ 	 & $0.9855$ 	 & $0.003885$\\
$m^2$ (${\rm GeV}^2$) 	 & $5.118\times 10^{4}$ 	 & $-5.597\times 10^{4}$ 	 & $2.189\times 10^{4}$ 	 & $1.173\times 10^{4}$ 	 & $-2.229\times 10^{4}$\\
$b_2$ (${\rm GeV}^2$) 	 & $-3.193\times 10^{4}$ 	 & $-5.147\times 10^{4}$ 	 & $-3.484\times 10^{4}$ 	 & $-3.811\times 10^{4}$ 	 & $1362$\\
$b_1$ (${\rm GeV}^2$) 	 & $9.434\times 10^{4}$ 	 & $5.864\times 10^{4}$ 	 & $1.623\times 10^{4}$ 	 & $1.599\times 10^{4}$ 	 & $3674$\\
$a_1$ (${\rm GeV}^3$) 	 & $-1.236\times 10^{7}$ 	 & $-2.169\times 10^{6}$ 	 & $-4.325\times 10^{6}$ 	 & $-2.735\times 10^{6}$ 	 & $-1.255\times 10^{6}$\\
\hline \hline
$\mu^{C}_{h_1}/\mu^T_{h_1}$ 	 & $0.0127$ 	 & $0.0407$ 	 & $0.365$ 	 & $0.117$ 	 & $0.687$\\ \hline
$\mu_{h_1}$ 	 & $0.836$ 	 & $0.771$ 	 & $0.0362$ 	 & $0.0958$ 	 & $0.00767$\\ \hline
$\sigma_1\equiv \sigma(g g\rightarrow h_1)$ 	 & $36.1$ [pb] 	 & $33.3$ [pb] 	 & $6.42$ [pb] 	 & $8.03$ [pb] 	 & $4.61$ [pb]\\
$\sigma_1\times {\rm BR}(h_1\rightarrow WW)$ 	 & $7.55$ [pb] 	 & $6.96$ [pb] 	 & $0.345$ [fb] 	 & $10.3$ [fb] 	 & $<0.01$ [fb]\\
$\sigma_1\times {\rm BR}(h_1\rightarrow ZZ)$ 	 & $944$ [fb] 	 & $871$ [fb] 	 & $0.106$ [fb] 	 & $2.44$ [fb] 	 & $<0.01$ [fb]\\
$\sigma_1\times {\rm BR}(h_1\rightarrow bb)$ 	 & $21.3$ [pb] 	 & $19.6$ [pb] 	 & $5.48$ [pb] 	 & $6.6$ [pb] 	 & $4.01$ [pb]\\
$\sigma_1\times {\rm BR}(h_1\rightarrow \tau \tau)$ 	 & $2.29$ [pb] 	 & $2.11$ [pb] 	 & $501$ [fb] 	 & $659$ [fb] 	 & $323$ [fb]\\
$\sigma_1\times {\rm BR}(h_1\rightarrow \gamma \gamma)$ 	 & $83.7$ [fb] 	 & $77.2$ [fb] 	 & $2.87$ [fb] 	 & $9.13$ [fb] 	 & $0.617$ [fb]\\
\hline \hline
$\mu^{C}_{h_2}/\mu^T_{h_2}$ 	 & $0.0958$ 	 & $0$ 	 & $0.0128$ 	 & $0.0104$ 	 & $0.353$\\\hline
$\mu_{h_2}$ 	 & $0.0752$ 	 & $0.0759$ 	 & $0.782$ 	 & $0.785$ 	 & $0.0106$\\\hline
${\rm BR}(h_2\rightarrow X_{\rm SM})$ \% 	 & $87.9$ 	 & $100$ 	 & $96.2$ 	 & $100$ 	 & $100$\\\hline
$\sigma_2 \equiv \sigma(g g\rightarrow h_2)$ 	 & $1.01$ [pb] 	 & $1.11$ [pb] 	 & $35.1$ [pb] 	 & $33.9$ [pb] 	 & $1.51$ [pb]\\
$\sigma_2\times {\rm BR}(h_2\rightarrow WW)$ 	 & $618$ [fb] 	 & $784$ [fb] 	 & $7.06$ [pb] 	 & $7.09$ [pb] 	 & $0.185$ [fb]\\
$\sigma_2\times {\rm BR}(h_2\rightarrow ZZ)$ 	 & $265$ [fb] 	 & $319$ [fb] 	 & $883$ [fb] 	 & $887$ [fb] 	 & $0.0553$ [fb]\\
$\sigma_2\times {\rm BR}(h_2\rightarrow bb)$ 	 & $0.83$ [fb] 	 & $1.66$ [fb] 	 & $19.9$ [pb] 	 & $20$ [pb] 	 & $1.27$ [pb]\\
$\sigma_2\times {\rm BR}(h_2\rightarrow \tau \tau)$ 	 & $0.103$ [fb] 	 & $0.201$ [fb] 	 & $2.14$ [pb] 	 & $2.15$ [pb] 	 & $120$ [fb]\\
$\sigma_2\times {\rm BR}(h_2\rightarrow \gamma \gamma)$ 	 & $0.0189$ [fb] 	 & $0.0373$ [fb] 	 & $78.3$ [fb] 	 & $78.6$ [fb] 	 & $0.873$ [fb]\\
\hline
${\rm BR}(h_2\rightarrow h_1h_1)$ \% 	 & $12.1$ 	 & $0$ 	 & $3.82$ 	 & $0$ 	 & $0$\\ 
\hline
$\sigma_2\times {\rm BR}(h_2\rightarrow h_1 h_1)$ 	 & $122$ [fb] 	 & $0$ 	 & $1.34$ [pb] 	 & $0$ 	 & $0$\\
$\sigma_2\times {\rm BR}(h_2\rightarrow h_1 h_1\rightarrow bbbb)$ 	 & $42.5$ [fb] 	 & $0$ 	 & $977$ [fb] 	 & $0$ 	 & $0$\\
$\sigma_2\times {\rm BR}(h_2\rightarrow h_1 h_1\rightarrow bb\tau\tau)$ 	 & $9.13$ [fb] 	 & $0$ 	 & $179$ [fb] 	 & $0$ 	 & $0$\\
$\sigma_2\times {\rm BR}(h_2\rightarrow h_1 h_1\rightarrow bbWW)$ 	 & $30.1$ [fb] 	 & $0$ 	 & $0.123$ [fb] 	 & $0$ 	 & $0$\\
$\sigma_2\times {\rm BR}(h_2\rightarrow h_1 h_1\rightarrow bb\gamma\gamma)$ 	 & $0.334$ [fb] 	 & $0$ 	 & $1.02$ [fb] 	 & $0$ 	 & $0$\\
$\sigma_2\times {\rm BR}(h_2\rightarrow h_1 h_1\rightarrow \tau\tau\tau\tau)$ 	 & $0.491$ [fb] 	 & $0$ 	 & $8.16$ [fb] 	 & $0$ 	 & $0$\\
\hline
\end{tabular}
\end{center}
\caption{{\em Benchmark points for the CxSM broken phase:}  The
  parameters of the theory that we take as input values are denoted
  with a star ($\star$). The cross sections are for $\sqrt{s}\equiv 13$~TeV. }
\label{tabTh:CxSM_Broken}
\end{table}
\begin{table}[t!]
\begin{center}
\footnotesize
\begin{tabular}{|c|c|c|c|c|c|}
\hline
 	 & CxSM.B1 	 & CxSM.B2 	 & CxSM.B3 	 & CxSM.B4 	 & CxSM.B5\\ \hline \hline
$\mu_{h_3}$ 	 & $0.0558$ 	 & $0.0791$ 	 & $0.0788$ 	 & $0.0491$ 	 & $0.855$\\
\hline
${\rm BR}(h_3\rightarrow X_{\rm SM})$ \% 	 & $71$ 	 & $51.6$ 	 & $52.2$ 	 & $41.2$ 	 & $87.1$\\
\hline
$\sigma_3 \equiv \sigma(g g\rightarrow h_3)$ 	 & $520$ [fb] 	 & $1.46$ [pb] 	 & $1.48$ [pb] 	 & $1.2$ [pb] 	 & $42.4$ [pb]\\
$\sigma_3\times {\rm BR}(h_3\rightarrow WW)$ 	 & $201$ [fb] 	 & $519$ [fb] 	 & $536$ [fb] 	 & $344$ [fb] 	 & $7.72$ [pb]\\
$\sigma_3\times {\rm BR}(h_3\rightarrow ZZ)$ 	 & $95$ [fb] 	 & $232$ [fb] 	 & $238$ [fb] 	 & $152$ [fb] 	 & $966$ [fb]\\
$\sigma_3\times {\rm BR}(h_3\rightarrow bb)$ 	 & $0.0569$ [fb] 	 & $0.401$ [fb] 	 & $0.468$ [fb] 	 & $0.323$ [fb] 	 & $21.8$ [pb]\\
$\sigma_3\times {\rm BR}(h_3\rightarrow \tau \tau)$ 	 & $<0.01$ [fb] 	 & $0.0513$ [fb] 	 & $0.0594$ [fb] 	 & $0.0408$ [fb] 	 & $2.34$ [pb]\\
$\sigma_3\times {\rm BR}(h_3\rightarrow \gamma \gamma)$ 	 & $<0.01$ [fb] 	 & $<0.01$ [fb] 	 & $0.0105$ [fb] 	 & $<0.01$ [fb] 	 & $85.6$ [fb]\\
\hline
${\rm BR}(h_3\rightarrow h_1h_1)$ \% 	 & $8.53$ 	 & $48.4$ 	 & $29.5$ 	 & $35.4$ 	 & $11.0$\\
\hline
$\sigma_3\times {\rm BR}(h_3\rightarrow h_1 h_1)$ 	 & $44.3$ [fb] 	 & $706$ [fb] 	 & $438$ [fb] 	 & $426$ [fb] 	 & $4.66$ [pb]\\
$\sigma_3\times {\rm BR}(h_3\rightarrow h_1 h_1\rightarrow bbbb)$ 	 & $15.4$ [fb] 	 & $246$ [fb] 	 & $319$ [fb] 	 & $289$ [fb] 	 & $3.52$ [pb]\\
$\sigma_3\times {\rm BR}(h_3\rightarrow h_1 h_1\rightarrow bb\tau\tau)$ 	 & $3.32$ [fb] 	 & $52.8$ [fb] 	 & $58.2$ [fb] 	 & $57.6$ [fb] 	 & $567$ [fb]\\
$\sigma_3\times {\rm BR}(h_3\rightarrow h_1 h_1\rightarrow bbWW)$ 	 & $10.9$ [fb] 	 & $174$ [fb] 	 & $0.0401$ [fb] 	 & $0.897$ [fb] 	 & $0.011$ [fb]\\
$\sigma_3\times {\rm BR}(h_3\rightarrow h_1 h_1\rightarrow bb\gamma\gamma)$ 	 & $0.121$ [fb] 	 & $1.93$ [fb] 	 & $0.334$ [fb] 	 & $0.798$ [fb] 	 & $1.08$ [fb]\\
$\sigma_3\times {\rm BR}(h_3\rightarrow h_1 h_1\rightarrow \tau\tau\tau\tau)$ 	 & $0.178$ [fb] 	 & $2.84$ [fb] 	 & $2.66$ [fb] 	 & $2.88$ [fb] 	 & $22.9$ [fb]\\
\hline
${\rm BR}(h_3\rightarrow h_1h_2)$ \% 	 & $20.5$ 	 & $0$ 	 & $5.98$ 	 & $17.2$ 	 & $1.93$\\
\hline
$\sigma_3\times {\rm BR}(h_3\rightarrow h_1 h_2)$ 	 & $107$ [fb] 	 & $0$ 	 & $88.8$ [fb] 	 & $207$ [fb] 	 & $820$ [fb]\\
$\sigma_3\times {\rm BR}(h_3\rightarrow h_1 h_2\rightarrow bbbb)$ 	 & $0.0518$ [fb] 	 & $0$ 	 & $43$ [fb] 	 & $100$ [fb] 	 & $603$ [fb]\\
$\sigma_3\times {\rm BR}(h_3\rightarrow h_1 h_2\rightarrow bb\tau\tau)$ 	 & $0.012$ [fb] 	 & $0$ 	 & $8.55$ [fb] 	 & $20.8$ [fb] 	 & $105$ [fb]\\
$\sigma_3\times {\rm BR}(h_3\rightarrow h_1 h_2\rightarrow bbWW)$ 	 & $38.6$ [fb] 	 & $0$ 	 & $15.2$ [fb] 	 & $35.8$ [fb] 	 & $0.0883$ [fb]\\
$\sigma_3\times {\rm BR}(h_3\rightarrow h_1 h_2\rightarrow bb\gamma\gamma)$ 	 & $<0.01$ [fb] 	 & $0$ 	 & $0.191$ [fb] 	 & $0.534$ [fb] 	 & $0.506$ [fb]\\
$\sigma_3\times {\rm BR}(h_3\rightarrow h_1 h_2\rightarrow \tau\tau\tau\tau)$ 	 & $<0.01$ [fb] 	 & $0$ 	 & $0.422$ [fb] 	 & $1.08$ [fb] 	 & $4.56$ [fb]\\
\hline
${\rm BR}(h_3\rightarrow h_2h_2)$ \% 	 & $0$ 	 & $0$ 	 & $12.3$ 	 & $6.24$ 	 & $0$\\
\hline
$\sigma_3\times {\rm BR}(h_3\rightarrow h_2 h_2)$ 	 & $0$ 	 & $0$ 	 & $182$ [fb] 	 & $75.2$ [fb] 	 & $0$\\
$\sigma_3\times {\rm BR}(h_3\rightarrow h_2 h_2\rightarrow bbbb)$ 	 & $0$ 	 & $0$ 	 & $58.7$ [fb] 	 & $26.2$ [fb] 	 & $0$\\
$\sigma_3\times {\rm BR}(h_3\rightarrow h_2 h_2\rightarrow bb\tau\tau)$ 	 & $0$ 	 & $0$ 	 & $12.6$ [fb] 	 & $5.63$ [fb] 	 & $0$\\
$\sigma_3\times {\rm BR}(h_3\rightarrow h_2 h_2\rightarrow bbWW)$ 	 & $0$ 	 & $0$ 	 & $41.6$ [fb] 	 & $18.5$ [fb] 	 & $0$\\
$\sigma_3\times {\rm BR}(h_3\rightarrow h_2 h_2\rightarrow bb\gamma\gamma)$ 	 & $0$ 	 & $0$ 	 & $0.462$ [fb] 	 & $0.206$ [fb] 	 & $0$\\
$\sigma_3\times {\rm BR}(h_3\rightarrow h_2 h_2\rightarrow \tau\tau\tau\tau)$ 	 & $0$ 	 & $0$ 	 & $0.679$ [fb] 	 & $0.303$ [fb] 	 & $0$\\
\hline\hline 
$\log_{10}\left(\frac{\mu}{\rm GeV}\right)$ 	 & $9.40$ 	 & $6.05$ 	 & $19.3$ 	 & $15.7$ 	 & $6.64$\\
\hline
\end{tabular}
\end{center}\vspace{-4mm}
\caption{{\em CxSM broken phase benchmarks (continuation of
    table~\ref{tabTh:CxSM_Broken})} \label{tabPh:CxSM_Broken}}
\end{table}
\begin{itemize}
\item \underline{CxSM.B1}: 
For this point  the SM-like Higgs is the lightest of the three Higgs
bosons and all Higgs-to-Higgs decay channels are open apart from
$h_3\rightarrow h_2 + h_2$.\footnote{In 
  this scenario we do not present a case with all channels open
  because the spectrum would be even heavier and more difficult to be
  tested.} The presented point has been chosen such that it has a maximal ratio $\mu_{h_{125}}^C/\mu_{h_{125}}^T$ for this kinematical situation within the imposed bounds. The production rates for Higgs-to-Higgs decays of the non-SM-like
Higgs bosons at the LHC run~2 are also maximised. With the direct
production of the heavier Higgs bosons in SM-like final states being on
the lower side\footnote{CxSM.B1 is the only benchmark point, where the
mass of the heavier Higgs boson is large enough to decay into
$t\bar{t}$. For completeness, we add here the rate into this final
state. It amounts to 72~fb for $h_3$.}, the additional Higgs-to-Higgs
decays with rates of 
e.g.~$h_{2,3} \to h_1 + h_1 \to bb + \tau\tau$ of a few fb suggest, that
all new scalars can be expected to be observed. 
\item \underline{CxSM.B2}: 
This benchmark point, also with $h_{125} \equiv h_1$, features an
overall lighter Higgs mass spectrum, so that in the Higgs-to-Higgs
decays only the decay channel $h_3\rightarrow h_{125} + h_{125}$ is
open. With a large branching ratio of $\sim 48\%$, the
corresponding rates are important and lead to a fraction of chain decays in the total
production of the SM-like Higgs boson of about 4\%, which is the
largest ratio achieved in all five benchmark points. 
The point has been chosen such that $h_3$ should be accessible
both in the chain decays through a pair of $h_{125}$ bosons in the 
$4b$, $2b2\tau$ or $2b2W$ final states as well as through direct
decays into SM particles (mostly into massive vector bosons), while $h_{2}$
would be visible in its direct decays (also mostly into
massive vector bosons). Both for $h_2$ and $h_3$ the decays
into top quark pairs are kinematically closed. 
\item \underline{CxSM.B3}: 
For this point, where $h_{125} \equiv h_2$ and $m_{h_1} < m_{h_{125}}/2$, all
kinematic situations for the scalar decays are available while the
spectrum remains light.  As can be inferred from 
Fig.~\ref{fig:chain_vs_pvalue_CxSM_broken} (lower), in this scenario
the decay chain fraction cannot become large. Still, the benchmark point has
been chosen to maximise it
reaching a value of 1.3\%. The $h_3$ Higgs-to-Higgs decay
rates in the $bb\tau\tau$ final state lie between about 10 and 50 fb
depending on the intermediate Higgs pair state. The $h_2 \to h_1 + h_1$
rate even goes up to 179 fb in this final state, so that
Higgs-to-Higgs decays present an interesting discovery option for the
heavy Higgs states. Furthermore, large production cross sections have
been required for the new light scalar $h_1$ so that it will be
visible in its direct decays in addition to chain production from heavier scalars.
\item \underline{CxSM.B4}: This scenario differs from the previous one 
in the larger $h_1$ mass so that the channel $h_2 \to h_1 + h_1$ is
kinematically closed. At the same time the direct $h_1$ production
rates are increased, allowing for its discovery through these
channels. The heavy Higgs boson $h_3$ decays with an overall branching
ratio of close to 60\% into lighter Higgs pairs. Compared to the
previous benchmark point the rates into SM final states via the $h_1
+h_2$ decay are approximately doubled and the ones via $h_2+h_2$ are
roughly halved, while those via $h_1 + h_1$ are about the same. Again
these decay chains allow for $h_3$ (and also $h_1$ and $h_2$)
discovery through Higgs decay chains, supplementing the discovery in
direct SM final state production.  
\item \underline{CxSM.B5}: 
Finally, we have also chosen a point that does not allow for SM-like
Higgs production through chain decays. With $h_3 \equiv h_{125}$ the
overall spectrum is very light. As $m_{h_1} > m_{h_2}/2$, the $h_2 \to
h_1 + h_1$ decay is kinematically closed. Instead $h_1$ production is
possible through $h_3 \to h_1 + h_1$ or $h_3 \to h_1 + h_2$. The
benchmark point has been required to have large branching fractions
for these Higgs-to-Higgs decays. These channels can reach in the $bb \tau \tau$ final state up to $\sim 560$~fb for the former and $\sim 100$~fb for the
latter of the Higgs-to-Higgs decays and are hence accessible at the
LHC run 2. The lightest Higgs $h_1$ can also be observed directly in
e.g.~$bb$ or $\tau\tau$ decays. The direct production rates of the heavier
$h_2$ are smaller, still large enough to be measurable in the
$bb$ or $\tau\tau$ final state.
\end{itemize}

%%%%%%%%%%%%%%%%%%%%%%%%%%%%%%%%%%%%%%%%%%%%%%%%%%%%%%%%%%
\subsubsection{CxSM Dark Phase}
In the phase of the CxSM that contains dark matter, the first requirement is
that the relic density predicted by the model agrees with the
measurements from the CMB. In Table~\ref{tab:CxSM_Dark} we have
selected four benchmark points that all obey this requirement. 
\begin{table}[t!]
\begin{center}
\footnotesize
\begin{tabular}{|c|c|c|c|c|}
\hline
 	 & CxSM.D1 	 & CxSM.D2 	 & CxSM.D3 & CxSM.D4 \\ \hline \hline
$\star$ $m_{h_1}$ (GeV) 	 & $125.1$ 	 & $125.1$ 	 & $56.12$ 	 & $121.2$\\
$\star$ $m_{h_2}$ (GeV) 	 & $335.2$ 	 & $341.4$ 	 & $125.1$ 	 & $125.1$\\
$\star$ $m_A$ (GeV) 	 & $52.46$ 	 & $93.97$ 	 & $139.3$ 	 & $51.96$\\
$\star$ $\alpha$ 	 & $0.4587$ 	 & $-0.4156$ 	 & $1.507$ 	 & $1.358$\\
$\star$ $v_S$ (GeV) 	 & $812.5$ 	 & $987.5$ 	 & $177.9$ 	 & $909.7$\\
$\lambda$ 	 & $1.142$ 	 & $1.059$ 	 & $0.5146$ 	 & $0.5149$\\
$\delta_2$ 	 & $-0.3839$ 	 & $0.3066$ 	 & $-0.0362$ 	 & $-0.001764$\\
$d_2$ 	 & $0.2669$ 	 & $0.164$ 	 & $0.1653$ 	 & $0.03508$\\
$m^2$ (${\rm GeV}^2$) 	 & $9.21\times 10^{4}$ 	 & $-1.816\times 10^{5}$ 	 & $-1.503\times 10^{4}$ 	 & $-1.488\times 10^{4}$\\
$b_2$ (${\rm GeV}^2$) 	 & $-6.838\times 10^{4}$ 	 & $-6.027\times 10^{4}$ 	 & $1.848\times 10^{4}$ 	 & $-1.154\times 10^{4}$\\
$b_1$ (${\rm GeV}^2$) 	 & $2570$ 	 & $1.132\times 10^{4}$ 	 & $-1.883\times 10^{4}$ 	 & $-2479$\\
$\star$ $a_1$ (${\rm GeV}^3$) 	 & $-3.057\times 10^{6}$ 	 & $-1.407\times 10^{7}$ 	 & $-7.362\times 10^{4}$ 	 & $-1.418\times 10^{5}$\\
\hline \hline
$\mu^{C}_{h_1}/\mu^T_{h_1}$ 	 & $0.019$ 	 & $0.0235$ 	 & $0.97$ 	 & $0$\\
\hline
$\mu_{h_1}$ 	 & $0.804$ 	 & $0.837$ 	 & $0.00404$ 	 & $0.0444$\\
\hline
$ {\rm BR}(h_1\rightarrow X_{\rm SM})$ \% 	 & $70.5$ 	 & $100$ 	 & $100$ 	 & $1.56$\\
\hline
$\sigma_1\equiv \sigma(g g\rightarrow h_1)$ 	 & $34.7$ [pb] 	 & $36.2$ [pb] 	 & $759$ [fb] 	 & $2.03$ [pb]\\
$\sigma_1\times {\rm BR}(h_1\rightarrow WW)$ 	 & $5.12$ [pb] 	 & $7.56$ [pb] 	 & $0.0331$ [fb] 	 & $4.81$ [fb]\\
$\sigma_1\times {\rm BR}(h_1\rightarrow ZZ)$ 	 & $640$ [fb] 	 & $945$ [fb] 	 & $0.0103$ [fb] 	 & $0.561$ [fb]\\
$\sigma_1\times {\rm BR}(h_1\rightarrow bb)$ 	 & $14.4$ [pb] 	 & $21.3$ [pb] 	 & $649$ [fb] 	 & $20.4$ [fb]\\
$\sigma_1\times {\rm BR}(h_1\rightarrow \tau \tau)$ 	 & $1.55$ [pb] 	 & $2.29$ [pb] 	 & $58.9$ [fb] 	 & $2.18$ [fb]\\
$\sigma_1\times {\rm BR}(h_1\rightarrow \gamma \gamma)$ 	 & $56.8$ [fb] 	 & $83.8$ [fb] 	 & $0.317$ [fb] 	 & $0.0723$ [fb]\\
\hline
$\sigma_1\times {\rm BR}(h_1\rightarrow A A)$ 	 & $10.2$ [pb] 	 & $0$ 	 & $0$ 	 & $2.00$ [pb]\\
\hline \hline
$\mu_{h_2}$ 	 & $0.138$ 	 & $0.108$ 	 & $0.710$ 	 & $0.834$\\
\hline
$ {\rm BR}(h_2\rightarrow X_{\rm SM})$ \% 	 & $70.3$ 	 & $66.1$ 	 & $71.3$ 	 & $87.3$\\
\hline
$\sigma_2 \equiv \sigma(g g\rightarrow h_2)$ 	 & $1.83$ [pb] 	 & $1.55$ [pb] 	 & $43$ [pb] 	 & $41.3$ [pb]\\
$\sigma_2\times {\rm BR}(h_2\rightarrow WW)$ 	 & $886$ [fb] 	 & $704$ [fb] 	 & $6.41$ [pb] 	 & $7.54$ [pb]\\
$\sigma_2\times {\rm BR}(h_2\rightarrow ZZ)$ 	 & $402$ [fb] 	 & $320$ [fb] 	 & $802$ [fb] 	 & $943$ [fb]\\
$\sigma_2\times {\rm BR}(h_2\rightarrow bb)$ 	 & $0.553$ [fb] 	 & $0.417$ [fb] 	 & $18.1$ [pb] 	 & $21.3$ [pb]\\
$\sigma_2\times {\rm BR}(h_2\rightarrow \tau \tau)$ 	 & $0.0717$ [fb] 	 & $0.0542$ [fb] 	 & $1.95$ [pb] 	 & $2.29$ [pb]\\
$\sigma_2\times {\rm BR}(h_2\rightarrow \gamma \gamma)$ 	 & $0.012$ [fb] 	 & $<0.01$ [fb] 	 & $71.1$ [fb] 	 & $83.6$ [fb]\\
\hline
${\rm BR}(h_2\rightarrow h_1h_1)$ \% 	 & $18.4$ 	 & $28$ 	 & $28.7$ 	 & $0$\\
\hline
$\sigma_2\times {\rm BR}(h_2\rightarrow h_1 h_1)$ 	 & $337$ [fb] 	 & $436$ [fb] 	 & $12.3$ [pb] 	 & $0$\\
$\sigma_2\times {\rm BR}(h_2\rightarrow h_1 h_1\rightarrow bbbb)$ 	 & $58.3$ [fb] 	 & $152$ [fb] 	 & $9.02$ [pb] 	 & $0$\\
$\sigma_2\times {\rm BR}(h_2\rightarrow h_1 h_1\rightarrow bb\tau\tau)$ 	 & $12.5$ [fb] 	 & $32.6$ [fb] 	 & $1.64$ [pb] 	 & $0$\\
$\sigma_2\times {\rm BR}(h_2\rightarrow h_1 h_1\rightarrow bbWW)$ 	 & $41.3$ [fb] 	 & $107$ [fb] 	 & $0.92$ [fb] 	 & $0$\\
$\sigma_2\times {\rm BR}(h_2\rightarrow h_1 h_1\rightarrow bb\gamma\gamma)$ 	 & $0.458$ [fb] 	 & $1.19$ [fb] 	 & $8.81$ [fb] 	 & $0$\\
$\sigma_2\times {\rm BR}(h_2\rightarrow h_1 h_1\rightarrow \tau\tau\tau\tau)$ 	 & $0.675$ [fb] 	 & $1.75$ [fb] 	 & $74.3$ [fb] 	 & $0$\\
\hline
$\sigma_2\times {\rm BR}(h_2\rightarrow A A)$ 	 & $207$ [fb] 	 & $91.3$ [fb] 	 & $0$ 	 & $5.23$ [pb]\\
\hline \hline
$\Omega_A h^2$ 	 & $0.118$ 	 & $0.123$ 	 & $0.116$ 	 & $0.125$\\
\hline \hline
$\log_{10}\left(\frac{\mu}{\rm GeV}\right)$ 	 & $14.9$ 	 & $17.1$ 	 & $6.69$ 	 & $6.69$\\
\hline
\end{tabular}
\end{center}
\caption{{\em Benchmark points for the CxSM dark phase:} The parameters of the theory that we take as input values are denoted with a star ($\star$). The cross-sections are for $\sqrt{s}\equiv 13$~TeV.
\label{tab:CxSM_Dark}}
\end{table}

For the first two points, CxSM.D1 and CxSM.D2, the lightest of the two
visible scalars is the SM-like Higgs. In point CxSM.D1, both visible
scalars can decay invisibly into the dark matter candidate $A$ whereas
in point CxSM.D2 $h_1$ cannot.   
Both benchmark points feature large invisible branching ratios, with
the largest one reaching 29\% for $h_1 \to A+ A$ in CxSM.D1. Also the
branching ratios for the Higgs-to-Higgs decays $h_2 \to h_1 + h_1$ are
large with 18.4\% for CxSM.D1 and 28\% for CxSM.D2. The signal rates
for these decays in the $bb\tau\tau$ final state are 13~fb and 
33~fb, respectively, so that chain decays contribute to the discovery
of $h_2$, and also $h_1$. Furthermore, both benchmark points 
have large cross sections for direct production of $h_2$ so that it
can also be discovered in its direct decays into SM particles.
Another attractive feature of these points is that the new heavy
scalar $h_2$ can stabilise the theory up to a high scale close to the
GUT scale, as can be inferred from $\log_{10}(\mu/{\rm GeV})$ in the last
row of the table for CxSM.D1 and CxSM.D2.

In the scenarios CxSM.D3 and CxSM.D4, where the SM-like Higgs boson is
the heaviest of the two visible Higgs bosons, the overall
spectrum is lighter and the theory must have a UV
completion above $\sim 10^3$~TeV. Point CxSM.D3 represents a case with
no invisible decays allowed, and in CxSM.D4  
decays of the SM-like Higgs $h_2$ into a lighter Higgs pair are
forbidden at the expense of allowing for a large 
invisible decay into the dark matter state $A$.  In CxSM.D3 the light
Higgs state $h_1$ can either be discovered directly or in the chain
decay of the SM-like Higgs $h_2$ into an $h_1$ pair.
 The production rates of 9.0~pb in the $4b$ final state and 1.6~pb in the $bb\tau\tau$ final
state are rather large and complement the discovery of
$h_1$ in direct production. 
At the same time the rather important branching ratio BR($h_2 \to h_1 h_1$) of 29\%, which is responsible for these rates, drives the overall rate of
the SM-like Higgs, $\mu_{h_2}$, to
a value at the edge of compatibility with the LHC data. The discovery
of the non-SM-like light $h_1$ in CxSM.D4, that has a mass close to the SM-like
Higgs, is extremely challenging.  With no contribution from chain decays, it is only
accessible in its decays into SM particles, with rates in the $bb$ and
$\tau\tau$ final states that are comparable to those of the SM-like
Higgs, but much smaller rates in the gauge boson final
states. 

%%%%%%%%%%%%%%%%%%%%%%%%%%%%%%%%%%%%%%%%%%%%%%%%%%%%%%%%%%%
\subsubsection{RxSM Broken Phase}
Table~\ref{tab:benchmarks_RxSM_Broken} contains four benchmark points
for the two possible kinematic configurations in this model. The
points RxSM.B1 and RxSM.B2 correspond to the case where the SM-like
Higgs boson is the lightest of the two scalar states. Benchmark
RxSM.B1 allows for the decay $h_2\rightarrow h_1 + h_1$ and we have
chosen a point with a relatively large cross section $\sigma_2 \times {\rm BR}(h_2
\to h_1 + h_1)$ for such a chain decay. It is comparable to the direct
$h_2$ production cross section. The $bb\tau\tau$ final state in the former even
reaches 72~fb. Thus $h_2$ could in principle be found directly or in
its chain decays. Note that the fraction of chain decays in SM-like
Higgs production amounts to 5\%. For 
complementarity, in RxSM.B2 we selected a point where the decay into scalars is
kinematically closed, but instead various direct decay channels of 
$h_2$ are enhanced compared to RxSM.B1, most notably the
$WW$ final state but also $bb$, $\tau\tau$ and $\gamma\gamma$. 
\begin{table}[t!]
\begin{center}
\footnotesize
\begin{tabular}{|c|c|c|c|c|}
\hline
 	 & RxSM.B1 	 & RxSM.B2 	 & RxSM.B3 	 & RxSM.B4\\ \hline \hline
$\star$ $m_{h_1}$ (GeV) 	 & $125.1$ 	 & $125.1$ 	 & $55.26$ 	 & $92.44$\\
$\star$ $m_{h_2}$ (GeV) 	 & $265.3$ 	 & $172.5$ 	 & $125.1$ 	 & $125.1$\\
$\star$ $\alpha$ 	 & $-0.4284$ 	 & $-0.4239$ 	 & $1.376$ 	 & $1.156$\\
$\star$ $v_S$ (GeV) 	 & $140.3$ 	 & $94.74$ 	 & $591$ 	 & $686.1$\\
$\lambda$ 	 & $0.828$ 	 & $0.595$ 	 & $0.5007$ 	 & $0.4782$\\
$\lambda_{HS}$ 	 & $0.599$ 	 & $0.2268$ 	 & $-0.01646$ 	 & $-0.01552$\\
$\lambda_S$ 	 & $9.294$ 	 & $9.149$ 	 & $0.03029$ 	 & $0.06182$\\
$m^2$ (${\rm GeV}^2$) 	 & $-3.688\times 10^{4}$ 	 & $-2.007\times 10^{4}$ 	 & $-9426$ 	 & $-7190$\\
$m^2_S$ (${\rm GeV}^2$) 	 & $-4.863\times 10^{4}$ 	 & $-2.056\times 10^{4}$ 	 & $-1265$ 	 & $-4380$\\
\hline \hline
$\mu^{C}_{h_1}/\mu^T_{h_1}$ 	 & $0.051$ 	 & $0$ 	 & $0.557$ 	 & $0$\\
\hline
$\mu_{h_1}$ 	 & $0.827$ 	 & $0.831$ 	 & $0.0376$ 	 & $0.163$\\
$\sigma_1\equiv \sigma(g g\rightarrow h_1)$ 	 & $35.7$ [pb] 	 & $35.9$ [pb] 	 & $7.26$ [pb] 	 & $12.2$ [pb]\\
$\sigma_1\times {\rm BR}(h_1\rightarrow WW)$ 	 & $7.47$ [pb] 	 & $7.5$ [pb] 	 & $0.285$ [fb] 	 & $35.4$ [fb]\\
$\sigma_1\times {\rm BR}(h_1\rightarrow ZZ)$ 	 & $935$ [fb] 	 & $938$ [fb] 	 & $0.0887$ [fb] 	 & $6.17$ [fb]\\
$\sigma_1\times {\rm BR}(h_1\rightarrow bb)$ 	 & $21.1$ [pb] 	 & $21.2$ [pb] 	 & $6.21$ [pb] 	 & $9.9$ [pb]\\
$\sigma_1\times {\rm BR}(h_1\rightarrow \tau \tau)$ 	 & $2.27$ [pb] 	 & $2.28$ [pb] 	 & $562$ [fb] 	 & $1$ [pb]\\
$\sigma_1\times {\rm BR}(h_1\rightarrow \gamma \gamma)$ 	 & $82.8$ [fb] 	 & $83.2$ [fb] 	 & $2.93$ [fb] 	 & $16.1$ [fb]\\
\hline \hline
$\mu_{h_2}$ 	 & $0.0887$ 	 & $0.169$ 	 & $0.857$ 	 & $0.837$\\
$\sigma_2 \equiv \sigma(g g\rightarrow h_2)$ 	 & $1.97$ [pb] 	 & $4.06$ [pb] 	 & $41.6$ [pb] 	 & $36.2$ [pb]\\
$\sigma_2\times {\rm BR}(h_2\rightarrow WW)$ 	 & $708$ [fb] 	 & $3.9$ [pb] 	 & $7.73$ [pb] 	 & $7.56$ [pb]\\
$\sigma_2\times {\rm BR}(h_2\rightarrow ZZ)$ 	 & $305$ [fb] 	 & $112$ [fb] 	 & $967$ [fb] 	 & $946$ [fb]\\
$\sigma_2\times {\rm BR}(h_2\rightarrow bb)$ 	 & $0.897$ [fb] 	 & $30.5$ [fb] 	 & $21.8$ [pb] 	 & $21.3$ [pb]\\
$\sigma_2\times {\rm BR}(h_2\rightarrow \tau \tau)$ 	 & $0.111$ [fb] 	 & $3.48$ [fb] 	 & $2.35$ [pb] 	 & $2.29$ [pb]\\
$\sigma_2\times {\rm BR}(h_2\rightarrow \gamma \gamma)$ 	 & $0.0204$ [fb] 	 & $0.582$ [fb] 	 & $85.8$ [fb] 	 & $83.9$ [fb]\\
\hline
${\rm BR}(h_2\rightarrow h_1h_1)$ \% 	 & $48.6$ 	 & $0$ 	 & $11$ 	 & $0$\\
\hline
$\sigma_2\times {\rm BR}(h_2\rightarrow h_1 h_1)$ 	 & $960$ [fb] 	 & $0$ 	 & $4.57$ [pb] 	 & $0$\\
$\sigma_2\times {\rm BR}(h_2\rightarrow h_1 h_1\rightarrow bbbb)$ 	 & $334$ [fb] 	 & $0$ 	 & $3.35$ [pb] 	 & $0$\\
$\sigma_2\times {\rm BR}(h_2\rightarrow h_1 h_1\rightarrow bb\tau\tau)$ 	 & $71.8$ [fb] 	 & $0$ 	 & $605$ [fb] 	 & $0$\\
$\sigma_2\times {\rm BR}(h_2\rightarrow h_1 h_1\rightarrow bbWW)$ 	 & $237$ [fb] 	 & $0$ 	 & $0.307$ [fb] 	 & $0$\\
$\sigma_2\times {\rm BR}(h_2\rightarrow h_1 h_1\rightarrow bb\gamma\gamma)$ 	 & $2.62$ [fb] 	 & $0$ 	 & $3.16$ [fb] 	 & $0$\\
$\sigma_2\times {\rm BR}(h_2\rightarrow h_1 h_1\rightarrow \tau\tau\tau\tau)$ 	 & $3.86$ [fb] 	 & $0$ 	 & $27.4$ [fb] 	 & $0$\\
\hline
\end{tabular}
\end{center}
\caption{{\em Benchmark points for the RxSM broken phase:} The
  parameters of the theory that we take as input values are denoted
  with a star ($\star$). The cross sections are for $\sqrt{s}\equiv
  13$~TeV. 
\label{tab:benchmarks_RxSM_Broken}}
\end{table}

The benchmarks RxSM.B3 and RxSM.B4 feature a SM-like Higgs boson
that is the heaviest of the two scalars. RxSM.B3 was again chosen such
that the non-SM-like Higgs $h_1$ can be found directly or in the decay
$h_2\rightarrow h_1 + h_1$. In particular note the large rates
for the direct $h_1$ production and decay into the $bb$ and also $\tau
\tau$ final states and compare with the indirect processes
$h_2\rightarrow h_1 + h_1\rightarrow bb+bb$ and $bb+\tau\tau$, where the
magnitude of the latter two is comparable to the former two. For
RxSM.B4, on the contrary, we have chosen a point to represent the
situation where the indirect channel is closed. Here we have required
a larger cross section for direct $h_1$ production allowing for larger
rates into the $bb$ and $\tau\tau$ final states. Still the discovery of
$h_1$ with a mass very close to the $Z$ boson peak will be
challenging. 

%%%%%%%%%%%%%%%%%%%%%%%%%%%%%%%%%%%%%%%%%%%%%%%%%%%%%%%%%
\section{Conclusions}
\label{sec:concl}

We have analysed in detail the phenomenology of two Higgs bosons
final states in a real (RxSM) and a complex (CxSM) singlet
extensions of the SM. Both models contain phases with one or two new
Higgs bosons, which may be found either directly or in Higgs-to-Higgs
decays in the next runs of the LHC. We have performed a comparison of
the achievable rates for Higgs-to-Higgs decays in the various singlet
models and with respect to the NMSSM. Finally, we have presented
benchmark points for the singlet models at the LHC run~2.

We started by presenting the models and the phenomenological
constraints that were imposed. For this purpose a new code based on
{\tt HDECAY} was developed for the calculation of the decay widths and
branching ratios of the scalar particles present in the RxSM and the
CxSM, both in the symmetric and in the broken phase. In this
publicly available tool, {\tt sHDECAY}, the SM higher order
electroweak corrections are consistently turned off. Including the
state-of-the-art higher order QCD corrections, the code will allow a
rigorous interpretation of the data in these singlet extensions of the SM.

In the numerical analysis, we first investigated how
  important the contribution from Higgs-to-Higgs decays to the
  signal of the 125~GeV Higgs boson can be compared to its direct production
  and decay. Depending on the singlet model and on the scenario, the
  indirect production of  
the SM-like Higgs boson, through the decay of a heavier scalar, can
attain a maximum value ranging between $2\%$($4\%$) and
$9\%$($17\%$) while remaining compatible with the Higgs signal
measurements within $2\sigma$($3\sigma$).

Subsequently, we performed a systematic comparison of Higgs-to-Higgs
decay rates at the LHC run~2 within the singlet models and 
with respect to the NMSSM. Among the singlet extensions,
the rates achieved in the broken phase of the RxSM and the dark matter
phase of the CxSM, in Higgs-to-Higgs decays with two identical scalars
in the final states, are larger than those in the broken phase of
the CxSM. In particular in the case where the non-SM-like scalar state
is lighter than $m_{h_{125}}/2$, the maximum rates of the former two 
exceed those of the latter by up to two orders of magnitude and can
reach about 10~pb in the $4b$-quark final state.

The comparative analysis between the maximum allowed rates in the singlet extensions and those of the NMSSM addressed the question: to which extent Higgs-to-Higgs decays allow for a distinction of different Higgs
sectors with more than two neutral Higgs bosons? The broken phase of the CxSM
features such a Higgs spectrum. We found that a clear distinction from
the NMSSM is possible based on Higgs-to-Higgs decay rates, but only in
final states with two different scalars, i.e.~in $pp \to h_i \to h_j +
h_k$, with $i \neq j \neq k$. The maximum rates obtained in the NMSSM
significantly exceed those of CxSM-broken. This means that,
if no direct hints of new physics outside the Higgs sector are found,
two-Higgs final states with different masses can play
the role of smoking gun signatures for the distinction of extended
Higgs sectors based on pure (complex) singlet extensions and those of
non-minimal SUSY sectors. Their analysis should therefore play a prime
role at the LHC run~2.

Finally, we proposed benchmark points for the RxSM and CxSM at the LHC
run~2. Our guiding principles in defining these benchmarks were: i) to maximize
the rates for different Higgs-to-Higgs processes, so that the new
scalars can be found in many different channels, ii) to look for
points where the theory is stable up to a high scale and, iii) that in
the dark phase of the CxSM the points conform with dark
matter observables and null searches. We provide full information on
the input parameters for each point together with all relevant cross
sections at the LHC run~2, their compatibility with dark matter
observables and the high-energy stability cutoff scale. The benchmarks
cover different kinematic situations and vary in their prospect of
discovering the related Higgs spectrum. They can therefore serve
as guidelines for the experiments in tuning their experimental
analyses.

%%%%%%%%%%%%%%%%%%%%%%%%%%%%%%%%%%%%%%%%%%%%%%%%%%%%%%%
\vspace*{0.5cm}
\section*{Acknowledgements}
The authors are grateful to the Mainz Institute for Theoretical
Physics (MITP) for its hospitality and its partial support during the
completion of this work. We thank Kathrin Walz for helpful discussions
related to the NMSSM. RC and RS are supported by FCT under contract PTDC/FIS/117951/2010. MS is funded by FCT through the grant SFRH/BPD/ 69971/2010. The work in this paper is also supported by the CIDMA project UID/MAT/04106/2013. 
\vspace*{0.5cm}
%%%%%%%%%%%%%%%%%%%%%%%%%%%%%%%%%%%%%%%%%%%%%%%%%%%%%%%

\appendix
%%%%%%%%%%%%%%%%%%%%%%%%%%%%%%%%%%%%%%%%%%%%%%%%%%%%%%%
\section*{Appendix}
\begin{appendix}
\section{Implementation of Singlet Models in \textsc{sHDECAY}}
\label{sec:app_SHDECAY}
For our analysis we have implemented the singlet extensions of the SM
in the Fortran code {\tt HDECAY}
\cite{Djouadi:1997yw,Djouadi:2006bz,Butterworth:2010ym}, version
6.50. The original code, {\tt HDECAY}, was initially created for the
computation of the partial decay widths and branching ratios in the
SM and its minimal supersymmetric extension (MSSM). In the meantime it
has been extended in {\tt eHDECAY} to include higher dimensional operators in the
non-linear and the linear parametrization of the Langrangian as well
as composite Higgs benchmark models
\cite{Contino:2013kra,Contino:2014aaa} and to other models like the 
2-Higgs-Doublet model \cite{Harlander:2013qxa} and also the NMSSM in
the program package {\tt NMSSMCALC}, both in the real and
in the complex case, \cite{Baglio:2013iia,King:2015oxa}. The
consistent combination of the MSSM Higgs decays with supersymmetric
particle decays is provided by the program package {\tt  SUSY-HIT}
\cite{Djouadi:2006bz}, of which {\tt HDECAY} is part.   
The new code for the singlet extensions is called {\tt sHDECAY} and is
self-contained like the original code {\tt HDECAY}. All changes related
to the singlet models have been implemented in the main source file
shdecay.f. Other linked routines are taken from the original
version. In the implementation of the model we took care to turn off
all electroweak corrections that cannot be taken over from the SM. The
QCD corrections on the other hand are not affected by the Higgs
coupling modifications and can be taken over.\footnote{For a detailed
  discussion of the implementation of SM extensions in {\tt HDECAY}
  and the treatment of higher order corrections, see also \cite{Contino:2014aaa}.}
Thus the Higgs decays into quarks include the fully massive NLO
corrections near threshold
\cite{Braaten:1980yq,Sakai:1980fa,Inami:1980qp,Drees:1989du,Drees:1990dq}
and massless ${\cal O}(\alpha_s^4)$ corrections far above threshold
\cite{Gorishnii:1990zu,Gorishnii:1991zr,Kataev:1993be,Gorishnii:1983cu,Surguladze:1994gc,Larin:1995sq,Chetyrkin:1995pd,Chetyrkin:1996sr,Baikov:2005rw}.
The resummation of large logarithms is assured by taking into account
the running of the quark masses and the strong coupling. In the decays
into gluons the QCD corrections have been included up to N$^3$LO in
the limit of heavy loop-particle masses \cite{Inami:1982xt,Djouadi:1991tka,Spira:1993bb,Spira:1995rr,Kramer:1996iq,Chetyrkin:1997iv,Chetyrkin:1997un,Schroder:2005hy,Chetyrkin:2005ia,Baikov:2006ch}. Additionally at NLO
QCD the mass effects in the top and bottom loops \cite{Spira:1995rr} have been
implemented. The decays into photons are computed at NLO QCD including
the full mass dependence of the quarks
\cite{Spira:1995rr,Zheng:1990qa,Djouadi:1990aj,Dawson:1992cy,Djouadi:1993ji,Melnikov:1993tj,Inoue:1994jq}. The
possibility of off-shell decays into massive gauge boson final states and
heavy quark pairs is also taken into account
\cite{Djouadi:1995gv}. The code has been extended to include all
possible Higgs-to-Higgs decays in the various singlet extensions.  The
leading order expressions for the Higgs-to-Higgs decay widths are presented in
appendix~\ref{sec:app_FeynmanRules}.   

The {\tt sHDECAY} input file {\tt shdecay.in} is an extension of
the original input file {\tt hdecay.in} of the code {\tt
  HDECAY}. In {\tt shdecay.in} the user can choose between the singlet
extensions of the SM. These are the real and the complex singlet extensions in both
phases of the models, i.e.~the broken and the dark one. The various
parameters of the different models are also specified in the input
file. From these {\tt sHDECAY} calculates, internally in {\tt shdecay.f}, the coupling
modifications $R_{i1}$ and the trilinear Higgs self-couplings needed
for the decay widths. 
For the computation of the decay widths in
the singlet extensions, the parameter {\tt isinglet} has to be set to
1. The model is chosen by setting the
input parameter {\tt icxSM} to 1 (RxSM-broken), 2 (RxSM-dark), 3
(CxSM-broken) or 4 (CxSM-dark). The input values {\tt COUPVAR, HIGGS, SM4,
  FERMPHOB} and {\tt ON-SH-WZ} have to be set to zero. Even if the user
does not set them to zero, by choosing {\tt isinglet} to be 1
they are set internally to zero in {\tt shdecay.f}. The input
parameters of the models that can be chosen in {\tt shdecay.in} are
the same as the ones specified in section \ref{sec:CxSM} for the CxSM
and in \ref{sec:RxSM} for the RxSM. 
Only $v$ is not an input value but obtained internally from the Fermi constant.

All files necessary for the program can be downloaded at the url: \\
\centerline{\tt http://www.itp.kit.edu/$\sim$maggie/sHDECAY} \\
The tar file provided at the webpage contains, besides the main routine
{\tt shdecay.f} and the input file {\tt shdecay.in}, several help
files, stemming already from the original code {\tt HDECAY}, and a
{\tt makefile} for compilation. The program is compiled with the file
{\tt makefile} by typing {\it make}, which provides an executable file
called {\it run}. Typing {\it run} executes the program and all
computed branching ratios for the various Higgs bosons are written out
together with their masses and total widths. For the CxSM-broken the
output files are called {\tt br.cbij} with {\tt i}=1,2,3 denoting one of
the three Higgs bosons and {\tt j} counting the three output files for
each of the Higgs bosons. The first one contains the
fermionic decays, the second the decays into gauge bosons and the total
width, and the third one the Higgs-to-Higgs decays. The corresponding
files for CxSM-dark are called {\tt br.cdkj} ({\tt j}=1,2,3) for the
two visible Higgs bosons {\tt k}=1,2. The output files for RxSM-broken
are named {\tt br.rbkj} with {\tt k}=1,2 for the two Higgs states and {\tt j}=1,2,3. In
RxSM-dark we have the output files {\tt br.rd1j} ({\tt j}=1,2,3) for
one visible Higgs state. Besides these files all input 
parameters are written out in a file called {\tt br.input}. On the
webpage, sample output files can be found for a given input. Updates
will be made available there as well. 

In the following we present an example output as obtained from the
following parameters for the singlet model in the input file and for
all the other parameters set at their standard values: 

\vspace*{0.2cm}
{\tt ********************** real or complex singlet Model ********************* 

Singlet Extension: 1 - yes, 0 - no (based on SM w/o EW corrections)

Model: 1 - real broken phase, 2 - real dark matter phase

\mbox{$\phantom{...............}$}3 - complex broken phase, 4 - complex dark matter phase

isinglet = 1

icxSM    = 3

*** real singlet broken phase ***

alph1    = -0.118574

mH1      = 125.1D0

mH2      = 306.361D0

vs       = 293.222D0

*** real singlet dark matter phase ***

mH1      = 125.1D0

mD       = 48.0215D0

m2s      = -463128.D0

lambdas  = 3.56328D0

*** complex singlet broken phase ***

alph1    = 0.160424D0

alph2    = -0.362128D0

alph3    = -0.552533D0

m1       = 125.518D0

m3       = 500.705D0

vs       = 510.922D0

*** complex singlet dark matter phase ***

alph1    = -0.317120D0

m1       = 125.3D0

m2       = 400.D0

m3       = 731.205D0

vs       = 522.181D0

a1       = -3.12115D07

}

\vspace*{0.2cm}
\noindent The produced output in the three files {\tt br.cb3j} for the heaviest
Higgs boson is given by 

\vspace*{0.2cm}
\begin{verbatim}
  MH3        BB       TAU TAU      MU MU         SS         CC        TT
---------------------------------------------------------------------------
500.705  0.7118E-04  0.9895E-05  0.3499E-07  0.2683E-07  0.3483E-05  0.1224

  MH3        GG        GAM GAM     Z GAM         WW         ZZ       WIDTH
---------------------------------------------------------------------------
500.705  0.3872E-03  0.1698E-06  0.4911E-05    0.3299     0.1576     4.794

   MH3      H1 H1       H1 H2       H2 H2 
-------------------------------------------
500.705    0.2175     0.7037E-04    0.1721    
\end{verbatim}

%%%%%%%%%%%%%%%%%%%%%%%%%%%%%%%%%%%%%%%%%%%%%%%%%%%%%%%%%%%
\section{Feynman Rules for the Triple Higgs Vertices}
\label{sec:app_FeynmanRules}

The widths for the decays $h_i \to h_j + h_j$ and $h_i \to h_j + h_k$
($k\ne j$) are given by 
\begin{equation} \label{eq:wid1}
\Gamma\left(h_i\rightarrow h_j h_j\right) = \dfrac{g^2_{ijj}}{32\pi
  m_{i}}\sqrt{1-\dfrac{4m_{j}^2}{m_{i}^2}}\; , 
\end{equation}
and 
\begin{equation} \label{eq:wid2}
\Gamma\left(h_i\rightarrow h_j h_k\right) = \dfrac{g^2_{ijk}}{16\pi m_{i}}
\sqrt{1-\dfrac{(m_{j}+m_k)^2}{m_{i}^2}}\, \sqrt{1-\dfrac{(m_{j}- m_k)^2}{m_{i}^2}}\; ,
\end{equation}
where $g_{ijk}$ is the coupling between the scalars $i,j,k$ defined in Eq.~\eqref{eq:cubic_norm} and $m_{j}$
is the mass of the scalar state $h_j$.\footnote{Note that in
  reference~\cite{Coimbra:2013qq} there is an extra 1/2 factor in  
Eqs.~\eqref{eq:wid1} and \eqref{eq:wid2}, due to a different convention
in $g_{ijk}$.}
%%%%%%%%%%%%%%%%%%%%%%%%%%%%%%%%%%%%%%%%%%%%%%%%%%%%%%%%%%%
\subsection{CxSM \label{sec:selfCxSM}}
The triple scalar vertices are shown below for the
broken phase, assuming the normalisation of
Eq.~\eqref{eq:cubic_norm}. Note that the $g_{ijk}$ are completely
symmetric under interchange of indices.
{\small
\begin{eqnarray}
g_{111} &=& \frac{3}{2} \Big[ d_2  \left( R_{13}^2+ R_{12}^2\right) ( R_{13}  v_A + R_{12}  v_S )+ \delta_2   R_{11} v \left( R_{13}^2+ R_{12}^2\right)+ \delta_2   R_{11}^2 ( R_{13}  v_A + R_{12}  v_S )+\lambda   R_{11}^3 v\Big] \nonumber \\
g_{112} &=& \frac{1}{2} \Big[ R_{13}^2 (3  d_2   R_{23}  v_A + d_2   R_{22}  v_S + \delta_2   R_{21} v)+ R_{12}^2 ( d_2   R_{23}  v_A +3  d_2   R_{22}  v_S + \delta_2   R_{21} v)+\nonumber \\
&&+ R_{11}^2 ( \delta_2  ( R_{23}  v_A + R_{22}  v_S )+3 \lambda   R_{21} v)+2  \delta_2   R_{11} ( R_{13}  R_{23} v+ R_{13}  R_{21}  v_A + R_{12}  R_{21}  v_S + R_{12}  R_{22} v) \nonumber \\ 
&&+2  d_2   R_{13}  R_{12} ( R_{23}  v_S + R_{22}  v_A )\Big]\nonumber
\end{eqnarray}
\begin{eqnarray}
g_{113} &=& \frac{1}{2} \Big[ R_{13}^2 (3  d_2   R_{33}  v_A + d_2   R_{32}  v_S + \delta_2   R_{31} v)+ R_{12}^2 ( d_2   R_{33}  v_A +3  d_2   R_{32}  v_S + \delta_2   R_{31} v)+\nonumber \\
&&+ R_{11}^2 ( \delta_2  ( R_{33}  v_A + R_{32}  v_S )+3 \lambda   R_{31} v)+2  \delta_2   R_{11} ( R_{13}  R_{33} v+ R_{13}  R_{31}  v_A + R_{12}  R_{31}  v_S + R_{12}  R_{32} v)\nonumber \\
&&+2  d_2   R_{13}  R_{12} ( R_{33}  v_S + R_{32}  v_A )\Big] \nonumber\\
g_{122} &=& \frac{1}{2} \Big[ R_{13} \left(3  d_2   R_{23}^2  v_A +2  d_2   R_{23}  R_{22}  v_S + d_2   R_{22}^2  v_A +2  \delta_2   R_{23}  R_{21} v+ \delta_2   R_{21}^2  v_A \right)+ \nonumber \\
&&+R_{12} \left( d_2  \left( R_{23}^2  v_S +2  R_{23}  R_{22}  v_A +3  R_{22}^2  v_S \right)+ \delta_2   R_{21}^2  v_S +2  \delta_2   R_{21}  R_{22} v\right)+ \nonumber \\
&&+R_{11} \left( \delta_2  v \left( R_{23}^2+ R_{22}^2\right)+2  \delta_2   R_{21} ( R_{23}  v_A + R_{22}  v_S )+3 \lambda   R_{21}^2 v\right)\Big] \nonumber \\
g_{123} &=& \frac{1}{2} \Big[ R_{13} ( R_{23} (3  d_2   R_{33}  v_A + d_2   R_{32}  v_S + \delta_2   R_{31} v)+ d_2   R_{22} ( R_{33}  v_S + R_{32}  v_A )+ \delta_2   R_{21} ( R_{33} v+ R_{31}  v_A )) \nonumber \\
&&+ R_{12} ( R_{22} ( d_2  ( R_{33}  v_A +3  R_{32}  v_S )+ \delta_2   R_{31} v)+ d_2   R_{23} ( R_{33}  v_S + R_{32}  v_A )+ \delta_2   R_{21} ( R_{31}  v_S + R_{32} v))+ \nonumber \\
&&+R_{11} ( R_{21} ( \delta_2  ( R_{33}  v_A + R_{32}  v_S )+3 \lambda   R_{31} v)+ \delta_2  ( R_{23}  R_{33} v+ R_{23}  R_{31}  v_A + R_{22}  R_{31}  v_S + R_{22}  R_{32} v))\Big] \nonumber \\
g_{133} &=& \frac{1}{2} \Big[ R_{13} \left(3  d_2   R_{33}^2  v_A +2  d_2   R_{33}  R_{32}  v_S + d_2   R_{32}^2  v_A +2  \delta_2   R_{33}  R_{31} v+ \delta_2   R_{31}^2  v_A \right)+ \nonumber \\
&&+R_{12} \left( d_2  \left( R_{33}^2  v_S +2  R_{33}  R_{32}  v_A +3  R_{32}^2  v_S \right)+ \delta_2   R_{31}^2  v_S +2  \delta_2   R_{31}  R_{32} v\right)+\nonumber \\
&&+ R_{11} \left( \delta_2  v \left( R_{33}^2+ R_{32}^2\right)+2  \delta_2   R_{31} ( R_{33}  v_A + R_{32}  v_S )+3 \lambda   R_{31}^2 v\right)\Big] \nonumber \\
g_{222} &=& \frac{3}{2} \Big[ d_2  \left( R_{23}^2+ R_{22}^2\right) ( R_{23}  v_A + R_{22}  v_S )+ \delta_2   R_{21} v \left( R_{23}^2+ R_{22}^2\right)+ \delta_2   R_{21}^2 ( R_{23}  v_A + R_{22}  v_S )+\lambda   R_{21}^3 v\Big] \nonumber \\ 
g_{223} &=& \frac{1}{2} \Big[ R_{23}^2 (3  d_2   R_{33}  v_A + d_2   R_{32}  v_S + \delta_2   R_{31} v)+ R_{22}^2 ( d_2   R_{33}  v_A +3  d_2   R_{32}  v_S + \delta_2   R_{31} v)+\nonumber \\
&&+ 2  d_2   R_{23}  R_{22} ( R_{33}  v_S + R_{32}  v_A )+ R_{21}^2 ( \delta_2  ( R_{33}  v_A + R_{32}  v_S )+3 \lambda   R_{31} v)+\nonumber \\
&&2  \delta_2   R_{21} ( R_{23}  R_{33} v+ R_{23}  R_{31}  v_A + R_{22}  R_{31}  v_S + R_{22}  R_{32} v)\Big] \nonumber \\ 
g_{233} &=& \frac{1}{2} \Big[ R_{23} \left(3  d_2   R_{33}^2  v_A +2  d_2   R_{33}  R_{32}  v_S + d_2   R_{32}^2  v_A +2  \delta_2   R_{33}  R_{31} v+ \delta_2   R_{31}^2  v_A \right)+ \nonumber \\
&&+R_{22} \left( d_2  \left( R_{33}^2  v_S +2  R_{33}  R_{32}  v_A +3  R_{32}^2  v_S \right)+ \delta_2   R_{31}^2  v_S +2  \delta_2   R_{31}  R_{32} v\right)+\nonumber \\
&&+ R_{21} \left( \delta_2  v \left( R_{33}^2+ R_{32}^2\right)+2  \delta_2   R_{31} ( R_{33}  v_A + R_{32}  v_S )+3 \lambda   R_{31}^2 v\right)\Big] \nonumber \\
g_{333} &=& \frac{3}{2} \Big[ d_2  \left( R_{33}^2+ R_{32}^2\right) ( R_{33}  v_A + R_{32}  v_S )+ \delta_2   R_{31} v \left( R_{33}^2+ R_{32}^2\right)+ \delta_2   R_{31}^2 ( R_{33}  v_A + R_{32}  v_S )+\lambda   R_{31}^3 v\Big]\; . \nonumber \\
&&
\end{eqnarray}
} Similarly we can define the quartic couplings
\begin{equation}\label{eq:quartic_norm}
V_{H_{\rm cubic}}=\frac{1}{4!}g_{ijkl}h_{i}h_{j}h_{k}h_{l} \; .
\end{equation}
Then for the broken phase one obtains:
{\small
\begin{eqnarray}
g_{1111} &=&\frac{3}{2} \Big[ d_2  \left( R_{13}^2+ R_{12}^2\right)^2+2  \delta_2   R_{11}^2 \left( R_{13}^2+ R_{12}^2\right)+\lambda   R_{11}^4\Big] \nonumber \\ 
g_{1112} &=&\frac{3}{2} \Big[ d_2  \left( R_{13}^2+ R_{12}^2\right) ( R_{13}  R_{23}+ R_{12}  R_{22})+ \delta_2   R_{11}  R_{21} \left( R_{13}^2+ R_{12}^2\right)+\nonumber \\
&&+ \delta_2   R_{11}^2 ( R_{13}  R_{23}+ R_{12}  R_{22})+\lambda   R_{11}^3  R_{21}\Big] \nonumber \\ 
g_{1113} &=&\frac{3}{2} \Big[  d_2  \left( R_{13}^2+ R_{12}^2\right) ( R_{13}  R_{33}+ R_{12}  R_{32})+ \delta_2   R_{11}  R_{31} \left( R_{13}^2+ R_{12}^2\right)+\nonumber \\
&&+ \delta_2   R_{11}^2 ( R_{13}  R_{33}+ R_{12}  R_{32})+\lambda   R_{11}^3  R_{31}\Big] \nonumber\\
g_{1122} &=&\frac{1}{2} \Big[  R_{13}^2 \left( d_2  \left(3  R_{23}^2+ R_{22}^2\right)+ \delta_2   R_{21}^2\right)+ R_{12}^2 \left( d_2  \left( R_{23}^2+3  R_{22}^2\right)+ \delta_2   R_{21}^2\right)+4  d_2   R_{13}  R_{12}  R_{23}  R_{22}+\nonumber \\
&&+ R_{11}^2 \left( \delta_2  \left( R_{23}^2+ R_{22}^2\right)+3 \lambda   R_{21}^2\right)+4  \delta_2   R_{11}  R_{21} ( R_{13}  R_{23}+ R_{12}  R_{22})\Big]\nonumber
\end{eqnarray}
\begin{eqnarray} 
g_{1123} &=&\frac{1}{2} \Big[  R_{13}^2 (3  d_2   R_{23}  R_{33}+ d_2   R_{22}  R_{32}+ \delta_2   R_{21}  R_{31})+ R_{12}^2 ( d_2   R_{23}  R_{33}+3  d_2   R_{22}  R_{32}+ \delta_2   R_{21}  R_{31})+\nonumber\\
&&+2  d_2   R_{13}  R_{12} ( R_{23}  R_{32}+ R_{22}  R_{33})+ R_{11}^2 ( \delta_2  ( R_{23}  R_{33}+ R_{22}  R_{32})+3 \lambda   R_{21}  R_{31})+\nonumber \\
&&+2  \delta_2   R_{11} ( R_{13}  R_{23}  R_{31}+ R_{13}  R_{21}  R_{33}+ R_{12}  R_{21}  R_{32}+ R_{12}  R_{22}  R_{31})\Big]\nonumber\\
g_{1133} &=&\frac{1}{2} \Big[  R_{13}^2 \left( d_2  \left(3  R_{33}^2+ R_{32}^2\right)+ \delta_2   R_{31}^2\right)+ R_{12}^2 \left( d_2  \left( R_{33}^2+3  R_{32}^2\right)+ \delta_2   R_{31}^2\right)+4  d_2   R_{13}  R_{12}  R_{33}  R_{32}+ \nonumber \\
&&+R_{11}^2 \left( \delta_2  \left( R_{33}^2+ R_{32}^2\right)+3 \lambda   R_{31}^2\right)+4  \delta_2   R_{11}  R_{31} ( R_{13}  R_{33}+ R_{12}  R_{32})\Big] \nonumber \\ 
g_{1222} &=&\frac{3}{2} \Big[ ( R_{13}  R_{23}+ R_{12}  R_{22}) \left( d_2  \left( R_{23}^2+ R_{22}^2\right)+ \delta_2   R_{21}^2\right)+ R_{11}  R_{21} \left( \delta_2  \left( R_{23}^2+ R_{22}^2\right)+\lambda   R_{21}^2\right)\Big] \nonumber \\ 
g_{1223} &=&\frac{1}{2} \Big[  R_{13} \left( d_2  \left(3  R_{23}^2  R_{33}+2  R_{23}  R_{22}  R_{32}+ R_{22}^2  R_{33}\right)+2  \delta_2   R_{23}  R_{21}  R_{31}+ \delta_2   R_{21}^2  R_{33}\right)+\nonumber \\
&&+ R_{12} \left( d_2  \left( R_{23}^2  R_{32}+2  R_{23}  R_{22}  R_{33}+3  R_{22}^2  R_{32}\right)+ \delta_2   R_{21}^2  R_{32}+2  \delta_2   R_{21}  R_{22}  R_{31}\right)+ \nonumber \\
&&+R_{11} \left( \delta_2   R_{31} \left( R_{23}^2+ R_{22}^2\right)+2  \delta_2   R_{21} ( R_{23}  R_{33}+ R_{22}  R_{32})+3 \lambda   R_{21}^2  R_{31}\right)\Big] \nonumber \\ 
g_{1233} &=&\frac{1}{2} \Big[  R_{13} \left( R_{23} \left(3  d_2   R_{33}^2+ d_2   R_{32}^2+ \delta_2   R_{31}^2\right)+2  d_2   R_{22}  R_{33}  R_{32}+2  \delta_2   R_{21}  R_{33}  R_{31}\right)+\nonumber \\
&&+ R_{12} \left(2  R_{32} ( d_2   R_{23}  R_{33}+ \delta_2   R_{21}  R_{31})+ R_{22} \left( d_2  \left( R_{33}^2+3  R_{32}^2\right)+ \delta_2   R_{31}^2\right)\right)+ \nonumber \\
&&+R_{11} \left( R_{21} \left( \delta_2  \left( R_{33}^2+ R_{32}^2\right)+3 \lambda   R_{31}^2\right)+2  \delta_2   R_{31} ( R_{23}  R_{33}+ R_{22}  R_{32})\right)\Big] \nonumber \\ 
g_{1333} &=&\frac{3}{2} \Big[ ( R_{13}  R_{33}+ R_{12}  R_{32}) \left( d_2  \left( R_{33}^2+ R_{32}^2\right)+ \delta_2   R_{31}^2\right)+ R_{11}  R_{31} \left( \delta_2  \left( R_{33}^2+ R_{32}^2\right)+\lambda   R_{31}^2\right)\Big] \nonumber \\ 
g_{2222} &=&\frac{3}{2} \Big[  d_2  \left( R_{23}^2+ R_{22}^2\right)^2+2  \delta_2   R_{21}^2 \left( R_{23}^2+ R_{22}^2\right)+\lambda   R_{21}^4\Big] \nonumber \\ 
g_{2223} &=&\frac{3}{2} \Big[  d_2  \left( R_{23}^2+ R_{22}^2\right) ( R_{23}  R_{33}+ R_{22}  R_{32})+ \delta_2   R_{21}  R_{31} \left( R_{23}^2+ R_{22}^2\right)+ \nonumber \\
&&+\delta_2   R_{21}^2 ( R_{23}  R_{33}+ R_{22}  R_{32})+\lambda   R_{21}^3  R_{31}\Big] \nonumber \\ 
g_{2233} &=&\frac{1}{2} \Big[  R_{23}^2 \left( d_2  \left(3  R_{33}^2+ R_{32}^2\right)+ \delta_2   R_{31}^2\right)+ R_{22}^2 \left( d_2  \left( R_{33}^2+3  R_{32}^2\right)+ \delta_2   R_{31}^2\right)+4  d_2   R_{23}  R_{22}  R_{33}  R_{32}+ \nonumber \\
&&+R_{21}^2 \left( \delta_2  \left( R_{33}^2+ R_{32}^2\right)+3 \lambda   R_{31}^2\right)+4  \delta_2   R_{21}  R_{31} ( R_{23}  R_{33}+ R_{22}  R_{32})\Big] \nonumber \\ 
g_{2333} &=&\frac{3}{2} \Big[ ( R_{23}  R_{33}+ R_{22}  R_{32}) \left( d_2  \left( R_{33}^2+ R_{32}^2\right)+ \delta_2   R_{31}^2\right)+ R_{21}  R_{31} \left( \delta_2  \left( R_{33}^2+ R_{32}^2\right)+\lambda   R_{31}^2\right)\Big] \nonumber \\ 
g_{3333} &=&\frac{3}{2} \Big[  d_2  \left( R_{33}^2+ R_{32}^2\right)^2+2  \delta_2   R_{31}^2 \left( R_{33}^2+ R_{32}^2\right)+\lambda   R_{31}^4\Big]\; .
\end{eqnarray}\\}
These expressions are all valid for the dark matter phase by replacing $\alpha_2=\alpha_3=0$ in the mixing matrix elements and setting $v_A=0$. 

\subsection{RxSM \label{sec:selfRxSM}}
For this model, the relevant cubic couplings are
{\small
\begin{eqnarray}
g_{111} &=& \frac{3}{2} \lambda   R_{11}^3 v+3  \lambda_{HS}   R_{11}^2  R_{12}  v_S +3  \lambda_{HS}   R_{11}  R_{12}^2 v+ \lambda_S   R_{12}^3  v_S  \nonumber \\
g_{112} &=&  R_{11}^2 \left(\frac{3 \lambda   R_{21} v}{2}+ \lambda_{HS}   R_{22}  v_S \right)+ R_{12}^2 ( \lambda_{HS}   R_{21} v+ \lambda_S   R_{22}  v_S )+2  \lambda_{HS}   R_{11}  R_{12} ( R_{21}  v_S + R_{22} v) \nonumber\\ 
g_{122} &=&  R_{11} \left(\frac{3}{2} \lambda   R_{21}^2 v+2  \lambda_{HS}   R_{21}  R_{22}  v_S + \lambda_{HS}   R_{22}^2 v\right)+ R_{12} \left( \lambda_{HS}   R_{21}^2  v_S +2  \lambda_{HS}   R_{21}  R_{22} v+ \lambda_S   R_{22}^2  v_S \right) \nonumber\\ 
g_{222} &=& \frac{3}{2} \lambda   R_{21}^3 v+3  \lambda_{HS}   R_{21}^2  R_{22}  v_S +3  \lambda_{HS}   R_{21}  R_{22}^2 v+ \lambda_S   R_{22}^3  v_S\; .
\end{eqnarray}\\}
The quartic couplings are
{\small
\begin{eqnarray}
g_{1111} &=&\frac{3 \lambda   R_{11}^4}{2}+6  \lambda_{HS}   R_{11}^2  R_{12}^2+ \lambda_S   R_{12}^4 \nonumber\\ 
g_{1112} &=&\frac{3}{2} \lambda   R_{11}^3  R_{21}+3  \lambda_{HS}   R_{11}^2  R_{12}  R_{22}+3  \lambda_{HS}   R_{11}  R_{12}^2  R_{21}+ \lambda_S   R_{12}^3  R_{22} \nonumber\\
g_{1122} &=& R_{11}^2 \left(\frac{3 \lambda   R_{21}^2}{2}+ \lambda_{HS}   R_{22}^2\right)+ R_{12}^2 \left( \lambda_{HS}   R_{21}^2+ \lambda_S   R_{22}^2\right)+4  \lambda_{HS}   R_{11}  R_{12}  R_{21}  R_{22} \nonumber\\ 
g_{1222} &=&\frac{3}{2} \lambda   R_{11}  R_{21}^3+3  \lambda_{HS}   R_{11}  R_{21}  R_{22}^2+3  \lambda_{HS}   R_{12}  R_{21}^2  R_{22}+ \lambda_S   R_{12}  R_{22}^3 \nonumber\\ 
g_{2222} &=&\frac{3 \lambda   R_{21}^4}{2}+6  \lambda_{HS}   R_{21}^2  R_{22}^2+ \lambda_S   R_{22}^4\; .
\end{eqnarray}\\}
The dark phase is obtained by setting $\alpha=0$ and $v_S=0$.
\end{appendix}

\bibliographystyle{unsrt}
\bibliography{refbench.bib}   % name your BibTeX data base
\end{document}